\documentclass[aps, prd,reprint,  amsfonts,
  amssymb, amsmath, showpacs, preprintnumbers, letterpaper,
  nofootinbib,]{revtex4-1}

\usepackage{array}

\usepackage{cancel}
\usepackage{braket,bm}
\usepackage[pass]{geometry} 
\usepackage{color}
\usepackage{mathrsfs,mathtools}
\usepackage{graphicx}
\usepackage{bbm}
\usepackage{enumerate}
  \DeclareMathAlphabet{\mathpzc}{OT1}{pzc}{m}{it}
\usepackage{subfigure}
  
\newcommand{\D}{\mathrm{d}}

\begin{document}

%%%%%%%%%%%%%%%%%%%%%%%%%%%%%%%%%%%%%%%%%%%%%%%%%%%%%%%%%%%%%%%%%%%%%%

\title{Radiative effects on 
  false vacuum decay in Higgs-Yukawa theory}
\author{Wen-Yuan Ai}
\email{ai.wenyuan@tum.de}

\author{Bj\"{o}rn Garbrecht}
\email{garbrecht@tum.de}

\affiliation{Physik Department T70, James-Franck-Stra\ss e,\\
Technische Universit\"{a}t M\"{u}nchen, 85748 Garching, Germany}

\author{Peter Millington}
\email{p.millington@nottingham.ac.uk}

\affiliation{School of Physics and Astronomy,\\ University of Nottingham,
Nottingham NG7 2RD, United Kingdom}

\pacs{03.70.+k, 11.10.-z, 66.35.+a}
% quantum field theory (first two), quantum tunneling of defects

\preprint{TUM-HEP-1151-18}

%%%%%%%%%%%%%%%%%%%%%%%%%%%%%%%%%%%%%%%%%%%%%%%%%%%%%%%%%%%%%%%%%%%%%%
\begin{abstract}

We derive fermionic Green's functions in the background of the Euclidean solitons describing false vacuum decay in a prototypal Higgs-Yukawa theory. In combination with
appropriate counterterms for the masses, couplings and wave-function normalization, these can be
used
to calculate radiative corrections to the soliton solutions and transition rates that fully account for the inhomogeneous background provided by the nucleated bubble.
We apply this approach to the archetypal example of transitions between the quasi-degenerate
vacua of a massive scalar field with a quartic self-interaction. The effect of fermion loops is compared
with those from additional scalar fields, and the
loop effects accounting for the spacetime inhomogeneity of the tunneling configuration are compared with those where gradients
are neglected. We find that scalar loops lead to an enhancement of the decay rate, whereas fermion loops lead to a suppression.
These effects get relatively amplified by a perturbatively small factor when gradients are
accounted for.
In addition, we observe that
the radiative corrections to the solitonic field profiles are smoother when the gradients are included. The method presented here for computing fermionic radiative corrections
should be applicable beyond the archetypal example of vacuum decay. In particular,
we work out methods that are suitable for calculations in the thin-wall limit,
as well as others that take
account of the full spherical symmetry of the solution. For the latter case, we construct the Green's functions based on spin hyperspherical harmonics, which are eigenfunctions
of the appropriate angular momentum operators that commute
with the Dirac operator in the solitonic background.

\end{abstract}

\maketitle

%%%%%%%%%%%%%%%%%%%%%%%%%%%%%%%%%%%%%%%%%%%%%%%%%%%%%%%%%%%%%%%%%%%%%%

\section{Introduction}
\label{sec:introduction}

The possible metastability of the electroweak vacuum is among the most important features of the Standard Model (SM) and may point to its embedding in the framework of more fundamental theories~\cite{Cabibbo:1979ay,Hung:1979dn,
Lindner:1985uk,Sher:1988mj,Sher:1993mf}. Due to the contributions from SM fermions to the beta function of the Higgs self-coupling, the latter turns negative at values of the Higgs field much larger than in the electroweak vacuum, generating a lower-lying minimum in the effective potential at high scale. The electroweak vacuum may then decay to this global minimum through quantum tunneling. Prior to the discovery of the Higgs boson, the beta function had been computed to next-to-leading order (NLO) accuracy~\cite{
Casas:1994qy,Casas:1996aq,Isidori:2001bm,
Burgess:2001tj,Isidori:2007vm,ArkaniHamed:2008ym,
Bezrukov:2009db,Hall:2009nd,Ellis:2009tp,
EliasMiro:2011aa,EliasMiro:2011aa}. The discovery of the 125~GeV Higgs boson~\cite{Aad:2012tfa,Chatrchyan:2012ufa} has motivated further assessments of the
metastability of the SM electroweak vacuum at next-to-next-to-leading order (NNLO)~\cite{Degrassi:2012ry,
Buttazzo:2013uya}. The Higgs mass of 125 GeV, together with the top-quark mass of 173 GeV~\cite{Lancaster:2011wr}, places the SM near criticality, with the Higgs having an almost vanishing quartic self-coupling at the Planck scale~\cite{Buttazzo:2013uya}. Nevertheless, the central values of these masses slightly favor the presence of an instability at around $10^{11}$~GeV, where the potential turns negative, and suggest a lifetime for the electroweak vacuum that is longer than the age of the universe, leading to the metastability scenario~\cite{Isidori:2001bm,DiLuzio:2015iua,Chigusa:2017dux,Andreassen:2017rzq}. This implies that the SM can be extrapolated up to the Planck scale with no problem of consistency in principle.

The dominant source of experimental uncertainty arises from the measurement of the top-quark mass~\cite{Bezrukov:2012sa,Masina:2012tz}. 
Theoretically, the lifetime of the electroweak vacuum can also be very sensitive to higher-dimensional operators, originating from new physics at around the Planck scale.
In certain areas of parameter space, this can dramatically reduce the lifetime of the electroweak vacuum, taking it below the current age of the universe, leading to strong constraints on physics beyond the SM~\cite{Branchina:2013jra,Branchina:2014usa,Branchina:2014rva,Branchina:2018xdh,Lalak:2014qua,
Eichhorn:2015kea}.

However, in comparison to the running couplings,
the radiative corrections to the actual tunneling problem~\cite{Langer:1967ax,Langer:1969bc,Kobzarev:1974cp,Coleman:1977py,
Callan:1977pt} (see also Ref.~\cite{Coleman:1988}) are known less accurately. In particular, in false vacuum decay, the so-called bounce provides an inhomogeneous background, corresponding to the profile of a nucleated critical bubble, and the beta functions for the coupling constants do
not account for the pertaining gradient effects. 
One-loop radiative corrections due to fluctuations about the bounce, which fully account for the inhomogeneity of the background, have been calculated
using the Gel'fand-Yaglom theorem~\cite{Gelfand:1959nq,Coleman:1988,Baacke:2003uw,Dunne:2005rt}. While the latter is a powerful way of evaluating functional determinants,
it has shortcomings: If the quantum-corrected bounce cannot be obtained by improving a classical solution via
perturbation theory, further elaboration is necessary. This problem applies to situations where
the scale of the nucleated
bubble depends on radiative effects, i.e.~as occurs in approximately scale invariant theories such as the SM~\cite{Chigusa:2017dux,Andreassen:2017rzq,Garbrecht:2018rqx}, but also
when the true vacuum only emerges radiatively in the first place through, e.g., the Coleman-Weinberg mechanism~\cite{Coleman:1973jx,Garbrecht:2015cla,Garbrecht:2015yza,Garbrecht:2017idb}. Furthermore, evaluating the functional determinant
using the Gel'fand-Yaglom theorem does not lead to a systematic method of computing higher-order
corrections.

These problems can be addressed by constructing Green's functions in the solitonic background, which can then be used to evaluate effective actions~\cite{Jackiw:1974cv,Cornwall:1974vz} and to construct self-consistent 
equations of motion in
perturbation theory or resummed variants thereof~\cite{Garbrecht:2015oea}. Compared to calculations
in homogeneous backgrounds, the reduced symmetry makes it harder, however, to advance calculations
to high orders. Nevertheless,
problems such as the perturbative
improvement of bounce solutions and decay rates~\cite{Garbrecht:2015oea,Bezuglov:2018qpq}, and
finding bounces in radiatively generated~\cite{Garbrecht:2015yza} or classically
scale invariant potentials~\cite{Garbrecht:2018rqx} can be addressed this way. So far, Green's functions and
loop effects have been calculated for scalar particles. In view of applications to well-motivated
scenarios, such as the SM and other situations where false vacuum decay may play a role~\cite{Witten:1984rs,Kosowsky:1991ua,Caprini:2009fx,Kuzmin:1985mm,Shaposhnikov:1987tw,Morrissey:2012db},  gauge and fermion loops should also be included. The fermionic case is
of particular relevance because of the pivotal role played by the top quark in electroweak symmetry breaking. Moreover, 
it corresponds to a simple example of a field with
spin in the background of a nucleated spherical bubble, and thus is a suitable setting to develop and test the corresponding methods.

In this article, we develop a method of calculating Green's functions for Dirac fermions coupling to spherically
symmetric bounce solutions, as well as to planar configurations. The latter limit allows the analysis to be simplified significantly in the case of quasi-degenerate vacua. As an application, we consider the archetypal model originally analyzed by Coleman and Callan~\cite{Coleman:1977py,Callan:1977pt} of tunneling between quasi-degenerate vacua in scalar field theory
and compute the leading radiative corrections to the bounce, as well as the tunneling action. Referring to the
tunneling degree of freedom as the Higgs field, these corrections
are induced by the self-interactions of the Higgs boson, as well as its Yukawa couplings to Dirac fermions and
dimensionless couplings to extra scalar fields, such that we can compare these different loop effects.

The paper is organized as follows. In Sec.~\ref{sec:falsevacdecay:radcor}, we review the calculation by Coleman and Callan~\cite{Coleman:1977py,Callan:1977pt} of the semi-classical tunneling rate and its first quantum corrections in $\phi^4$ theory, upon which we aim to add fermion loop corrections in this work. In particular, we compute the corrections
to the classical bounce solution and the decay rate by means of the Green's functions for the fluctuations about the tree-level bounce. In Sec.~\ref{sec:Planar-wall}, we present a simple way of computing the
Green's functions and functional determinants of the Dirac field in the planar-wall approximation that is applicable
to the thin-wall limit. A method of deriving fermionic Green's functions
for problems where the vacua are not quasi-degenerate, such that
the thin-wall approximation does not apply, is presented in App.~\ref{app:greenfunction}. In particular,
working in four-dimensional Euclidean space,
we construct these Green's functions based on the appropriate set of angular momentum eigenstates of
spinors on a three-dimensional hypersphere, such that the problem of finding the radial Green's function separates
from the angular problem, as is familiar from calculations in three space dimensions
with two-dimensional spherical symmetry. The planar-wall limit from Sec.~\ref{sec:Planar-wall} can be obtained
as a limiting case of this construction accounting for the full hyperspherical symmetry. Evaluating the loop corrections based on these Green's functions
leads to ultraviolet divergences. These need to be canceled by counterterms that must be chosen to
satisfy renormalization conditions on the Higgs mass, self-coupling and wave-function normalization, as we
work out in Sec.~\ref{sec:renormalization}. Based on these developments, in Sec.~\ref{sec:numericalresults}, we present numerical results, comparing the contributions from the Dirac and scalar fields to the effective action and the equation of motion
for the bounce. Our conclusions are stated in Sec.~\ref{sec:conc}.

%%%%%%%%%%%%%%%%%%%%%%%%%%%%%%%%%%%%%%%%%%%%%%%%%%%%%%%%%%%%%%%%%%%%%%

\section{Radiative corrections to false vacuum decay}
\label{sec:falsevacdecay:radcor}

In this section, we review the pertinent details of the calculation of the decay rate of a metastable vacuum state. The fundamentals for approaching this problem have been developed in the seminal works by Langer, Coleman and Callan~\cite{Langer:1967ax,Langer:1969bc,Coleman:1977py,Callan:1977pt,Coleman:1988}, up to the inclusion of one-loop corrections to the decay rate in perturbation theory. The calculation of scalar-field loops in the case of vacuum decay in field theory, and based on Green's functions evaluated in the inhomogeneous background of the tunneling soliton, has been introduced in Ref.~\cite{Garbrecht:2015oea} and carried out to two-loop order in the decay rate in Ref.~\cite{Bezuglov:2018qpq}. In the present work, we extend this methodology to consider the radiative corrections from fermion loops, making direct comparison with the corresponding effects from scalars. We focus, in particular, on the amplitude for the vacuum transition, as well as the deformations of the classical soliton, and present the main elements of the approach.

The formulation in terms of the two-particle-irreducible effective action is presented and applied in Refs.~\cite{Garbrecht:2015yza,Garbrecht:2017idb}. That approach is useful when the bounce solutions or even the vacua
are dominated by radiative effects in the first place, such as in Coleman-Weinberg~\cite{Coleman:1973jx} scenarios of symmetry breaking or in classically scale-invariant models~\cite{Garbrecht:2015yza,Garbrecht:2018rqx,Garbrecht:2017idb,Garbrecht:2015cla}. The main technical developments in that case apply to
the negative and zero modes of the symmetry-breaking scalar field. Since fermion fluctuations
are the main objective, we develop their computation here in a standard perturbation expansion. Based on this, the
generalization to effective-action techniques can be implemented straightforwardly, and this may be presented elsewhere.

\subsection{Prototypal Higgs-Yukawa model}

\begin{figure}

  \centering
  
  \includegraphics[scale=0.6]{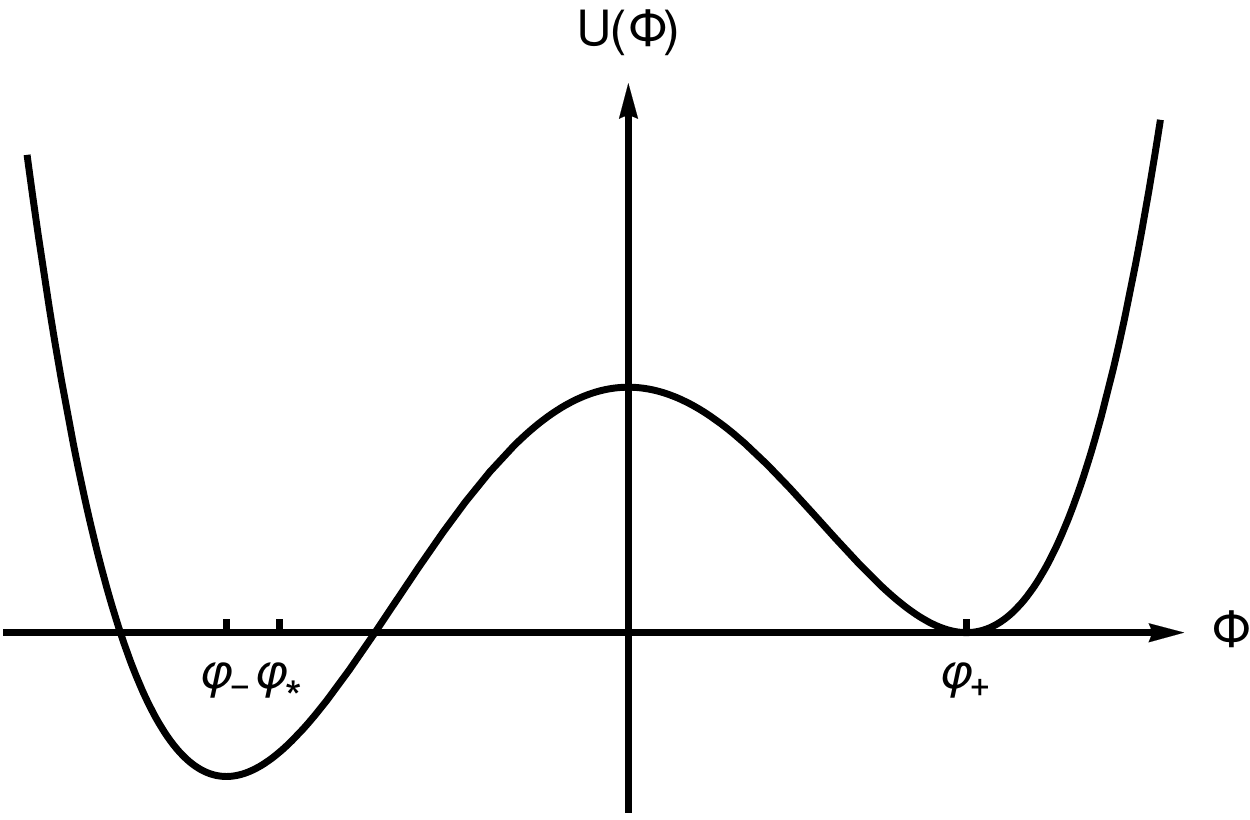}
  
  \caption{The classical  potential $U(\Phi)$ for for the archetypal scalar theory with false vacuum decay, described in Eqs.~\eqref{themodel} and~\eqref{classicalpotential}. \label{fig:potential}}

\end{figure}

We consider a  Higgs-Yukawa model based on the Euclidean Lagrangian
\begin{align}
\label{themodel}
\mathcal{L}\ =\ \bar{\Psi}\gamma_\mu\partial_\mu\Psi\:+\:\kappa\:\bar{\Psi}\Phi\Psi\:+\:\frac{1}{2}\,(\partial_\mu\Phi)^2\:+\:U(\Phi)\;,
\end{align}
where
\begin{align}
\label{classicalpotential}
U(\Phi)\ =\ -\:\frac{1}{2}\,\mu^2\,\Phi^2\:+\:\frac{1}{3!}\,g\,\Phi^3\:+\:\frac{1}{4!}\,\lambda\,\Phi^4\:+\:U_0
\end{align}
and $\kappa$ is the dimensionless Higgs-Yukawa coupling.
The Euclidean gamma matrices are obtained from their Minkowskian counterparts through
the replacement $\gamma^k\to  i\gamma_{k}$, for $k=1,2,3$, and $\gamma^0\to \gamma_4$.
For definiteness, we have chosen the same potential as
in the archetypal example of tunneling in field theory considered by Coleman and Callan~\cite{Coleman:1977py,
Callan:1977pt}, cf.~Fig.~\ref{fig:potential}.
We choose $\mu^2, \lambda, g>0$, such that there are false and true vacua located
at $\varphi_\pm\approx\pm\, v + {\cal O}(g/\sqrt\lambda)$, where
$v=\sqrt{6\mu^2/\lambda}$. The constant $U_0=(\mu v)^2/2-gv^3/3!$ is chosen such that the false vacuum has vanishing energy density.

\subsection{Leading-order bounce and tunneling rate}

The dominant contribution to the tunneling rate is due to the lowest-lying \emph{bounce}
solution $\varphi$ (and the fluctuations about it).
The (tree-level) bounce is a solution to the classical equation of motion
\begin{align}
\label{eommmmmmm}
-\,\partial^2\varphi\:+\:U'(\varphi)\ =\ 0
\end{align}
that satisfies the boundary conditions
$\varphi|_{x_4\rightarrow\pm\infty}=\varphi_+$ and $\dot{\varphi}|_{x_4=0}=0$, where the dot denotes the derivative with respect to $x_4$. The $'$ denotes the derivative of the classical potential in Eq.~\eqref{classicalpotential} with respect to the field $\varphi$. Notice that we are interested in a field configuration that starts \emph{and ends} (in Euclidean time) in the false vacuum --- hence the name \emph{bounce}. For the bounce action to be finite, we also require that $\varphi|_{|\bf{x}|\rightarrow\infty}=\varphi_+$. 
Given the anticipated $O(4)$ invariance of the bounce, it is convenient to work in four-dimensional hyperspherical coordinates, wherein the equation of motion takes the form
\begin{align}
\label{equationofmotion}
-\,\frac{\D^2\varphi}{\D r^2}\:-\:\frac{3}{r}\:\frac{\D\varphi}{\D r}\:+\:U'(\varphi)\ =\ 0\;,
\end{align}
with $r^2={\bf x}^2+x_4^2$. The boundary conditions become $\varphi|_{r\rightarrow \infty}=\varphi_+$. The solution must be regular at the origin, and we therefore require that $\D\varphi/\D r|_{r\,=\,0}=0$. The solution is a soliton that interpolates between the `escape-point' field value $\varphi_*$ (which lies close to the true vacuum $\varphi_-$) at the origin of Euclidean space and the false vacuum $\varphi_+$ at infinity, see Fig.~\ref{fig:potential}. Hence, the bounce describes a four-dimensional hyperspherical bubble nucleated within the false vacuum.

The Euclidean amplitude for transitions from the
false vacuum (at $x_4= -\,T/2$) to the false vacuum (at $x_4= +\,T/2$) is given in terms of the partition function (cf.~Eq.~\eqref{partitionfunctional})
\begin{align}
\label{partition}
Z[0,0,0]\ &=\ \langle \varphi_+| e^{-HT/\hbar}|\varphi_+\rangle \notag\\
&=\ \mathscr{N}\int\mathcal{D}\Phi\,\mathcal{D}\Psi\,\mathcal{D}\bar\Psi\;e^{-S[\Phi,\bar{\Psi},\Psi]/\hbar}\;,
\end{align}
where $H$ is the full Hamiltonian and $S[\Phi,\bar{\Psi},\Psi]$ is the Euclidean action.
The partition function is to be evaluated by expanding around the bounce solution,
and $\mathscr{N}$ is a normalization factor given by the inverse of the partition function evaluated in the false vacuum.
In terms of $Z[0,0,0]$, the decay rate is given by~\cite{Callan:1977pt}
 \begin{align}
\label{decayrate2}
\varGamma\ =\ 2\:|\textrm{Im}\,Z[0,0,0]|/T\;.
\end{align}

In the thin-wall approximation~\cite{Callan:1977pt} (see also Ref.~\cite{Brown:2017cca}), applicable when the minima are quasi-degenerate, i.e.~when the cubic coupling $g$ is very small, we may neglect the damping term in Eq.~\eqref{equationofmotion}, as well as the contribution from the cubic self-interaction $g\varphi^3$. In this approximation, Eq.~\eqref{equationofmotion} has the well-known kink solution~\cite{Dashen:1974ci}
\begin{align}
\label{kink}
\varphi^{(0)}(r)\ \equiv\ v\:\tanh\,[\gamma\,(r-R)]\equiv v\,u\;,
\end{align}
where $\gamma=\mu/\sqrt{2}$. Since we will later use $\varphi$ generally to denote the background scalar field variable, we add a superscript ``(0)"
to have it uniquely denote the classical bounce $\varphi^{(0)}$. The radius of the critical bubble $R$ is obtained by extremizing the bounce action 
\begin{align}
\label{bounceaction}
B\ \equiv\ S[\varphi^{(0)}]\ =\ \int {\rm d}^4x\:\bigg[\frac{1}{2}\:\bigg(\frac{\D\varphi^{(0)}}{\D r}\bigg)^{\!2}\:+\:U(\varphi^{(0)})\bigg]\;
\end{align}
with respect to $R$, which gives 
\begin{align}
\label{radius:bubble}
R\ =\ \frac{12\gamma}{gv}\;
\end{align}
and
\begin{align}
\label{B:tree}
B\ =\ 8\pi^2R^3\gamma^3/\lambda\;.
\end{align}
Note that this result for the bounce action contains contributions
from both the surface tension of the bubble and the latent heat released by the transition to the true vacuum.

\subsection{One-loop corrections to the action}

For non-vanishing external sources, the Euclidean partition function is given by
\begin{align}
\label{partitionfunctional}
&Z[J,\bar\eta,\eta]\  =\ \int\!\mathcal{D}\Phi\,\mathcal{D}\Psi\,\mathcal{D}\bar\Psi\;\exp\,\bigg\{-\,\frac{1}{\hbar}\,\bigg[S[\Phi,\bar\Psi,\Psi]\nonumber\\&\qquad -\:\int{\rm d}^4x\;\Big(J(x)\Phi(x)
\:+\:\bar\eta(x)\Psi(x)\:+\:\bar\Psi(x)\eta(x)\Big)\bigg]\bigg\}\;.
\end{align} 
It can be used to compute the one-point functions
\begin{subequations}
\begin{align}
\label{onepointphi}
\varphi(x)&\ =\ \langle\Omega|\Phi(x)|\Omega\rangle|_{J,\bar\eta,\eta} \ =\ +\:\hbar\:\frac{\delta\ln Z[J,\bar\eta,\eta]}{\delta J(x)}\;,\\
\label{onepointbarpsi}
\bar\psi(x)&\ =\ \langle\Omega|\bar{\Psi}(x)|\Omega\rangle|_{J,\bar\eta,\eta} \ =\ -\:\hbar\:\frac{\delta\ln Z[J,\bar\eta,\eta]}{\delta\eta(x)}\;,\\
\label{onepointpsi}
\psi(x)&\ =\ \langle\Omega|\Psi(x)|\Omega\rangle|_{J,\bar{\eta},\eta} \ =\ +\:\hbar\:\frac{\delta\ln Z[J,\bar\eta,\eta]}{\delta\bar\eta(x)}\,.
\end{align}
\end{subequations}

We proceed by expanding $\Phi(x)=\varphi^{}(x)+\hbar^{1/2}\hat{\Phi}^{}(x)$, $\bar\Psi(x)=\bar{\psi}^{}(x)+\hbar^{1/2}\hat{\bar{\Psi}}^{}(x)$ and $\Psi(x)=\psi^{}(x)+\hbar^{1/2}\hat{\Psi}^{}(x)$,\footnote{We work throughout in natural units; all factors of $\hbar$ are included only for bookkeeping purposes.} such
that the action $S[\Phi,\bar{\Psi},\Psi]$ is given to quadratic order by
\begin{align}
\label{expansionofS}
&S[\Phi,\bar{\Psi},\Psi]\ =\ S[\varphi^{},\bar{\psi}^{},\psi^{}]\:+\:\hbar^{1/2}\int \D^4x\,J(x)\,\hat{\Phi}(x)\notag\\
&\ \ \ \ \ +\ \hbar^{1/2}\int\D^4x\,\bar{\eta}(x)\hat{\Psi}(x)\:+\:\hbar^{1/2}\int\D^4x\,\hat{\bar{\Psi}}(x)\eta(x)\notag\\
&\ \ \ \ \ +\ \frac{\hbar}{2}\int\D^4x\,\D^4y\;  \hat{\Phi}(x)\,G^{-1}(\varphi^{},\bar{\psi}^{},\psi^{};x,y)\,\hat{\Phi}(y)\notag\\
&\ \ \ \ \ +\ \frac{\hbar}{2}\int\D^4x\,\D^4y\;  \hat{\bar{\Psi}}(x)\,D^{-1}(\varphi^{},\bar{\psi}^{},\psi^{};x,y)\,\hat{\Psi}(y)\;.
\end{align}
The linear terms in the above expressions have been chosen so as to cancel those appearing in the exponent of the partition function~\eqref{partitionfunctional}. Since the one-point functions $\bar{\psi}$, $\psi$ are necessarily vanishing, we denote $S[\varphi^{},\bar{\psi}^{}=0,\psi^{}=0]\equiv S[\varphi^{}]$, $G^{-1}(\varphi^{},\bar{\psi}^{}=0,\psi^{}=0;x,y)\equiv G^{-1}(\varphi^{};x,y)$ and, similarly, $D^{-1}(\varphi^{},\bar{\psi}^{}=0,\psi^{}=0;x,y)\equiv D^{-1}(\varphi^{};x,y)$.
Equation~\eqref{expansionofS} then defines the tree-level inverse Green's functions
\begin{subequations}
\label{operator:twopoint}
\begin{align}
\label{operator:KG:twopoint}
G^{-1}(\varphi^{}\,;x,y)\ &\equiv\ \frac{\delta^2S[\Phi,\bar\Psi,\Psi]}{\delta\Phi(x)\delta\Phi(y)}\bigg|_{\Phi\,=\,\varphi^{},\bar\Psi\,=\,0,\,\Psi\,=\,0}\nonumber\\ & =\ \delta^4(x-y)\big[-\,\partial_x^2\:+\:U''(\varphi^{}_x)\big]\;,
\\
\label{operator:Dirac:twopoint}
D^{-1}(\varphi^{}\,;x,y)\ &\equiv\ \frac{\delta^2S[\Phi,\bar\Psi,\Psi]}{\delta\Psi(x)\delta\bar\Psi(y)}\bigg|_{\Phi\,=\,\varphi^{},\bar\Psi\,=\,0,\,\Psi\,=\,0}\nonumber\\& =\ \delta^4(x-y)\big[\gamma_\mu\partial_\mu\:+\:\kappa\varphi^{}_x\big]\;,
\end{align}
\end{subequations}
with $\delta^4(x-y)$ being the four-dimensional Dirac delta function. The Klein-Gordon and Dirac operators in the background of $\varphi$ can then be obtained as
\begin{subequations}
\label{diffops}
\begin{align}
G^{-1}(\varphi^{}\,;x)\ &\equiv\ \int \D^4y\ G^{-1}(\varphi^{}\,;x,y)\;,\\
D^{-1}(\varphi^{}\,;x)\ &\equiv\ \int \D^4y\ D^{-1}(\varphi^{}\,;x,y)\;.
\end{align}
\end{subequations}

Substituting Eq.~\eqref{expansionofS} into Eq.~\eqref{partition}, we encounter functional determinants of the operators~\eqref{diffops} at quadratic order in the fluctuations. At the level of first quantum corrections, the operators~\eqref{diffops} need to be evaluated at the classical bounce $\varphi^{(0)}$.  
The spectrum of the Klein-Gordon operator $G^{-1}(\varphi^{(0)};x)$ requires special treatment~\cite{Callan:1977pt}. It contains four eigenmodes $\phi_\mu=B^{-1/2}\partial_\mu\varphi^{(0)}$ with zero eigenvalues. These are the Goldstone modes
resulting from the spontaneous breakdown of spacetime translational invariance.
The functional integral over the four zero eigenmodes is traded for an integral over the collective coordinates of the bounce, giving a factor~\cite{Gervais:1974dc}
\begin{align}
VT\left(\frac{B}{2\pi\hbar}\right)^2\;,
\end{align}
where $V$ is a three-space volume. In addition, there is one negative eigenvalue 
\begin{align}
\lambda_0\ =\ \frac{1}{B}\:\frac{\delta^2B}{\delta R^2}\ =\ -\:\frac{3}{R^2}\;,
\end{align} 
because the bounce actually corresponds to a \emph{maximum} of the Euclidean action with respect to dilatations of the critical bubble. Along the direction of the negative mode, the path integral can be evaluated by analytic continuation~\cite{Callan:1977pt,Coleman:1988}, leading to an overall factor of $-i|\lambda_0|^{-1/2}/2$ from the square root of the determinant. Putting everything together, we can therefore express the factor arising from the
Gaussian integrations as
\begin{align}
&\Big[\det G^{-1}(\varphi^{(0)})\Big]^{-\frac12}
\notag\\
\ &=\ -\:\frac{i}{2}\,|\lambda_0|^{-\frac12}\, VT \bigg(\frac{B}{2\pi\hbar}\bigg)^{\!2} \Big[ {\det}^{(5)} G^{-1}(\varphi^{(0)})\Big]^{-\frac12}\;,
\end{align}
where the superscript ``$(5)$'' indicates that the
five lowest eigenvalues discussed above are omitted.

We still need to normalize by the factor $\mathscr{N}$, that is we must divide by
the square root of the determinant of the Klein-Gordon operator evaluated at the homogeneous false vacuum $\varphi_+\approx v$~\cite{Callan:1977pt}.
For this purpose, we explicitly
extract the five lowest eigenvalues $4\gamma^2$ from the determinant of $G^{-1}(v;x)$ that are not canceled when building the quotient of the determinants~\cite{Konoplich:1987yd}; namely,
\begin{align}
\Big[\det G^{-1}(v)\Big]^{-\frac12}
\ =\ (2\gamma)^{-5} \Big[ {\det}^{(5)} G^{-1}(v)\Big]^{-\frac12}\;.
\end{align}

The logarithms of these determinants appear as additive corrections
to the classical action $B[\varphi^{(0)}]$ and, within the present approximations,
they can be interpreted as the one-loop contribution to the effective
action. Together with the corresponding fermion terms, we denote these as
\begin{subequations}
\label{B1:det}
\begin{align}
{\hbar}\, B_S^{(1)}\; \equiv\;{\hbar}\, B_S^{(1)}[\varphi^{(0)}]\ &\equiv \ \frac{{\hbar}}{2}\,\ln\,\frac{\det^{(5)} G^{-1}(\varphi^{(0)})}{\det^{(5)} G^{-1}(v)}\;,\\
\label{B1:Dirac}
{\hbar}\, B_D^{(1)}\; \equiv\;{\hbar}\, B_D^{(1)}[\varphi^{(0)}]\ &\equiv\ -\:{\hbar}\,\ln\,\frac{\det\,D^{-1}(\varphi^{(0)})}{\det\,D^{-1}(v)}\;,
\end{align}
\end{subequations}
where, and throughout this paper, the determinant of the Dirac operators is understood to be taken over both the coordinate and spinor spaces.
Here, and in the remainder of this work, we suppress the argument in $B^{(1)}_{S,D}$ when evaluated at the classical bounce, in accordance with what is implied for $B$.
We note that these expressions still need to be renormalized, which we will
carry out in Sec.~\ref{sec:renormalization}.

\subsection{One-loop corrections to the bounce}

Before we assemble the preceding results into an expression for the decay rate,
we consider, in addition, the first quantum corrections to the bounce. To this end, we write
$\varphi=\varphi^{(0)}+\hbar\,\delta\varphi$~and expand Eq.~\eqref{onepointphi} to next-to-leading order. We then obtain
\begin{align}
\label{correctedbounce}
\delta\varphi(x)\ &=\ -\,\int{\rm d}^4 y\; G(\varphi^{(0)};x,y)\, \Pi_S(\varphi^{(0)};y)\,\varphi^{(0)}(y)\,\notag\\
 &\ \ \ \ \,-\,\int{\rm d}^4 y\; G(\varphi^{(0)};x,y)\, \Pi_D(\varphi^{(0)};y)\,\varphi^{(0)}(y)\notag\\
 &\equiv\ -\,\int{\rm d}^4 y\; G(\varphi^{(0)};x,y)\, \Pi(\varphi^{(0)};y)\,\varphi^{(0)}(y)\;,
\end{align}
where we define the tadpole functions from scalar and fermion loops as
\begin{subequations}
\label{tadpoles}
\begin{align}
 \Pi_S(\varphi^{(0)};x)\,\varphi^{(0)}(x)\
 &=\ \frac{\lambda}{2}\:G(\varphi^{(0)};x,x)\,\varphi^{(0)}(x)\;,
\\
\label{tadpole:fermi}
\Pi_D(\varphi^{(0)};x)\,\varphi^{(0)}(x)
\ &=\ -\,\kappa\:\mbox{tr}_{\mbox{s}}\,
D(\varphi^{(0)};x,x)\;,
\end{align}
\end{subequations}
in which $\rm{tr}_{\mbox{s}}$ indicates the trace over the (suppressed) spinor indices.

Alternatively, rather than the expansion of
Eq.~\eqref{onepointphi} based on the partition function,
the one-loop tadpole functions also descend from the
functional determinants by functional differentiation; specifically,
\begin{align}
\label{Pinewform}
\Pi(\varphi^{(0)};x)\,\varphi^{(0)}(x) \ &=\ \left.\frac{{\delta} B_{S}^{(1)}[\varphi]}{{\delta}\varphi(x)}\right|_{\varphi^{(0)}}\:+\:\left.\frac{{\delta} B_{D}^{(1)}[\varphi]}{{\delta}\varphi(x)}\right|_{\varphi^{(0)}}\;.
\end{align}
We note that fermion loops at coincident points do not appear in diagrammatic expansions
of Green's functions in Yukawa theory but they \emph{do} appear in the expansion of the one-point function, cf.~Eq.~\eqref{tadpole:fermi}. To clarify how the coincident fermion propagator emerges in the tadpole function, we carry out the variation of the one fermion-loop effective action
in App.~\ref{app:fermitadpole} explicitly.
At one-loop accuracy, the bounce satisfies\footnote{This equation is exact when the tadpole self-energies are derived from the two-particle-irreducible effective action.}
\begin{align}
\label{eom:bounce:corrected}
-\:\partial^2\varphi(x)\:+\:U^\prime(\varphi(x))\:+\:
\hbar\,\Pi(\varphi^{(0)},x)\,\varphi^{(0)}(x)\ =\ 0\;,
\end{align}
i.e.~the tadpoles can be interpreted as corrections to the equation of motion.

When substituting the quantum-corrected bounce into the action $S[\varphi]$
and the one-loop terms \smash{$B^{(1)}_{S,D}[\varphi]$}, there appear extra contributions at order $\hbar^2$.
Expanding first the action $S[\varphi]$ about $\varphi^{(0)}$, we have
\begin{align}
S[\varphi]\ =\ S[\varphi^{(0)}]\:+\:\hbar^2 \delta S\:+\:\mathcal{O}(\hbar^3)\;,
\end{align}
where
\begin{align}
\label{deltaS}
\,\delta S \ &=\ \frac 12 \int \D^4x\; \,\delta\varphi(x)\,G^{-1}(\varphi^{(0)};x)\,\delta\varphi(x)\notag\\
&=\ -\,\frac{1}{2}\,\int \D^4x\;\delta\varphi(x)\,\Pi(\varphi^{(0)};x)\,\varphi^{(0)}(x)\;{,}
\end{align}
and where, for the first identity, we have used the fact that the classical bounce $\varphi^{(0)}$ is the stationary point of the classical action and, for the second identity, we have used Eq.~\eqref{correctedbounce}.
Expanding
\begin{align}
\hbar\, B_{S,D}^{(1)}[\varphi]\ =\ \hbar\, B_{S,D}^{(1)}\:+\:\hbar^2\, \delta B_{S,D}^{(1)}\;,
\end{align}
we obtain from Eq.~\eqref{B1:det}
\begin{subequations}
\begin{align}
\,\delta B^{(1)}_S\ &=\ {\frac{1}{2}\,\int \D^4x\,\delta\varphi(x)\,\frac{\delta}{\delta\varphi(x)}\,\ln\,\mathrm{det}^{(5)}
G^{-1}(\varphi)\bigg|_{\varphi^{(0)}}\;,}\\
\,\delta B^{(1)}_D\ &=\ -{\int\D^4x\,\delta\varphi(x)\,\frac{\delta}{\delta\varphi(x)}\,\ln\det D^{-1}(\varphi)\bigg|_{\varphi^{(0)}}\;.}
\end{align}
\end{subequations}
Comparing with Eq.~\eqref{deltaS} and using Eq.~\eqref{Pinewform}, we see that
the total correction to the expansion in $\delta\varphi$ is~\cite{Garbrecht:2015oea}
\begin{align}
\label{B2}
\,B^{(2)}\ &\equiv\  \delta S\:+\:{\delta B^{(1)}_S\:+\: \delta B^{(1)}_D}\nonumber\\ &=\ -\:\delta S\nonumber\\& =\ \frac{1}{2}\,\Big(\delta B^{(1)}_S\:+\:\delta B^{(1)}_D\Big)\;.
\end{align}

Diagrammatically, the contributions to $B^{(2)}$ correspond to (one-particle reducible) dumbbell graphs, cf.~Fig.~\ref{fig:feynman}~(b)
for the fermion contributions. They therefore constitute a subset of the two-loop corrections.
However, this subset
can be the dominant two-loop contribution for theories with a large number of degrees of freedom propagating in the loop~\cite{Garbrecht:2015oea}.
For the numerical examples in Sec.~\ref{sec:numericalresults}, we consider such a setup with a
large number of fermion and scalar fields coupling to the Higgs degree of freedom $\Phi$.

\subsection{Radiatively corrected decay rate}

We can now summarize all the contributions that we
include in the approximation of the tunneling rate per unit volume as
\begin{align}
\label{decayrate}
 \varGamma/V&\ =\ \bigg(\frac{B}{2\pi\hbar}\bigg)^{\!2}(2\gamma)^5\,|\lambda_0|^{-\frac{1}{2}}\notag\\
 &\times\exp\bigg[-\,\frac{1}{\hbar}\,\Big(B\,+\,\hbar\,B_S^{(1)}\,+\,\hbar\,B_D^{(1)}\,+\,\hbar^2\,B^{(2)}\Big)\bigg]\;,
\end{align}
where $B$, $B_S^{(1)}$ and $B^{(1)}_D$ are evaluated at the classical bounce.
Note that the exponent in this expression corresponds to an approximation of the
effective action, up to the modes that are not positive definite but lead to a vanishing contribution in
the planar-wall limit, cf.~Eq.~\eqref{B1S:sinteg} and the associated comments.

In summary, given the leading-order approximation $\varphi^{(0)}$ to the bounce [Eq.~\eqref{kink}]
and $B$ for the action [Eq.~\eqref{bounceaction}] 
we apply the following procedure in order to calculate the radiative corrections to the bounce and
to the decay rate:

\begin{itemize}

\item
We first invert Eqs.~\eqref{operator:KG:twopoint} and~\eqref{operator:Dirac:twopoint} to find the Green's functions $G(\varphi^{(0)};x,y)$ and
 $D(\varphi^{(0)};x,y)$.
\item
The Green's functions are used in order to
calculate $B_S^{(1)}$ and $B_D^{(1)}$ according to Eq.~\eqref{B1:det}
and the
discussion in Sec.~\ref{sec:renorm:eff}.
\item
The Green's functions are also used to obtain the
tadpole functions~\eqref{tadpoles}, which, in turn, yield the corrections to the bounce~\eqref{correctedbounce}.
\item
When substituted into the tree and one-loop
actions, the radiative corrections to the bounce yield the quadratic correction $B^{(2)}$ in Eq.~\eqref{B2} by means of Eq.~\eqref{deltaS}, 
corresponding to dumbbell graphs. We account for these contributions in the calculation of the
decay rate because they can be relevant when a large number of fields is running in the loops.
\item
Finally, the pieces $B$, $B_{S,D}^{(1)}$ and $B^{(2)}$ can be put together to obtain the decay rate
per unit volume~\eqref{decayrate}.

\end{itemize}

\section{Green's functions, loop-improved effective action and bounce in the planar-wall approximation}
\label{sec:Planar-wall}

In this section, we calculate the Green's functions for the scalar and fermion fields in the background of the classical bounce and in the planar-wall limit.
A method of calculating the Green's function of the Dirac operator in a spherically symmetric
background in terms of all of its spinor
components is worked out in App.~\ref{app:greenfunction}. In the approximation of a
planar wall, it is calculationally simpler, however, to obtain the functional determinant,
as well as the fermion tadpole $\Pi_D$, from the Green's functions of the corresponding Klein-Gordon-like operators, obtained from ``squaring'' the Dirac operator, as we discuss below.

\subsection{Planar-wall limit of Green's function and functional determinants}

\begin{figure}

  \includegraphics[scale=0.95]{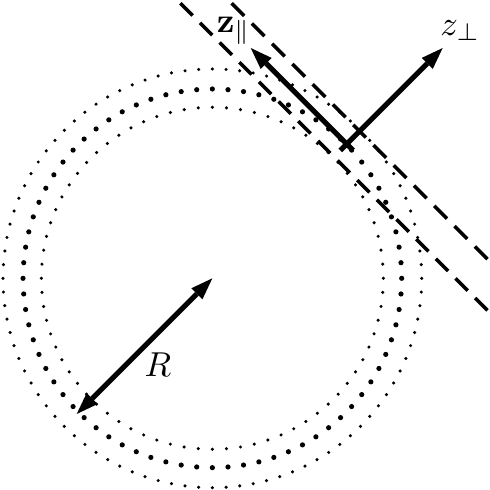}
  
  \caption{Planar-wall approximation to the bounce.
  Dotted: lines of constant values of the bounce $\varphi$, where the middle line indicates the
  center of the bounce $\varphi=0$. We choose the coordinates
  $z_{\perp}\equiv z$  perpendicular and $\bf{z}_{\parallel}$ parallel to the three-dimensional planar hypersurface tangential to the bubble wall.
  \label{fig:planar}}
\end{figure}

First, we review the calculation of the Green's function for the Klein-Gordon operator, as presented in Ref.~\cite{Garbrecht:2015oea}.
Equation~\eqref{operator:KG:twopoint} for the scalar Green's function can be recast to
the standard form
\begin{align}
\big(-\:\partial^2\:+\:U''(\varphi^{(0)};x)\big)\,G(\varphi^{(0)};x,x')\ =\ 
\delta^4(x-x')\;.
\end{align}
In a spherically symmetric background, this equation can be solved by separating the
angular part by means of a partial-wave decomposition. This reduces the problem to one of finding the  hyperradial function $G_j(\varphi^{(0)};r,r')$, where $j$ is the quantum number of angular momentum, see, e.g., Ref.~\cite{Garbrecht:2015oea}. However, to keep things simple, we apply here the planar-wall approximation: When the radius of the bubble wall~\eqref{radius:bubble} is very large
compared to $\mu^{-1}$, i.e.~when \smash{$g\ll\sqrt{\lambda}\mu$}, we can treat the geometry as planar. The bounce is then a function of the coordinate $z_{\perp}$ perpendicular to the wall only and has no dependence on the parallel coordinates $\bf{z}_{\parallel}$ on the three-dimensional hypersurface, cf.~Fig.~\ref{fig:planar}. In the remainder of this article (save App.~\ref{app:greenfunction}), we employ this planar-wall approximation.

Without loss of generality, we choose $z_{\perp}\equiv z=x_4-R$ and $\bf{z}_{\parallel}=\bf{x}$. The Fourier transform of the Green's function with respect to  the coordinates $\bf{x}$,
\begin{align}
\label{Fouriertrans}
&G(\varphi^{(0)};z,z',\mathbf{k})\nonumber\\ &\qquad=\ \int\D^3(\mathbf{x}-\mathbf{x'})\;e^{-i\mathbf{k}\cdot(\mathbf{x}-\mathbf{x}')}\,G(\varphi^
{(0)};x,x')\;,
\end{align}
satisfies the equation
\begin{align}
\label{kdependentGreenfunction}
(-\:\partial_z^2\:+\:{\bf k}^2\:+\:U''(\varphi^{(0)};z)\big)G(\varphi^{(0)};z,z',{\bf k})\ =\ \delta(z-z')\;,
\end{align}
and we refer to it as the {\it three-plus-one representation}. It can be found numerically or analytically, and the analytic solution is derived in Refs.~\cite{Garbrecht:2018rqx} and~\cite{Garbrecht:2015oea}.

Starting from the representation of the Green's function as a spectral sum, one can show~\cite{Baacke:1993jr,Baacke:1993aj,Baacke:1994ix,Baacke:2008zx} that it can be used to calculate the functional determinant directly, and this approach was used in the planar-wall limit for scalar fluctuations in Ref.~\cite{Garbrecht:2015yza}.
Specifically, the functional determinant can be evaluated as
\begin{align}
\label{B1S:sinteg}
B_S^{(1)}\ &=\ -\:\frac12 \int\limits_{-\infty}^\infty {\rm d} z
\int\D^3\mathbf{x}
\int\limits_0^{\infty}{\rm d}s\int\limits_0^\Lambda\frac{{\mathbf{k}}^2\, {\rm d} {|\mathbf{k}|}}{2\pi^2}\notag\\
\times&\Big[G\big(\varphi^{(0)};z,
z,\sqrt{{\bf k}^2+s}\big)
-G\big(v;z,z,\sqrt{{\bf k}^2+s}\big)
\Big]\;,
\end{align}
where, making use of the isotropy parallel to the bubble wall, we have written the coincident Green's function 
$G(\varphi^{(0)};z,z,|{\bf k}|)\equiv G(\varphi^{(0)};z,z,{\bf k})$. The $\D^3 \mathbf{x}$
integration is parallel to the bubble wall{,} and
$\Lambda$ is a three-momentum cutoff. Note that, in these integrals, the negative
and zero modes correspond to $|\mathbf k|=0$ and therefore lead to a vanishing contribution.

In analogy with the scalar case, the Green's function for the Dirac operator
in a spherically symmetric background can be solved by a separation ansatz. The angular solutions 
in this approach are spin hyperspherical harmonics
of definite total angular momentum, as is explained in
detail in App.~\ref{app:greenfunction}. Considerable simplifications
are, however, possible in the
planar-wall approximation.

In the planar-wall approximation, $\varphi^{(0)}$ depends on $z$ (or $x_4$) only, and we must define the
local Dirac mass $m_D(z)\equiv\kappa\,\varphi^{(0)}(z)$. We therefore need to generalize the well-known method of evaluating the determinant
of the Dirac operator to the situation where the mass may vary in one direction of
spacetime. Noticing that the  determinant
\begin{align}
I\ &=\ \ln\det D^{-1}(\varphi^{(0)};x)\nonumber\\ &=\ \ln\det\big(\gamma_\mu\partial_\mu+m_D(z)\big)
\end{align}
is invariant under chiral conjugation, we
multiply by $\gamma_5$ from the left and right of the inhomogeneous Dirac operator. This yields
\begin{align}
\det\big(\gamma_\mu\partial_\mu+m_D(z)\big)
\ &=\ \det\big[\gamma_5\big(\gamma_\mu\partial_\mu+m_D(z)\big)
\gamma_5\big]\notag\\
&=\ \det\big(-\gamma_\mu\partial_\mu+m_D(z)\big)\;,
\end{align} 
where we have used the anticommutation relations between $\gamma_5$ and $\gamma_\mu$ and the multiplicativity property of the determinant.
We can thus express
\begin{align}
I\ &=\ \frac{1}{2}\,\big[\ln\det\big(\gamma_\mu\partial_\mu+m_D(z)\big)\nonumber\\&\qquad \qquad +\:\ln\det\big(
-\gamma_\mu\partial_\mu+m_D(z)\big)\big]\notag\\
&=\ \frac{1}{2}\,\ln\det\big[-\partial^2+m_D^2(z)+\gamma_4\big(\partial_4 m_D(z)\big)\big]\;.
\end{align}

Now, after employing the representation of gamma matrices
\begin{align}
\gamma_4\ =\ 
\begin{bmatrix}
\label{gamma}
0 & 1 \\
1 & 0 \\
\end{bmatrix}
\;, \ \ \ 
\gamma_i\ =\ 
\begin{bmatrix}
0 & i\sigma_i \\
-i\sigma_i & 0 \\
\end{bmatrix}\;,
\end{align}
where $\sigma_i$ are the Pauli matrices, we obtain
\begin{align}
I\ =\ \frac{1}{2}\,\ln\begin{Vmatrix}
\big(-\partial^2+m_D^2(z)\big)\cdot\mathbb{I}_2 & \big(\partial_4 m_D(z)\big)\cdot\mathbb{I}_2\\
\big(\partial_4 m_D(z)\big)\cdot\mathbb{I}_2 & \big(-\partial^2+m_D^2(z)\big)\cdot\mathbb{I}_2
\end{Vmatrix}\;.
\end{align}
This is a determinant in block form, and we can make use of the relation
\begin{align}
\det\begin{bmatrix}
A & B\\
B & A
\end{bmatrix}\ =\ \det (A-B)\det(A+B)\;,
\end{align}
which applies even when $A$ and $B$ do not commute with each other, provided $A$ and $B$ are square matrices of the same dimensions. We therefore arrive at
\begin{align}
I\ &=\ \frac{1}{2}\,\ln\,\big\{\!\det\big[ 
\big(-\partial^2+m_D^2(z)- \partial_zm_D(z)\big)\cdot\mathbb{I}_2\big]\notag\\
&\qquad \times\det\big[
\big(-\partial^2+m_D^2(z)+\partial_zm_D(z)\big)\cdot\mathbb{I}_2\big]\big\}\notag\\
&=\ \ln\det\big[ 
-\partial^2+m_D^2(z)- \partial_zm_D(z)\big]\notag\\
&\qquad +\:\ln\det\big[
-\partial^2+m_D^2(z)+\partial_zm_D(z)\big]\;,
\end{align}
where, in the second step, the determinant over the remaining $2\times 2$ block has been performed. When $m_D$ is constant, this reduces to the well-known result for the fermion
determinant. The result of this computation can be substituted into Eq.~\eqref{decayrate}
for the decay rate.

With the above results, we can now rewrite the fermion radiative correction~\eqref{B1:Dirac} to the
tunneling action as
\begin{align}
B^{(1)}_D\ &=\ B^{(1)}_{D+}\:+\: B^{(1)}_{D-}\;,
\end{align}
where
\begin{subequations}
\label{eq:BDpm}
\begin{gather}
\label{B1:Dirac:newform}
B^{(1)}_{D\pm}\ \equiv\ B^{(1)}_{D\pm}[\varphi^{(0)}]\ =\ -\:\ln\,\frac{\det\,D_{\pm}^{-1}(\varphi^{(0)})}{\det\,D_{\pm}^{-1}(v)}\;,\\
D_{\pm}^{-1}(\varphi^{(0)}\,;x)\ =\ -\:\partial^2\:+\:\kappa^2
{\varphi^{(0)2}}(z)\:\pm\:\kappa\,\partial_z\varphi^{(0)}(z)\;.\label{eq:Dplusminus}
\end{gather}
\end{subequations}
The problem has thus been reduced to one of evaluating the
determinants of scalar fluctuation operators. This can be accomplished
by first calculating the Green's functions $D_{\pm}(\varphi^{(0)};x)$, as well as their
Fourier transforms with respect to the
coordinates parallel to the bubble wall, as in Eq.~\eqref{Fouriertrans}.
From these, the quantities \smash{$B^{(1)}_{D\pm}$} can be obtained in analogy with
the determinant for the scalar field in Eq.~\eqref{B1S:sinteg}, i.e.
\begin{align}
\label{B1D:sinteg}
B_{D\pm}^{(1)}\ &=\ \int\limits_{-\infty}^\infty {\rm d} z\int\D^3 \mathbf{x} \int\limits_0^{\infty}{\rm d}s\int\limits_0^\Lambda\frac{{\mathbf{k}}^2\, {\rm d}{|\mathbf{k}|}}{2\pi^2}\notag\\
&\hskip-.9cm
\times\Big[D_{\pm}\big(\varphi^{(0)};z
,z,\sqrt{{\bf k}^2+s}\big)
\:-\:D_{\pm}\big(v;z,z,\sqrt{{\bf k}^2+s}\big)
\Big]\;.
\end{align}

The fermionic tadpole is now most straightforwardly obtained by functionally differentiating
the one-loop determinant as in Eq.~\eqref{Pinewform}:
\begin{align}
\label{PiDnewform}
\Pi_D(\varphi^{(0)};x)\,\varphi^{(0)}(x) \ &=\ \left.\frac{{\delta} B_{D+}^{(1)}[\varphi]}{{\delta}\varphi(x)}\right|_{\varphi^{(0)}}\:+\:\left.\frac{{\delta} B_{D-}^{(1)}[\varphi]}{{\delta}\varphi(x)}\right|_{\varphi^{(0)}}\;.
\end{align}
We emphasize that $\delta/\delta\varphi$ denotes the functional derivative and, after making use of the Jacobi formula, we obtain from Eq.~\eqref{eq:BDpm} that
\begin{align}
\label{PiD1}
\left.\frac{{\delta} B_{D\pm}^{(1)}[\varphi]}{{\delta}\varphi(x)}\right|_{\varphi^{(0)}}\ &=\ -\:2\,\kappa^2 D_{\pm}(\varphi^{(0)};x)\,\varphi^{(0)}\nonumber\\&\qquad\qquad\pm\:\kappa\,\partial_z D_{\pm}
(\varphi^{(0)};x)\;.
\end{align}
These expressions can be substituted into Eqs.~\eqref{correctedbounce} and~\eqref{tadpole:fermi}
in order to obtain the fermion contribution to the one-loop correction $\delta \varphi$ to the bounce.
We also note from Eqs.~\eqref{tadpole:fermi}, \eqref{PiDnewform} and~\eqref{PiD1} {that} there
follows the following identity for the spinor trace:
\begin{align}
\label{trD:identity}
&{\rm tr}_{\mbox{s}}D(\varphi^{(0)};x,x) \hskip4.3cm\notag\\
&\qquad =\ \sum\limits_\pm\Big(
2\,\kappa D_{\pm}(\varphi^{(0)};x)\,\varphi^{(0)}\:\mp\: \partial_z D_{\pm}(\varphi^{(0)};x)\Big)
\;.
\end{align}
In the homogeneous limit $\varphi^{(0)}={\rm const}.$, $D_+$ and $D_-$ coincide, and we quickly recover the familiar result for the one-loop fermion determinant --- including the factor of $4$ from the spinor trace --- after summing over $\pm$.

We conclude this part of the discussion by noting that it is also possible
to compute $D(\varphi^{(0)};x,x^\prime)$ and all of its spinor components directly
by solving the equations derived in App.~\ref{app:greenfunction}. While this is
calculationally more cumbersome, we have checked the analytical and
numerical results for the fermion loop effects reported on in this work using both methods.

\subsection{One-loop correction to the bounce in the planar-wall limit}

The one-loop corrections to the bounce in Eq.~\eqref{correctedbounce} satisfy the equation of motion
\begin{align}
\left[\frac{\D^2}{\D z^2}\,+\,\mu^2\,-\,\frac{\lambda}{2}\,{\varphi^{(0)2}}\right]\,\delta\varphi\ =\ \Pi(\varphi^{(0)};x)\,\varphi^{(0)}\;,
\end{align}
as we can quickly confirm by acting on Eq.~\eqref{correctedbounce}  with the tree-level Klein-Gordon operator.
This can be solved using the Green's function $G(\varphi^{(0)};z,z',{\bf k})$ at ${\bf k}={\bf 0}$,
for which there is an analytic solution~\cite{Garbrecht:2015oea}:
\begin{align}
\label{Guunu}
&G(\varphi^{(0)};z,z',{\bf k})\ \equiv\  G(u,u',\nu)\notag\\
&=\ \frac{1}{2\gamma\, \nu}\bigg[\theta(u-u')\left(\frac{1-u}{1+u}\right)^{\frac{\nu}{2}}\left(\frac{1+u'}{1-u'}\right)^{\frac{\nu}{2}}\notag\\
&\times\left(1-3\frac{(1-u)(1+\nu+u)}{(1+\nu)(2+\nu)}\right)\notag\\
&\times\left(1-3\frac{(1-u')(1-\nu+u')}{(1-\nu)(2-\nu)}\right)+(u\leftrightarrow u')\bigg]\;,
\end{align}
where
\begin{align}
u\ =\ \tanh(\gamma z)
\end{align}
and $\nu=2\sqrt{1+{\bf k}^2/(4\gamma^2)}$. Proceeding in this way, we obtain
\begin{align}
\label{deltavarphi}
\delta\varphi(u)\ =\ -\,\frac{v}{\gamma}\,\int^1_{-1}{\rm d}u'\;\frac{u'\,G(u,u',2)\,{\Pi}(\varphi;u')}{1-u'^2}\;.
\end{align}
As can be seen from Eq.~\eqref{Guunu}, $G(u,u',\nu)$ is singular as $\nu\rightarrow 2$, i.e.~$|\mathbf k|\to 0$. However, since $\Pi(u^\prime)$ is an even function, the part of the integrand multiplied by $G(u,u',2)$ is an odd function in $u^\prime$. The integral remains finite because the singularity of $G(u,u',2)$ turns out to reside in the even part.
We can therefore replace $G(u,u^\prime,2)$ in Eq.~\eqref{deltavarphi} with the odd part of the Green's function~\cite{Garbrecht:2015oea}
\begin{align}
G^{\rm odd}(u,u')\ \equiv&\ \frac{1}{2}\,\left(G(u,u',2)\,-\,G(u,-u',2)\right)\,,
\end{align}
which can be expressed as~\cite{Garbrecht:2015oea}
\begin{align}
&G^{\rm odd}(u,u')\ =\ \vartheta(u-u')\,\frac{1}{32\,\gamma}\,\frac{1-u^2}{1-u'^2}\notag\\
&\,\times\,\left[2u'\,(5-3\,u'^2)+3\,(1-u'^2)^2\,\ln\,\frac{1+u'}{1-u'}\right]\,+\,(u\leftrightarrow u')
\end{align} 
in the domain $0\leq u,u'\leq 1$. For given $\Pi(u)$, we can thus compute the first correction
to the bounce.

\section{Renormalization}
\label{sec:renormalization}

The calculation of radiative corrections to the decay rate and to the bounce solutions requires the renormalization of the scalar operators. We therefore add the following counterterms
to the Lagrangian in Eq.~\eqref{themodel}:
\begin{align}
\label{CTLagrangian}
{\cal L}\ \to\ {\cal L}\:+\:\frac{1}{2}\,\delta Z\,(\partial_\mu\varphi)^2\:+\:\frac{1}{2}\,\delta\mu^2\,\varphi^2\:+\:\frac{1}{4!}\,\delta\lambda\,\varphi^4\;.
\end{align}
The one-loop corrections entail, in particular, the usual ultraviolet divergences from {the} quartic scalar and Yukawa interactions.
In position space, these appear when taking the coincident limit of the Green's functions and, in momentum space, when performing the loop integrals. In the present setup,
we use the mixed three-plus-one representation of momentum space parallel to the three-dimensional hypersurface
and position space in the perpendicular direction.
As a regulator, we introduce a three-momentum cutoff $\Lambda$.

The counterterms must be uniquely specified by certain renormalization conditions.
For the problem of metastable vacua, we find it most useful to impose these conditions
on the derivatives of the effective action  evaluated at the false vacuum. In the one-loop
approximation, the effective action is given by
\begin{equation}
\label{efact}
\Gamma[\varphi]\ =\ S[\varphi]\:+\:\frac{\hbar}{2}\,\ln\,\frac{\det\,G^{-1}(\varphi)}{\det\,G^{-1}(v)}\:-\:\hbar\,\ln\,\frac{\det\,D^{-1}(\varphi)}{\det\,D^{-1}(v)}\;.
\end{equation}
When
ignoring derivative operators, the effective action in the false vacuum
coincides with the Coleman-Weinberg effective potential. We therefore
use the derivatives of the latter to define the renormalization
conditions for $\mu^2$ and $\lambda$, and these are worked out in Sec.~\ref{subsec:CW}.

The radiative effects also lead to corrections to the wave-function normalization. These
are logarithmically divergent for the fermion loops and finite for the scalar loops. In Sec.~\ref{subsec:wv}, we extract
these terms analytically by a gradient expansion of the Green's functions.
From the latter, we can calculate the leading gradient corrections to the effective
action, which can then be identified with the wave-function normalization,
where we impose as
a renormalization condition the standard unit residue at the single-particle pole.
In Sec.~\ref{sec:renorm:eff}, we then summarize these results by combining the counterterms with
the one-loop determinants and tadpole functions. In this way, renormalized results for $B^{(1)}$, $B^{(2)}$ and the tadpole function are obtained, which lead directly to the renormalized decay rate and the correction $\delta\varphi$ to the bounce.

\subsection{Renormalization of the mass and the quartic coupling constant using the Coleman-Weinberg potential}
\label{subsec:CW}

The Coleman-Weinberg effective potential
is obtained by evaluating the effective action~\eqref{efact}
for configurations $\varphi$ that are constant throughout
spacetime. The latter is given by
\begin{align}
\label{cwefact}
\Gamma_{\text{CW}}[\varphi]\ &=\ S[\varphi]+\frac{\hbar}{2}\int \D^4x\int\!\frac{\D^4 k}{(2\pi)^4}\;\ln\,\frac{k^2+U''(\varphi)}{k^2+U''(v)}\notag\\
&-2\,\hbar\,\int\D^4x\int\!\frac{\D^4k}{(2\pi)^4}\;\ln\,\frac{k^2+\kappa^2\varphi^2}{k^2+\kappa^2v^2}\;.
\end{align}
The spacetime and four-momentum variables $x$ and $k$ are understood to have natural units; the overall factor of $\hbar$ in the one-loop corrections appears only for bookkeeping purposes.\footnote{Were we to proceed otherwise, we would need to include additional inverse powers of $\hbar$ in the logarithms to ensure the mass terms have the correct dimensions. Note that these factors would then cancel the overall factor of $\hbar$ when one arrives at, e.g., Eq.~\eqref{UCW}.} It can be directly read from the above expression that
\begin{subequations}
\begin{align}
&B_S^{(1){\rm hom}}[\varphi]\ =\ \frac{1}{2}\int \D^4x\int\!\frac{\D^4 k}{(2\pi)^4}\;\ln\,\frac{k^2+U''(\varphi)}{k^2+U''(v)}\;,\\
&B_D^{(1){\rm hom}}[\varphi]\ =\ -\:2\int\D^4x\int\!\frac{\D^4k}{(2\pi)^4}\;\ln\,\frac{k^2+\kappa^2
\varphi^2}{k^2+\kappa^2v^2}\;.
\end{align}
\end{subequations}
We note that we could just as well compute these quantities from the Green's functions in the three-plus-one
representation for the limit of a constant background field, using Eqs.~\eqref{B1S:sinteg} and~\eqref{B1D:sinteg} (cf.~Eqs.~\eqref{Ghom} and~\eqref{Dhom} later).
Factoring out the integral over the spacetime four-volume, one gets the Coleman-Weinberg effective potential
\begin{align}
\label{cwpotential}
U_{\text{CW}}(\varphi)\ &=\ U(\varphi)\:+\:\frac{\hbar}{2}\int\!\frac{\D^4 k}{(2\pi)^4}\;\ln\,\frac{k^2+U''(\varphi)}{k^2+U''(v)}\notag\\
&-2\,\hbar\,\int\!\frac{\D^4k}{(2\pi)^4}\;\ln\,\frac{k^2+\kappa^2\varphi^2}{k^2+\kappa^2v^2}\;,
\end{align}
where we reiterate that the factor of $\hbar$ again appears only for bookkeeping purposes.

In order to apply a regularization scheme that is congruent with our calculations of the decay rate and the corrections
of the bounce (both of which are performed in the background that is spatially inhomogeneous in the direction perpendicular to
the bubble wall), we make a three-plus-one decomposition of momentum space.
Performing the integral over $k_0$, we obtain 
\begin{align}
\label{cwefact2}
&U_{\text{CW}}(\varphi)\ =\ U(\varphi)\notag\\
&+\:\frac{\hbar}{2} \int\!\frac{\D^3 {\bf k}}{(2\pi)^3}\Big(\sqrt{{\bf k}^2+U''(\varphi)}\:-\:\sqrt{{\bf k}^2+U''(v)}\Big)\notag\\
&-\:2\,\hbar\,\int\!\frac{\D^3 {\bf k}}{(2\pi)^3}\Big(\sqrt{{\bf k}^2+\kappa^2\varphi^2}\:-\:\sqrt{{\bf k}^2+\kappa^2v^2}\Big)\;.
\end{align}
Evaluating the remaining integral up to a cutoff $\Lambda$, this yields (dropping the bookkeeping factors) 
\begin{align}
\label{UCW}
 &U_{\rm CW}(\varphi)\ =\ U(\varphi)\:+\:\bigg\{\bigg[\frac{\Lambda^2}{16\pi^2}\bigg(-\mu^2+\frac{\lambda}{2}\,\varphi^2\bigg)\notag\\
 &\ \ +\:\frac{1}{64\pi^2}\bigg(-\mu^2+\frac{\lambda}{2}\,\varphi^2\bigg)^{\!2}\,\bigg(\ln\,\frac{-\,\mu^2\,+\,\frac{\lambda}{2}\,\varphi^2}{4\Lambda^2}+\frac{1}{2}\bigg)\notag\\
 &\ \ -\:\frac{\Lambda^2\kappa^2\varphi^2}{4\pi^2}\:-\:\frac{\kappa^4\varphi^4}{16\pi^2}\bigg(\ln\,\frac{\kappa^2\varphi^2}{4\Lambda^2}+\frac{1}{2}\bigg)\bigg]\:-\:(\varphi\rightarrow v)\bigg\}\;.
\end{align}

The renormalized Coleman-Weinberg potential is obtained when adding the counterterms
$\delta\mu^2$ and $\delta\lambda$ in accordance with Eq.~\eqref{CTLagrangian} as
\begin{align}
U_{\text{CW}}\ \to\  U_{\text{CW}}\:+\:\delta\mu^2\,(\varphi^2-v^2)/2\:+\:\delta\lambda\,(\varphi^4-v^4)/4!\;,
\end{align}
where $\delta\mu^2$ and $\delta\lambda$ are specified by the following renormalization conditions:
\begin{subequations}
\begin{align}
 \frac{\partial^2U_{\rm CW}(\varphi)}{\partial\varphi^2}\Big|_{\varphi\,=\,v}\ &=\ -\:\mu^2\:+\:\frac{\lambda}{2}\,v^2\ =\ 2\,\mu^2\;,\\
 \frac{\partial^4U_{\rm CW}(\varphi)}{\partial\varphi^4}\Big|_{\varphi\,=\,v}\ &=\ \lambda\;.
\end{align}
\end{subequations}
These counterterms are given explicitly in Eq.~\eqref{counterterms} below,
along with the remaining one for the wave-function correction.

\subsection{Wave-function renormalization through adiabatic expansion of the Green's functions}
\label{subsec:wv}

While the effective potential offers a convenient way to define the counterterms for the
coupling constants, it does not lead to conditions on the renormalization
of the derivative operator, i.e.~the wave-function normalization. Our objective is to
express this additional counterterm in an analytic form. This can be achieved by performing a gradient
expansion of the Green's functions around a constant background field configuration. One may also refer to
this calculation as an adiabatic or Wentzel-Kramers-Brillouin (WKB) expansion.

We first construct the adiabatic expansion of the scalar Green's function, which satisfies
\begin{align}
\label{scalarGreensequationk}
\big(-\partial_z^2+M^2(z)\big)\,G(\varphi^{(0)};z,z',{\bf k})\ =\ \delta(z-z')\;,
\end{align}  
{with}
\begin{align}
M^2(z)\ =\ {\bf k}^2\:+\:U''(\varphi^{(0)}(z))\;.
\end{align}
This can be solved by the ansatz
\begin{align}
G(\varphi^{(0)};z,z',{\bf k})\ &=\ \theta(z-z')A^{>}(z')f^{>}(z)\notag\\
&\qquad +\:\theta(z'-z)A^{<}(z')f^{<}(z)\;,
\end{align}
where
\begin{align}
\label{eq:fgrle}
\big(-\partial_z^2+M^2(z)\big)\,f^{\gtrless}(z)\ =\ 0
\end{align}
and where we impose
\begin{subequations}
\begin{align}
&f^{>}(z)\rightarrow 0,\ \text{for}\  z\rightarrow +\:\infty\;;\\
&f^{<}(z)\rightarrow 0,\ \text{for}\ z\rightarrow -\:\infty\;.
\end{align}
\end{subequations}
The latter enforce the boundary condition that the Green's function vanishes at infinity.

In order to isolate the leading gradient effects, we make the WKB ansatz
\begin{align}
f^{\gtrless}(z)\ =\ \frac{1}{\sqrt{2W(z)}}\,e^{\mp\int^z \D z'\,W(z')}\;.
\end{align}
When substituted into Eq.~\eqref{eq:fgrle}, this leads to
\begin{align}
\label{eq:W}
W^2\ =\ M^2\:-\:\frac{3}{4}\frac{W'^2}{W^2}\:+\:\frac{1}{2}\frac{W''}{W}\;,
\end{align}
where a prime $'$ on $W(z)$ (and in the following on $m_{S}(z)$) denotes a derivative with respect to $z$. In the gradient expansion, the zeroth{-}order approximation to $W$ is given by
\smash{${W^{(0)}}^2=M^2$}. Substituting this back into Eq.~\eqref{eq:W}, we obtain
\begin{align}
{W^{(1)}}^2\ =\ -\:\frac{3}{4}\frac{{{W^{(0)}}'}^2}{{W^{(0)}}^2}\:+\:\frac{1}{2}\frac{{W^{(0)}}''}{W^{(0)}}\;.
\end{align}

Demanding continuity and the correct jump in the first derivative
to reproduce the delta function in Eq.~\eqref{scalarGreensequationk} at the coincident point $z=z^\prime$, we obtain the matching conditions
\begin{subequations}
\begin{gather}
A^{>}(z)f^{>}(z)\ =\ A^{<}(z)f^{<}(z)\;,\\
A^{>}(z){f^{>}}'(z)\:-\:A^{<}(z){f^{<}}'(z)\ =\ -\:1\;,
\end{gather}
\end{subequations}
such that
\begin{align}
G(\varphi^{(0)};z,z,{\bf k})\ =\ \frac{f^{>}(z)f^{<}(z)}{W[f^{>}(z),f^{<}(z)]}\;,
\end{align}
where
\begin{align}
W[f^{>}(z),f^{<}(z)]\ =\ f^{>}(z){f^{<}}'(z)\:-\:{f^{>}}'(z)f^{<}(z)
\end{align}
is the Wronskian. Putting these results together, we find that the Green's function is approximated to
second order in gradients by
\begin{align}
G(\varphi^{(0)};z,z,{\bf k})\ \approx\ \frac{1}{2}\,\frac{1}{M(z)}\:+\:\frac{3}{16}\frac{{M'^2(z)}}{M^5(z)}\:-\:\frac{1}{8}\,\frac{M''(z)}{M^4(z)}\;.
\end{align}
In terms of the local squared mass $m_S^2(z)=U''(\varphi^{(0)}(z))$, this reads
\begin{align}
\label{WKBGreenscalar}
&G(\varphi^{(0)};z,z,{\bf k})\ \approx\ \frac{1}{2}\,\frac{1}{\sqrt{{\bf k}^2+m_S^2(z)}}\notag\\
&+\:\frac{5}{16}\,\frac{m_S^2(z)\cdot{m}_S^{\prime 2}(z)}{({\bf k}^2+m_S^2(z))^{\frac{7}{2}}}\:-\:\frac{1}{8}\,\frac{m_S(z)\cdot m_S''(z)+{m}_S^{\prime 2}(z)}{({\bf k}^2+m_S^2(z))^{\frac{5}{2}}}\;.
\end{align}
The first term on the right-hand side can be identified with the Green's function in the three-plus-one representation and a homogeneous background, i.e.~for field values that are constant throughout spacetime.
We explicitly define this as
\begin{align}
\label{Ghom}
G^{\rm hom}(\varphi^{(0)};z,z^\prime,\mathbf k)\ &=\
\frac{1}{2}\,\frac{1}{\sqrt{{\bf k}^2+m_S^2(z)}}\notag\\
&\times\:\Big(\vartheta(z-z^\prime){\rm e}^{\sqrt{{\bf k}^2+m_S^2(z)}(z^\prime-z)}\notag\\
&+\:\vartheta(z^\prime-z){\rm e}^{\sqrt{{\bf k}^2+m_S^2(z)}(z-z^\prime)}\Big)\;,
\end{align}
where we show the general expression that also holds away from the coincident points in order to
highlight the exponentially decaying behavior of the Green's functions for large separations.
When discussing parametric examples in Sec.~\ref{sec:numericalresults},
we will use $G^{\rm hom}$ in order to compare loop contributions without gradient effects with
those that include them.
The last two terms in Eq.~\eqref{WKBGreenscalar} are the leading gradient corrections.
The{se} corrections are $\sim 1/|\mathbf k|^5$ such that the three-momentum integral leads to a finite
correction to the derivative operators of the field $\Phi$, i.e.~the wave-function {normalization}.
It is most straightforward to determine the correction to this operator from
its contribution to the effective action, which we can compute by substituting
the result~\eqref{WKBGreenscalar} into Eq.~\eqref{B1S:sinteg}:
\begin{align}
\label{B1S:derivtive}
B^{(1)}_S\ \supset\ \int\limits_{-\infty}^{\infty}{\rm d}z\int\D^3 \mathbf{x}\;\frac{1}{384 \, \pi^2} \,\frac{\lambda^2 \varphi^2 {\varphi^\prime}^2}{-\mu^2+\frac{\lambda}{2}\varphi^2}\;,
\end{align}
where we recall that $B^{(1)}_{S}$ is related to the functional determinants through the definitions~\eqref{B1:det}.
While this result is finite, i.e.~independent of the momentum cutoff, the corresponding contribution from the Dirac field is not, as we will see below.

We compute the fermionic contributions to the wave-function {normalization} analogously.
The propagator{s} $D_{\pm}(\varphi^{(0)};z,z',{\bf k})$ satisfy the same equation as $G(\varphi^{(0)};z,z',{\bf k})$,  i.e.~Eq.~\eqref{scalarGreensequationk}, but with 
\begin{align}
M^2(z)\ =\ {\bf k}^2\:+\:\kappa^2{\varphi^{(0)2}}(z)\:\pm\:
\kappa\,\partial_z{\varphi^{(0)}}(z)\;.
\end{align}
Solving that equation in the WKB approximation, we find
\begin{align}
\label{Dpm}
D_+(\varphi^{(0)};z,z,{\bf k})\:+\:D_-(\varphi^{(0)};z,z,{\bf k})\ \approx\
\frac{1}{\sqrt{\mathbf{k}^2+m^2_D(z)}}\notag\\
+\:\frac{5}{8}\,\frac{m_D^2(z)\;{m}_D^{\prime 2}(z)}{({\bf k}^2+m_D^2(z))^{\frac{7}{2}}}\:+\:\frac{1}{8}\,\frac{{m}_D^{\prime 2}(z)-2m_D(z)\; m_D''(z)}{({\bf k}^2+m_D^2(z))^{\frac{5}{2}}}\;.
\end{align} 
Again, in order to isolate the gradient effects in Sec.~\ref{sec:numericalresults}, we
define the contributions that arise in homogeneous backgrounds without gradients as
\begin{align}
\label{Dhom}
D_\pm^{\rm hom}(\varphi^{(0)};z,z,{\bf k})\ =\ \frac{1}{2\sqrt{\mathbf{k}^2+m^2_D(z)}}\;.
\end{align}
We note that, while the three-momentum integrals over the derivative terms in Eq.~\eqref{Dpm} are finite,
the trace of the coincident Dirac propagator~\eqref{trD:identity}, and therefore the
tadpole correction, cf.~Eqs.
\eqref{tadpole:fermi}, \eqref{PiDnewform} and \eqref{PiD1}, contains an extra
derivative with respect to $z$. This generates a logarithmically divergent contribution that
can be identified with the divergent part of the wave-function {normalization}.
Computing the corrections to the derivative operators in the effective action using
Eq.~\eqref{B1D:sinteg}, we find
\begin{align}
\label{B1D:derivative}
B^{(1)}_D\ \supset\ \int\limits_{-\infty}^{\infty}{\rm d}z \int \D^3 \mathbf{x} \;\frac{\kappa^2{\varphi^\prime}^2}{8 \pi^2 } \bigg(-\ln\,\frac{\kappa^2 \varphi^2}{\Lambda^2}+2 \ln 2 -\frac83 \bigg)\;.
\end{align}
We use this result along with Eq.~\eqref{B1S:derivtive} in order to specify the renormalization
condition
\begin{align}
\frac{\partial^2}{\partial(\partial_\nu\varphi)^2}\bigg[{\cal L}\:+\:
\frac{1}{384 \, \pi^2}\, \frac{\lambda^2 \varphi^2 (\partial_\mu\varphi)^2}{-\mu^2+\frac{\lambda}{2}\varphi^2}\hspace{3.2cm}\notag\\
+\:\frac{\kappa^2}{8 \pi^2} (\partial_\mu\varphi)^2\bigg(-\ln\,\frac{\kappa^2 \varphi^2}{\Lambda^2}+2\ln2 -\frac83 \bigg)\bigg]_{\varphi\,=\,v}\ =\ 1\;,
\end{align}
where the Lagrangian is implied to contain the counterterms as per the definition~\eqref{CTLagrangian}. This relation fixes the counterterm $\delta Z$.

\subsection{Renormalized bounce, effective action and decay rate}
\label{sec:renorm:eff}

We can now summarize the one-loop counterterms as
\begin{subequations}
\label{counterterms}
\begin{align}
\label{deltaZ}
\delta Z \ &=\ \frac{\kappa^2}{8\pi^2}\bigg(\ln\, \frac{\kappa^2 v^2}{\Lambda^2}-2\ln 2+{\frac{8}{3}} \bigg)\:-\:\frac{\lambda}{64\pi^2}\;,\\
\label{deltamu2}
 \delta\mu^2\ &=\ -\:\frac{\lambda\mu^2}{32\pi^2}\bigg(\frac{2\Lambda^2}{\mu^2}-\ln\,\frac{\mu^2}{2\Lambda^2}-31\bigg)\notag\\
 &\qquad+\:\frac{\Lambda^2\kappa^2}{2\pi^2}\bigg(-\frac{27\kappa^2\mu^2}{\lambda\Lambda^2}\,+\,1\bigg)\;,\\
\label{deltalambda}
\delta\lambda\ &=\ -\:\frac{3\lambda^2}{32\pi^2}\bigg(\ln\,\frac{\mu^2}{2\Lambda^2}+5\bigg)\notag\\
 &\qquad +\:\frac{3\kappa^4}{2\pi^2}\,\bigg(\ln\,\frac{3\kappa^2\mu^2}{2\lambda\Lambda^2}+\frac{14}{3}\bigg)\;.
\end{align}
\end{subequations}
The renormalized tadpole correction can then be defined as
\begin{align}
\label{renormpitilde}
{\Pi}^{\text{ren}}\,\varphi^{(0)}\ &=\ 
{\Pi}\,\varphi^{(0)}\: +\:\delta\mu^2\,\varphi^{(0)}\nonumber\\&\qquad+\:\frac{\delta
\lambda}{3!}\,{\varphi^{(0)3}}\:-\:\delta Z\,{\partial_z^2\varphi^{(0)}}\;,
\end{align}
which should replace $\Pi \varphi^{(0)}$ in the equation of motion~\eqref{eom:bounce:corrected}
for the bounce.
Finally, to the one-loop contributions to the effective action [i.e.~to the exponent of Eq.~\eqref{decayrate}], we consistently add
\begin{align}
\delta B^{(1)}\ &=\ \int \D^4x\;\bigg[\frac{1}{2}\,\delta\mu^2\,({\varphi^{(0)2}}-v^2)+\frac{1}{4!}\,\delta\lambda\,({\varphi^{(0)4}}-v^4)\notag\\
&\qquad\qquad +\frac{1}{2}\,\delta Z\,(\partial_z\varphi^{(0)})^2\bigg]\;,
\end{align}
and the renormalized result for $B^{(2)}$ is obtained when replacing $\Pi$ with $\Pi^{\rm ren}$
in Eq.~\eqref{B2}.

%%%%%%%%%%%%%%%%%%%%%%%%%%%%%%%%%%%%%%%%%%%%%%%%%%%%%%%%%%%%%%%%%%%%%%

\section{Quantum corrections to the bounce and decay rate from fermion and scalar loops}
\label{sec:numericalresults}

We now apply the methods elaborated in the previous sections in order to compute
the leading radiative corrections to the bounce and to the decay rate numerically. In presenting the
results, we isolate the contributions from the scalar and the fermion loops.

\begin{figure}

\centering

\subfigure[]{\includegraphics[scale=0.48]{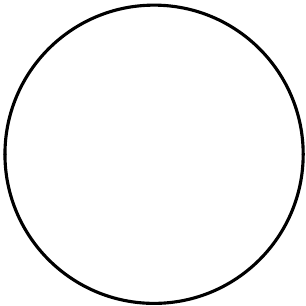}}\qquad
\subfigure[]{\includegraphics[scale=1.0]{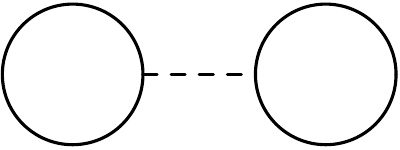}}\\
\subfigure[]{\includegraphics[scale=0.62]{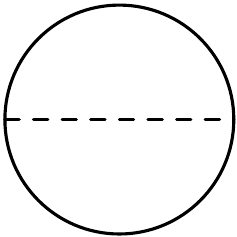}}

\caption{Diagrammatic representation of the fermionic contributions to the effective action: (a) is the one-loop term $B_D^{(1)}$, of order $\mathcal{O}[(\kappa^4/\lambda^2) N_\Psi]$, (b) is the $\mathcal{O}[(\kappa^6/\lambda^2) N_\Psi^2]$ contribution to $B^{(2)}_D$ and (c) is the term of $\mathcal{O}[(\kappa^6/\lambda^2) N_\Psi]$. Solid lines denote the Dirac propagator $D(\varphi^{(0)};x,x')$; dashed lines denote the scalar propagator $G(\varphi^{(0)};x,x')$.}
\label{fig:feynman}
\end{figure}

The term $B_D^{(2)}$ turns out to contain the dominant two-loop contributions from fermions to the effective
action when enhanced by multiple copies of the fermion fields, i.e.~$\Psi_i$ with $i=1,\ldots,N_{\Psi}$. To illustrate this,
we present a diagrammatic representation of the fermionic corrections to the bounce action in Fig.~\ref{fig:feynman}. Diagram (a) is the one-loop term \smash{$B^{(1)}_D$} of order $(\kappa^4/\lambda^2)N_\Psi$. Note that the dependence on $\lambda$ comes from the background bounce $\varphi^{(0)}\sim 1/\sqrt{\lambda}$ appearing in the Dirac operator.  Diagram (b) is the main contribution to \smash{$B_D^{(2)}$}, which is of order
$(\kappa^6/\lambda^2) N_{ \Psi}^2$, cf.~Eqs.~\eqref{B2} and~\eqref{deltavarphi}. In addition to diagram (b), there is also a contribution of order $(\kappa^6/\lambda^2) N_{\Psi}$ represented by diagram (c), which we do not calculate. Diagram (c) is therefore suppressed by a relative factor of $1/N_{\Psi}$ compared to diagram (b) and may hence be neglected, cf.~also Ref.~\cite{Garbrecht:2015oea}.

In contrast to the Higgs field $\Phi$,
the vacuum expectation values of the fermion fields do not change through the bubble wall,
such that their role may be referred to as {\it spectator}. In order to directly compare
effects from fermion and scalar loops,  we introduce additional
scalar species $\chi_i$ with $i=1,\ldots,N_{\chi}$ that also couple as spectators to the Higgs field. In this setup, the Lagrangian without counterterms is
\begin{align}
\mathcal{L}&=\sum\limits_{i\,=\,1}^{N_\Psi}\bigg\{\bar{\Psi}_i\gamma_\mu\partial_\mu\Psi_i+
\kappa\,\bar{\Psi}_i\Phi\Psi_i\bigg\}\:+\:\frac{1}{2}(\partial_\mu\Phi)^2\:+\:U(\Phi)\notag\\
&\ \ \ +\sum\limits_{i\,=\,1}^{N_\chi}\left\{\frac{1}{2}(\partial_\mu\chi_i)^2+\frac{1}{2}m_{\chi_i}^2 \chi_i^2+\frac{\alpha}{4}\Phi^2\chi_i^2\right\}.
\end{align}
For simplicity and in order to have the scalar and fermionic
spectators have similar properties, we take $m_{\chi_i}=0$.
The developments for the effective action and the tadpole corrections
of the previous sections
generalize straightforwardly to loops of the fields $\chi_i$, yielding one-loop corrections \smash{$B^{(1)}_\chi$} to the action and tadpole functions
$\Pi_\chi$.

Since it is of interest to compare the loop contributions from the various
species, we decompose the tadpole function and the one-loop corrections to the action as
\begin{align}
\Pi\ &=\ \Pi_S\:+\:\Pi_D\:+\:\Pi_{\rm spec}\;,\\
B^{(1)}\ &=\ B^{(1)}_S\:+\:B^{(1)}_D\:+\:B^{(1)}_{\rm spec}\;,
\end{align}
where the subscript $S$ indicates the Higgs scalar field $\Phi$, $D$ indicates the Dirac fermion spectators $\Psi_i$ and
``spec'', the scalar spectator fields $\chi_i$. We note that all results presented in this section for the tadpole corrections
$\Pi$, as well as the corrections to the effective action $B^{(1,2)}$, are understood to
be renormalized, i.e.~to include contributions from counterterms.

In order to identify the impact of the gradient effects, we can use the propagators $G^{\rm hom}(\varphi^{(0)};x,x)$ and $D^{\rm hom}(\varphi^{(0)};x,x)$ in place of the Green's functions $G(\varphi^{(0)};x,x)$ and $D(\varphi^{(0)};x,x)$ in Eqs.~\eqref{B1S:sinteg} and~\eqref{B1D:sinteg} to compute the  one-loop action terms in the homogeneous background. We can compute the corresponding spectator corrections, leading altogether to the results
\begin{align}
B^{(1)\rm hom}\ &=\ B^{(1)\rm hom}_S\:+\:B^{(1)\rm hom}_D\:+\:B^{(1)\rm hom}_{\rm spec}\;.
\end{align}
Accordingly, substituting the homogeneous Green's functions in Eq.~\eqref{tadpoles} (and analogously for the scalar
spectators), we also obtain
\begin{align}
\Pi^{\rm hom}\ &=\ \Pi^{\rm hom}_S\:+\:\Pi^{\rm hom}_D\:+\:\Pi^{\rm hom}_{\rm spec}\;.
\end{align}

\subsection{Tadpoles and corrections to the bounce}

\begin{figure}
\centering
\includegraphics[scale=0.6]{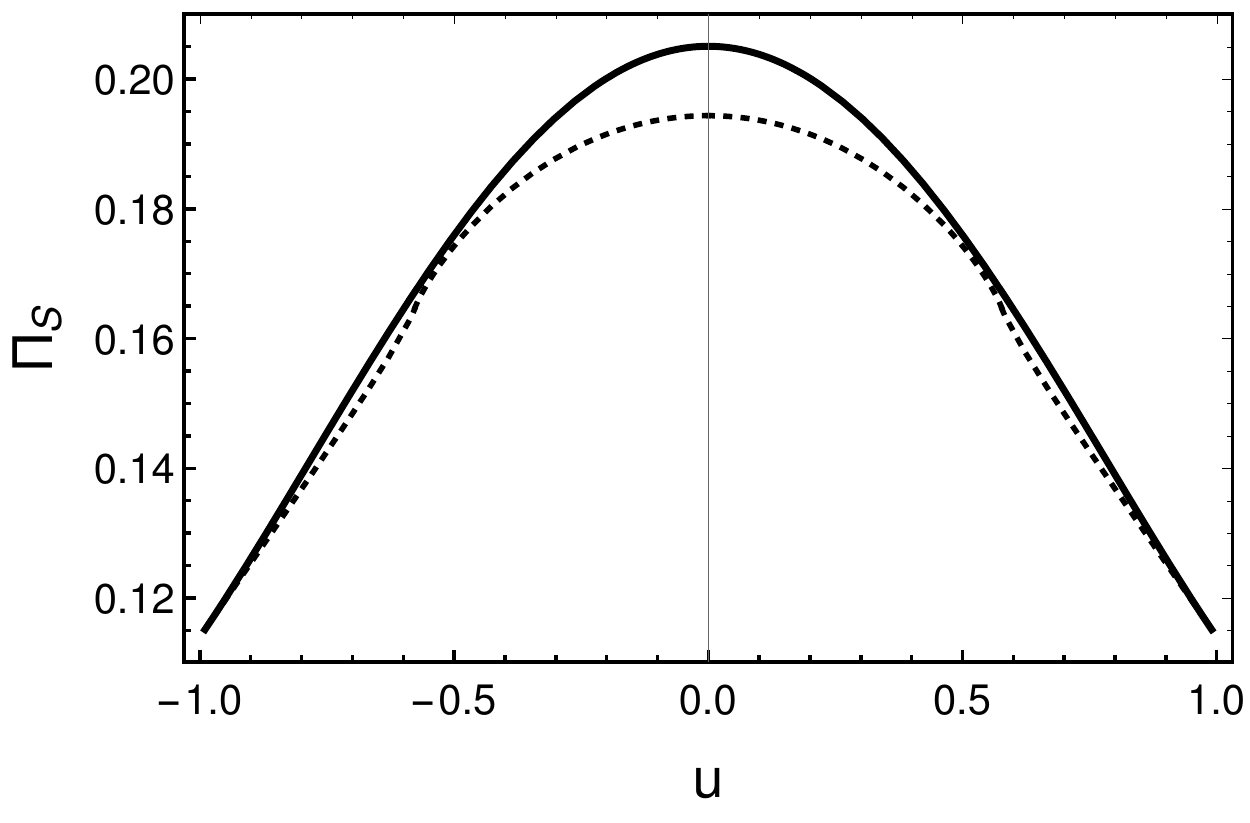}

\includegraphics[scale=0.6]{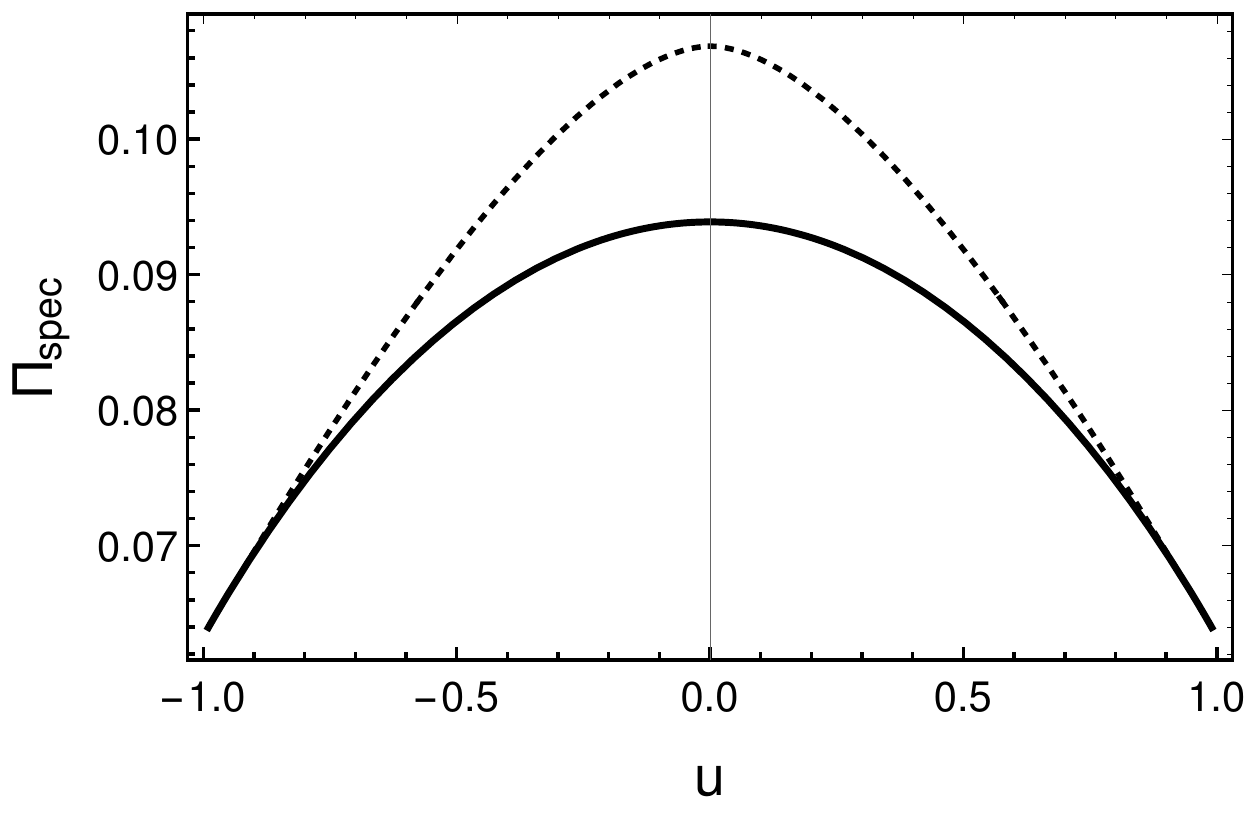}

\includegraphics[scale=0.6]{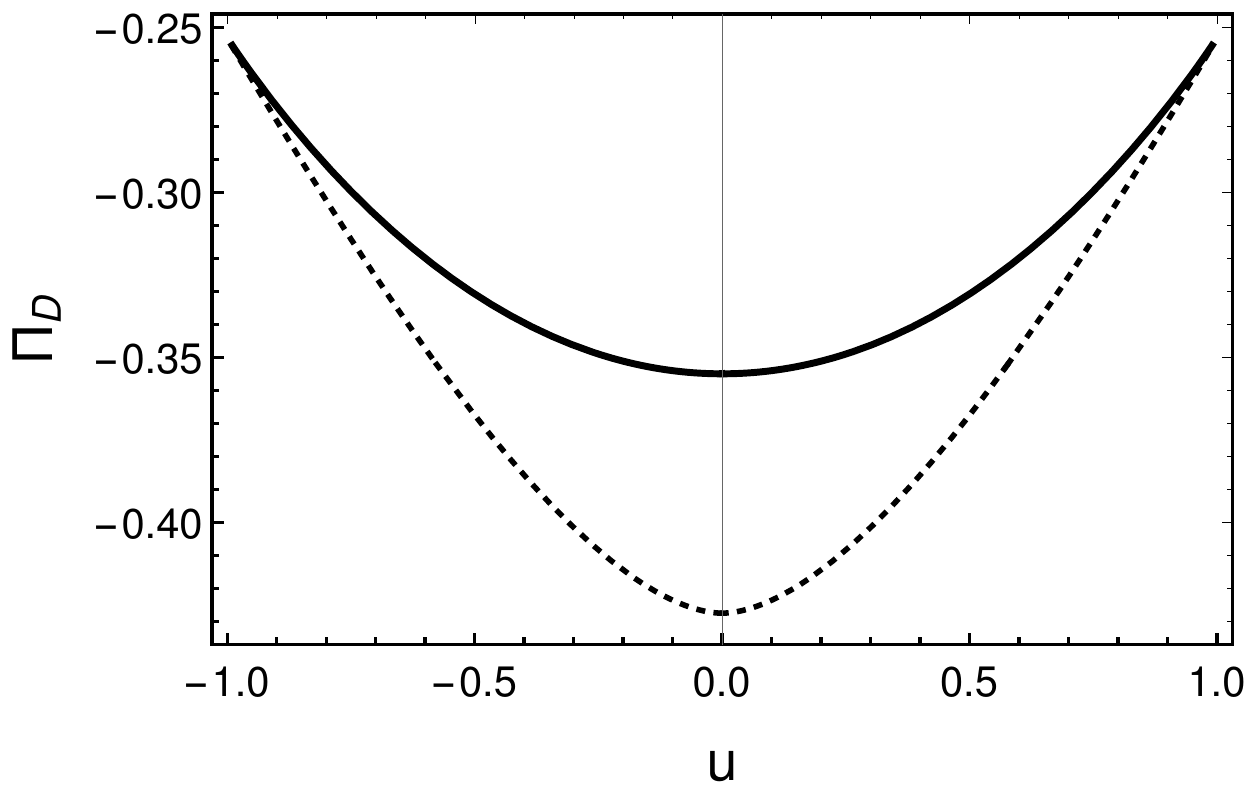}

\includegraphics[scale=0.6]{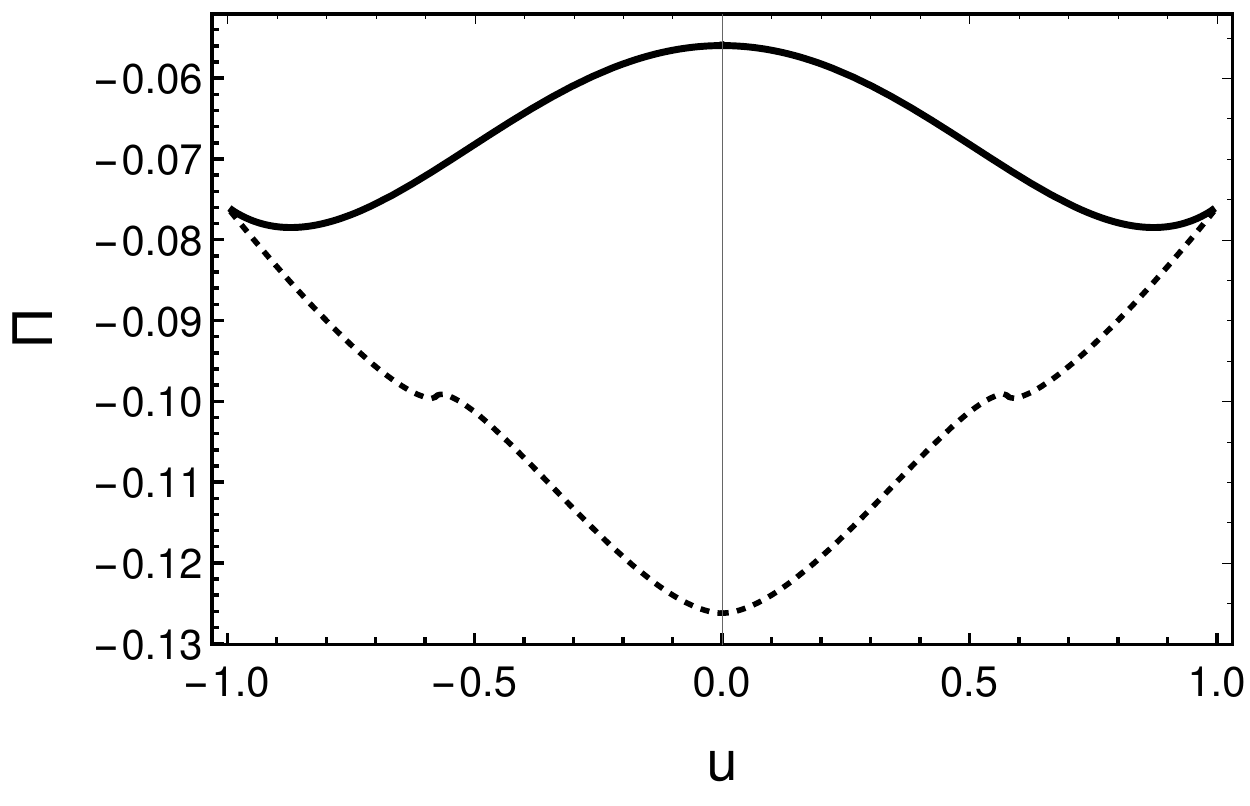}

\caption{The renormalized tadpole contributions $\Pi$ as a function of $u$.
The tadpole $\Pi$ is the total result, $\Pi_S$ is the contribution
from the scalar field $\Phi$, $\Pi_{\rm spec}$ is from scalar
spectator fields $\chi_i$ and $\Pi_D$ is from Dirac spectator fields $\Psi_i$.
Graphs with solid lines show the tadpole corrections $\Pi$ that include gradient effects;
dotted graphs show the corrections  $\Pi^{\rm hom}$, where these effects are ignored.}
\label{fig:Pi} 
\end{figure}

The tadpole correction is ${\Pi}\,\varphi^{(0)}$, where the classical bounce $\varphi^{(0)}$ is an
odd function and $\Pi$ is an even function about the center of the bubble wall.
In Fig.~\ref{fig:Pi}, we show the total ${\Pi}$ {along with} the contributions
from the different species that run in the loops.

We compare these full one-loop results with the tadpole
functions without gradient corrections. For each value of $u$, we calculate these
assuming a constant background $\varphi=\varphi^{(0)}(u)$. For the field $\Phi$, the
squared mass may become negative for a certain range of $u$ such that the resulting tadpole function acquires an
imaginary part that we do not show in the diagrams. The imaginary part arises if we assume a constant field configuration, because there is a continuum of negative modes contributing to the Gaussian integrals
in the tachyonic region where $-\mu^2+(\lambda/2)\varphi^2<0$. It is, however, an artefact, since constant configurations
do not correspond to extremal points (or saddle points) of the effective action away from the minima of the potential at $\pm\,v$.
In turn, as explained in Ref.~\cite{Garbrecht:2015oea}, imaginary parts, save the one associated with the negative mode, do not appear
when calculating the loop diagrams for fluctuations around the full bounce solutions that also
account for gradients, i.e.~when expressing the loops in terms of the
Green's function solutions in the bounce background.

The parameters chosen for Fig.~\ref{fig:Pi}  are
\begin{gather}
\mu\ =\ 1\;,\ \lambda\ =\ 2\;,\ \kappa\ =\ 0.5\;,\nonumber\\
\alpha\ =\ 0.5\;,\ N_{\Psi}\ =\ N_{\chi}\ =\ 10\;,
\label{benchmarkpoint}
\end{gather}
a set which we also use as a benchmark point for the remaining numerical results
presented in this section.
The bumps around $u=\pm\,0.4$ in the graph of $\Pi_S$ without gradient contributions
are due to the transition to the tachyonic region in the classical potential.
This causes a divergence in the derivative of $\Pi_S$.
Note that, in the graph of the full one-loop result for $\Pi_S$, which accounts for gradient corrections,
these bumps are absent because the field gradients counteract the tachyonic instability.
For the fermionic and scalar spectator fields, there are no tachyonic regions across the
bubble wall.

For the fermionic and scalar spectator fields, we can see from Fig.~\ref{fig:Pi} that the gradient effects
suppress the tadpole corrections when compared with the corrections without gradients.
For the Higgs field, the tadpole correction appears to be enhanced compared to the real part
of the loop correction without gradients. One should note, however, that this is not directly comparable with corrections from the spectator fields because of the tachyonic modes and the imaginary
part in the Coleman-Weinberg potential that has no direct physical interpretation.

\begin{figure}
\centering
\includegraphics[scale=0.6]{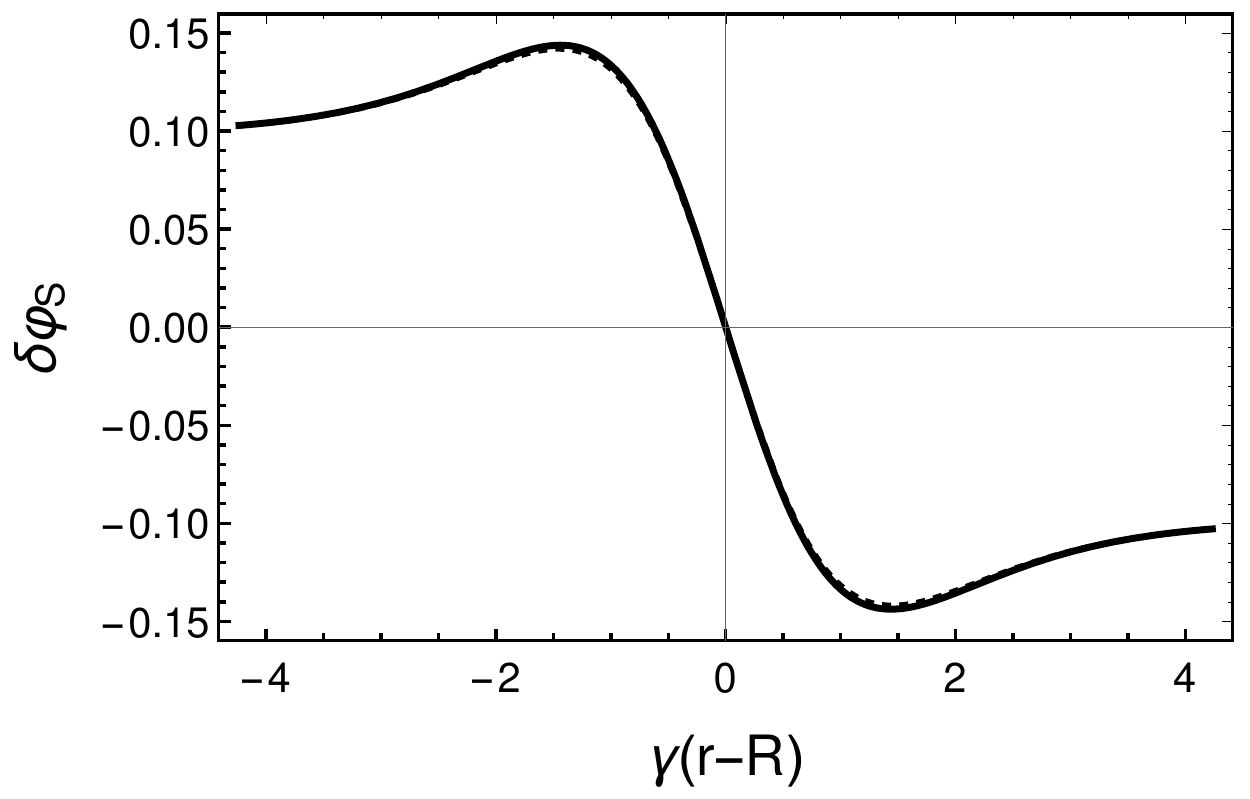}

\includegraphics[scale=0.6]{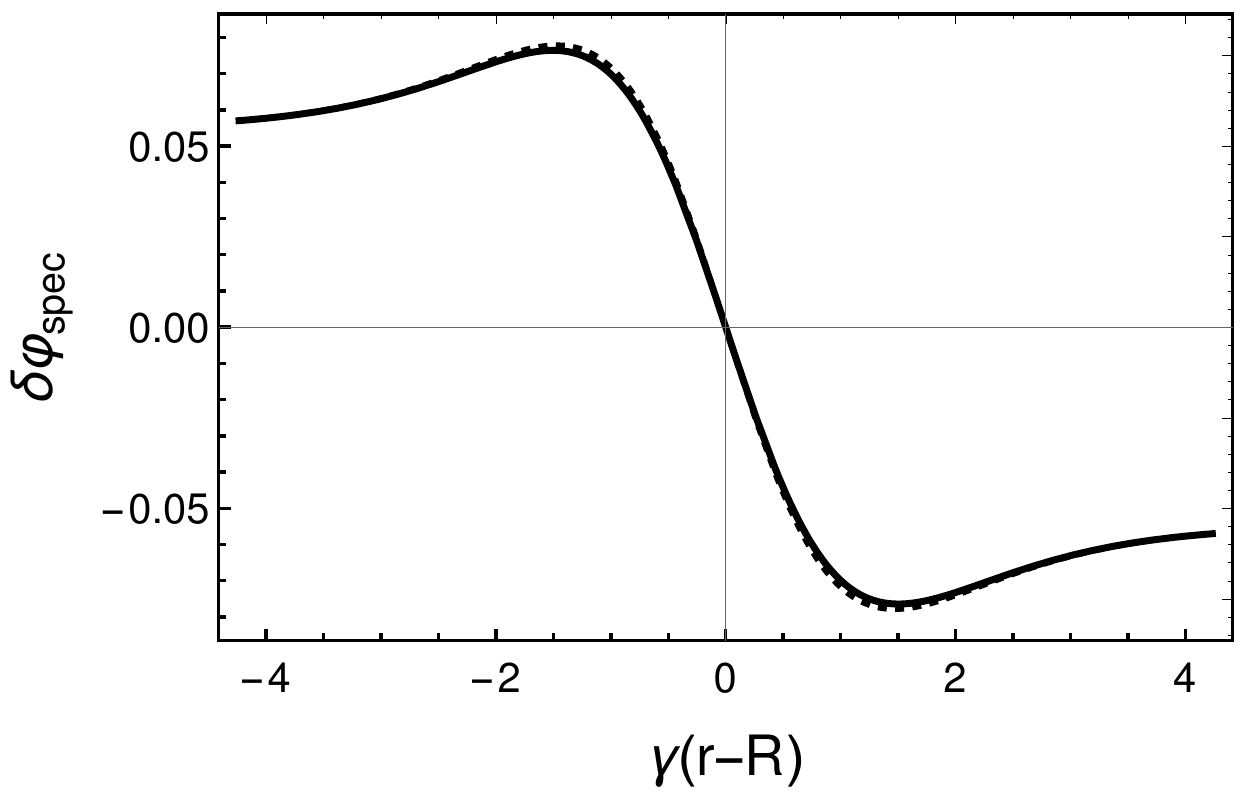}

\includegraphics[scale=0.6]{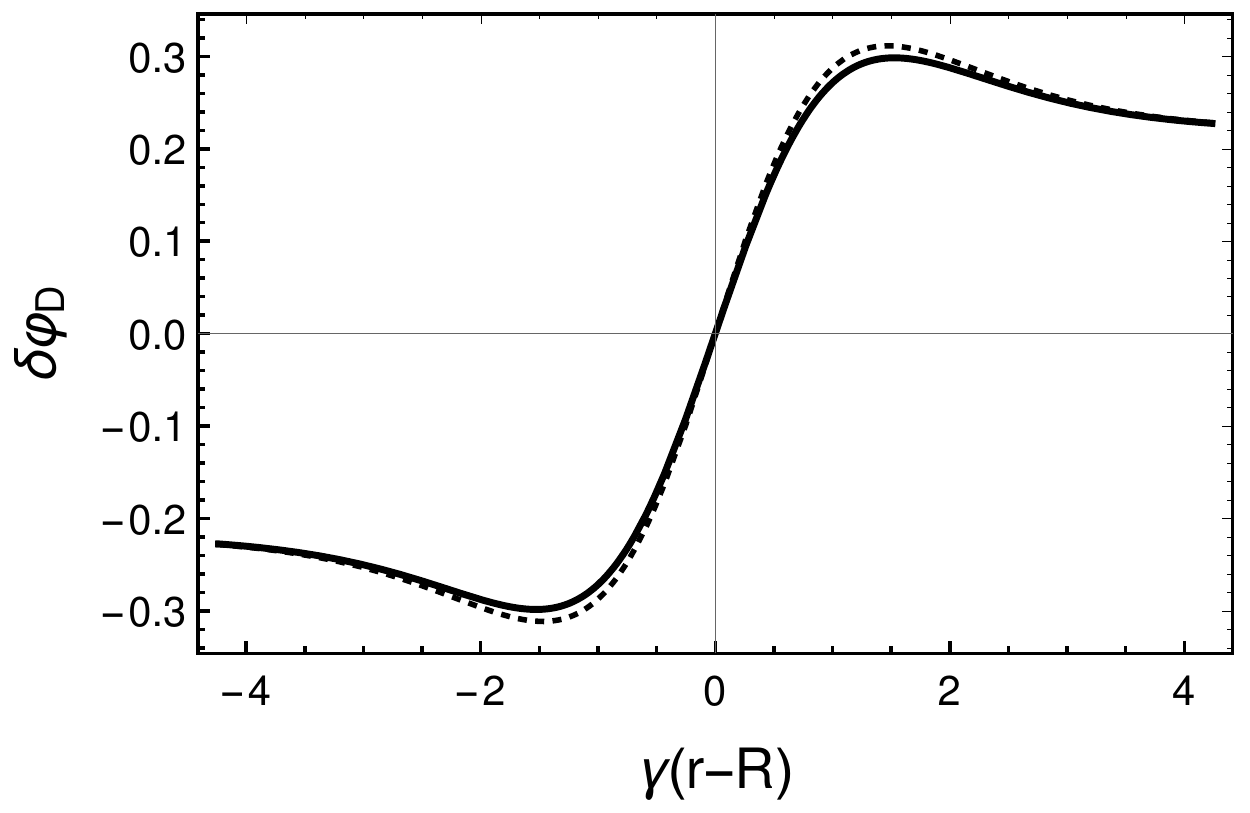}

\includegraphics[scale=0.6]{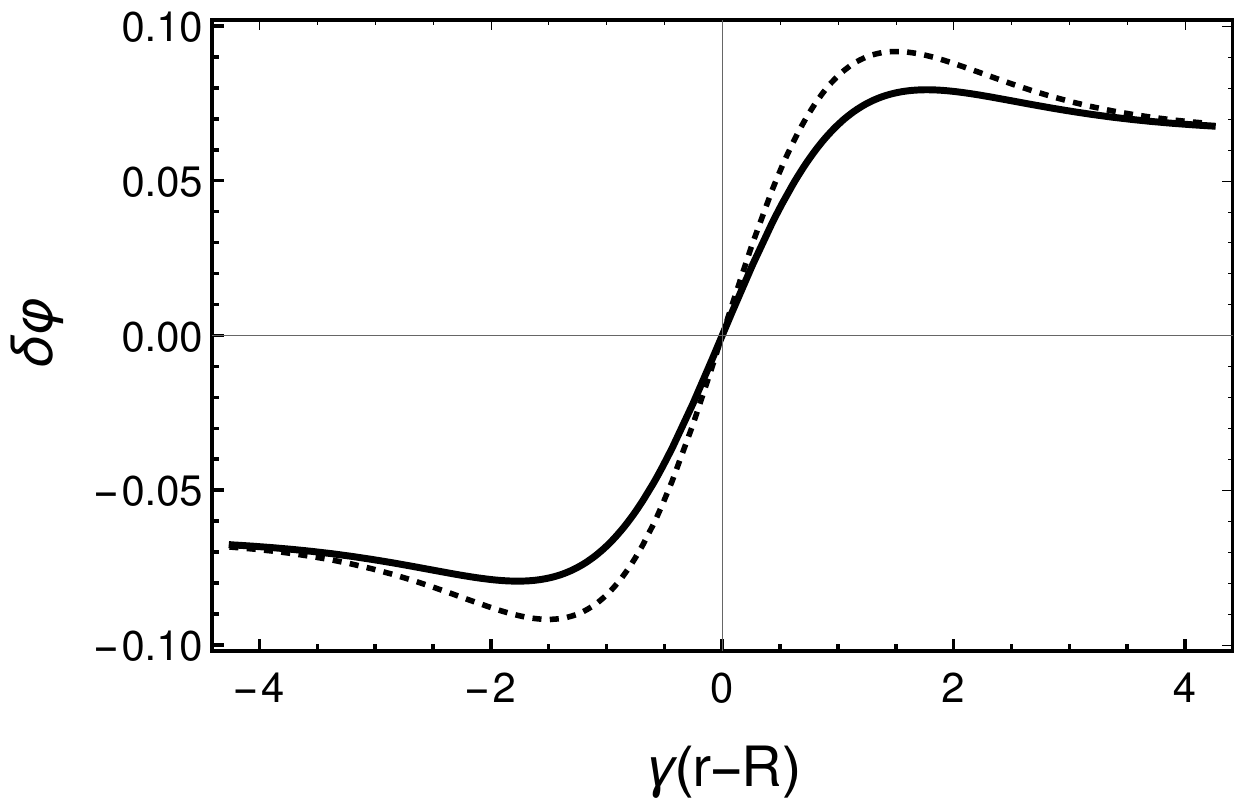}

\caption{The correction $\delta\varphi$ to the bounce as a function of $\gamma(r-R)$ with ($\delta\varphi$, solid) and without ($\delta\varphi^{\rm hom}$, dotted) gradients in separate panels for the contributions $\delta\varphi_S$
from the Higgs field, $\delta\varphi_{\rm spec}$ from the scalar spectators, $\delta\varphi_D$ from the Dirac spectators and for the total correction $\delta\varphi$.} 
\label{fig:withorwithoutgradient} 
\end{figure}

\begin{figure}[t!]
\centering
\vspace{1.5em}

\includegraphics[scale=0.6]{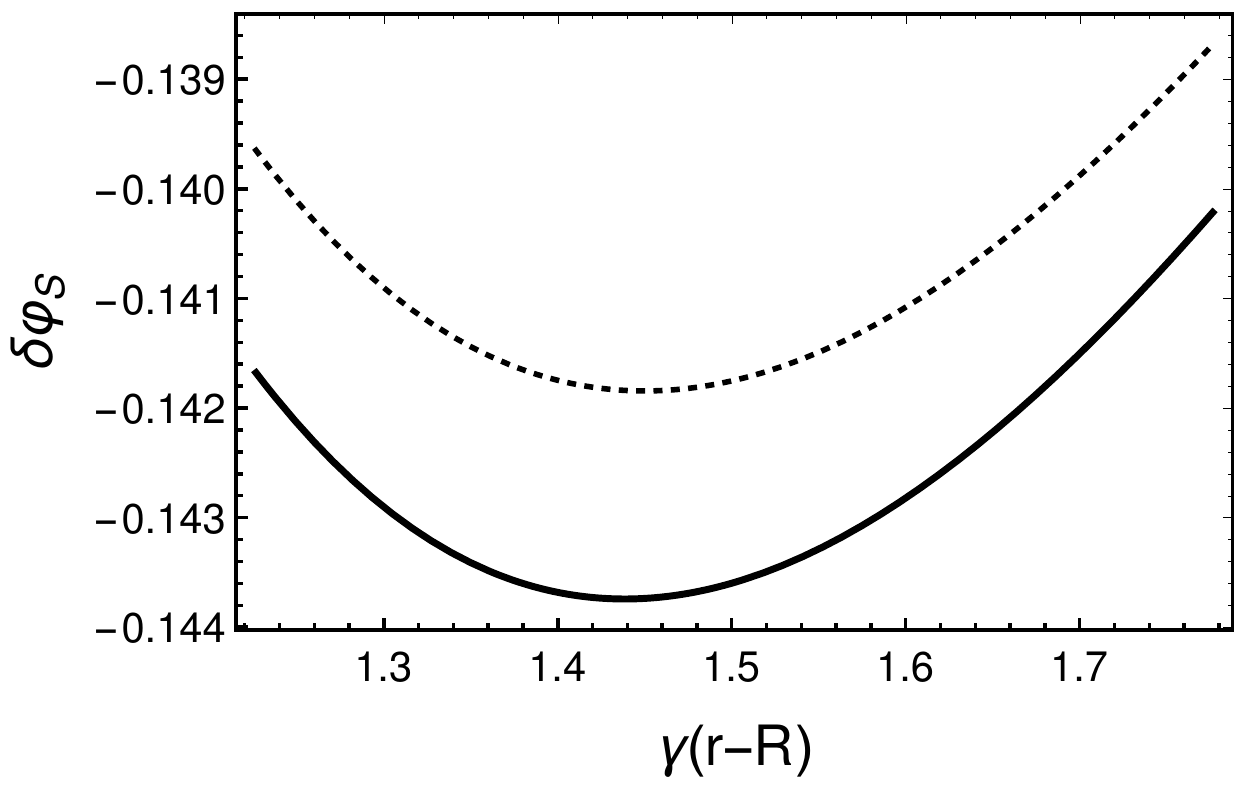}\\ \vspace{1.5em}

\includegraphics[scale=0.6]{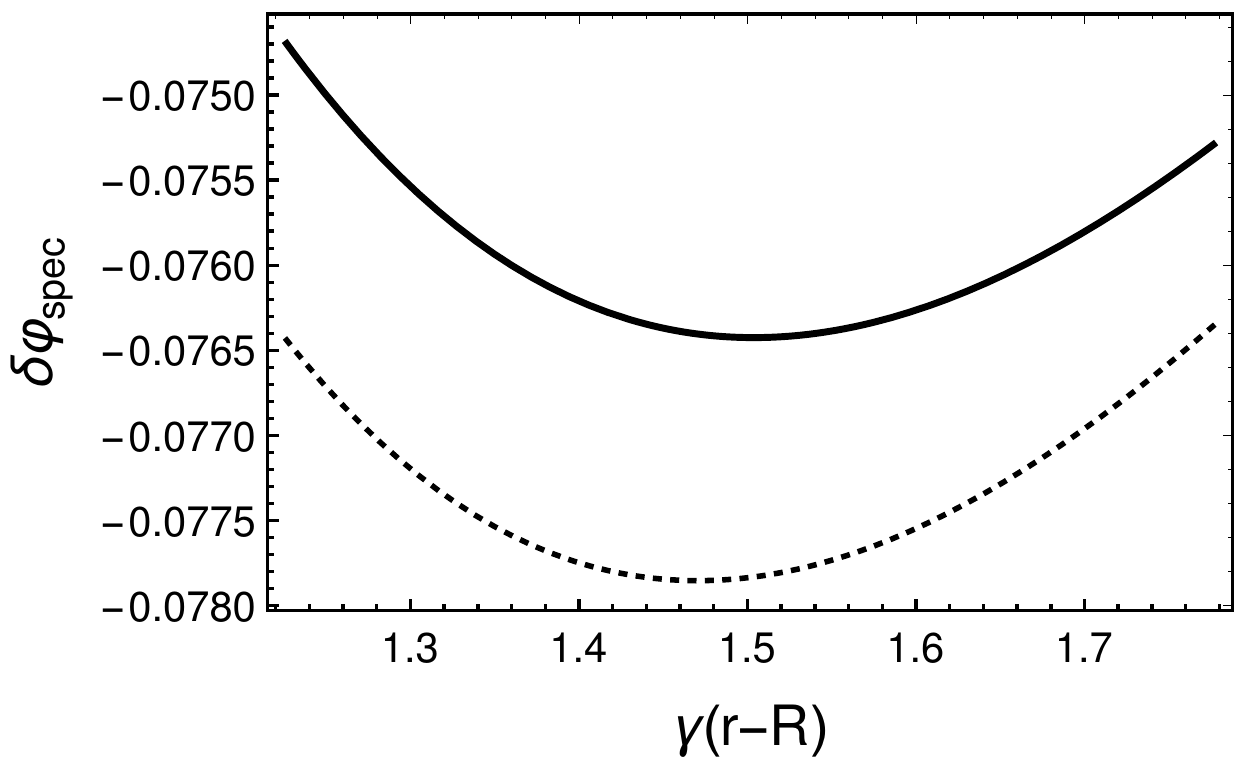}\\ \vspace{1.5em}

\includegraphics[scale=0.6]{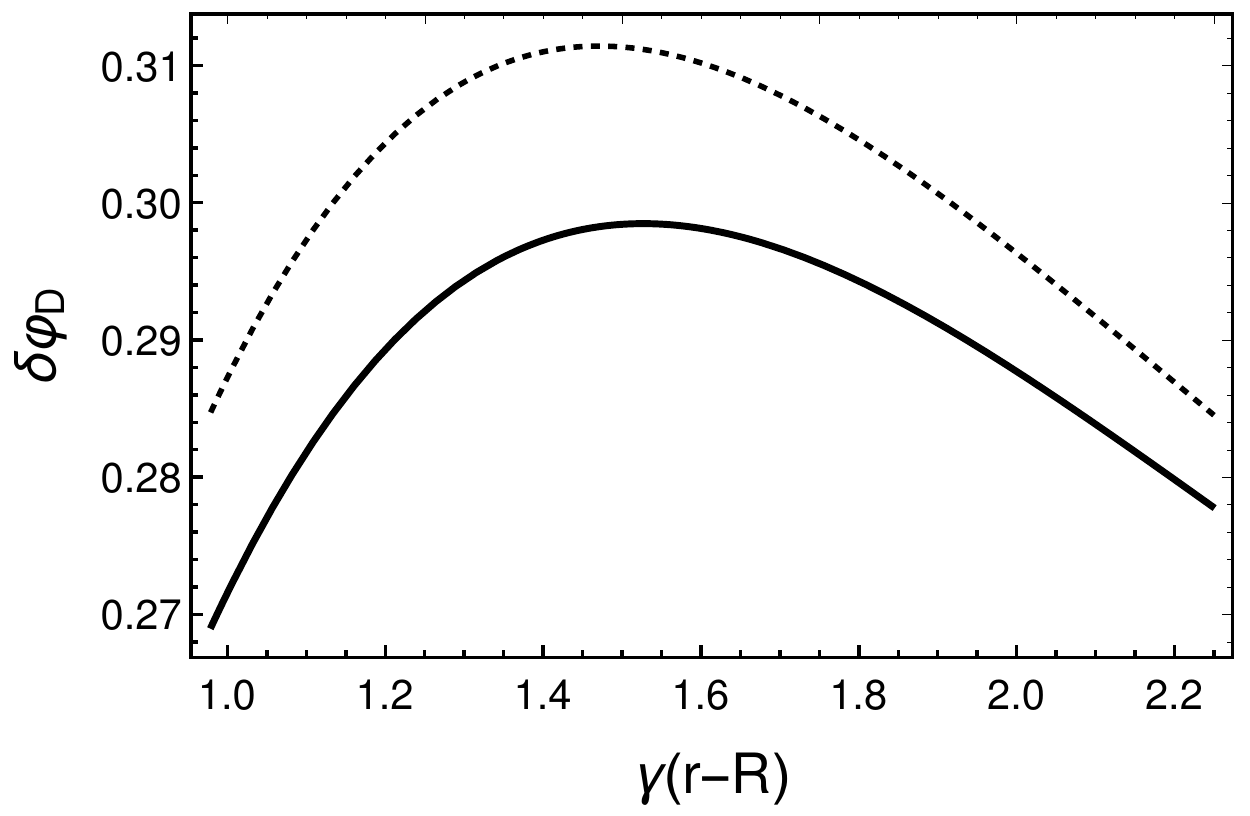}\\ \vspace{1.5em}

\includegraphics[scale=0.6]{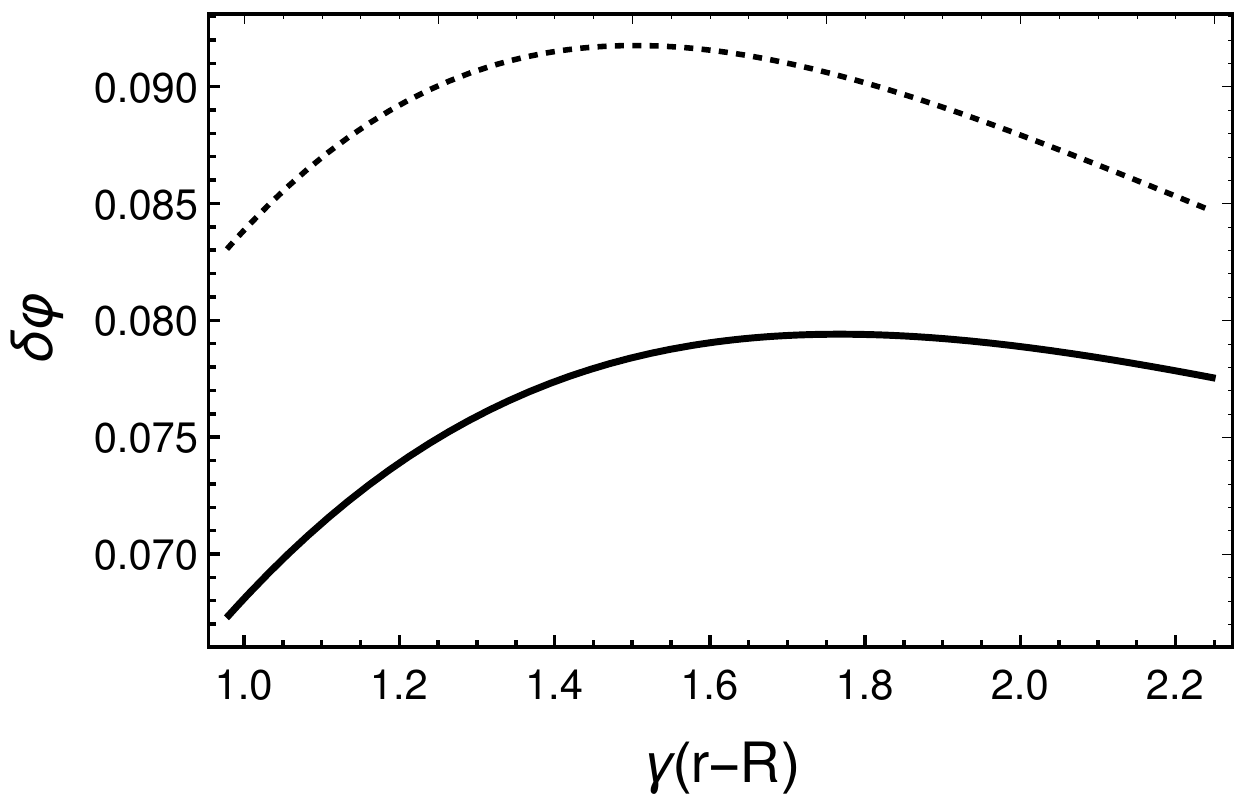}

\caption{Same as Fig.~\ref{fig:withorwithoutgradient}, showing the detail around the point where the radiative correction
to the bounce is maximal.} 
\label{fig:withorwithoutgradientzoomin} 
\end{figure}

The parameters in Eq.~\eqref{benchmarkpoint} have been chosen such that there
is a substantial amount of accidental cancellation between the fermion and scalar loop contributions.
In Fig.~\ref{fig:Pi}, we see this in the plot of the total tadpole correction $\Pi$,
from which we also conclude that, in such a situation, the relative impact of the
gradient corrections can be of order one.
For all tadpole corrections $\Pi$, we note that these are largest around the point $u=0$, corresponding
to $\varphi^{(0)}=0$. From Eq.~\eqref{correctedbounce}, we also observe that it is the combination $\varphi^{(0)}\Pi(\varphi^{(0)};y)$ that acts as a source for the radiative correction. We can therefore
expect that the relative impact on the radiative correction $\delta\varphi$ to the bounce is
suppressed.

In Figs.~\ref{fig:withorwithoutgradient} and~\ref{fig:withorwithoutgradientzoomin}, we compare the radiative corrections
to the bounce with and without gradient effects.
These corrections are calculated using Eq.~\eqref{deltavarphi}, where we substitute \smash{$\Pi_{S,D,{\rm spec}}$}
and \smash{$\Pi^{\rm hom}_{S,D,{\rm spec}}$} in order to obtain the particular contributions \smash{$\delta \varphi_{S,D,{\rm spec}}$} and \smash{$\delta \varphi^{\rm hom}_{S,D,{\rm spec}}$}. These are linear responses to the particular one-loop tadpoles, such that
$\delta \varphi=\delta \varphi_{S}+\delta \varphi_{D}+\delta \varphi_{\rm spec}$ and accordingly
for the corrections with the superscript ``hom'' that exclude gradient effects.
We observe that, for the Dirac spectator fields $\Psi_i$ (subscript ``$D$'') and the scalar spectators $\chi_i$ (subscript ``spec''), inclusion of the gradient effects smooths the field profile, that is the turning points in the functions $\delta\varphi_{D,{\rm spec}}$ are softened relative to those in $\delta\varphi^{\rm hom}_{D,{\rm spec}}$. For the correction 
$\delta\varphi_S$ from Higgs loops, there appears to be an opposite effect. However, we recall that we have dropped the imaginary parts from $\Pi^{\rm hom}_S$ such that this
contribution is not directly comparable with the one from the spectator fields. We also note the larger
total correction from the Dirac spectators when compared with their scalar counterparts. The apparent relative
factor of minus four can, of course, be accounted for by the number of Dirac spinor degrees of freedom and the opposite
sign of the fermion loop. [The squared mass of the Dirac spectator fields across the wall is $\kappa^2 \varphi^2(z)$,
for the scalar spectators it is $\alpha\, \varphi^2(z)/2$. For the values of $\alpha$ and $\kappa$ chosen in Eq.~\eqref{benchmarkpoint},
these are therefore coincident, such that the relative factor of minus four is actually exact when ignoring gradients.] The relative impact of the gradient corrections on $\delta\varphi$ turns out to be about twice as large for the Dirac spectators when compared with the scalar spectators.

\subsection{Corrections to the action}

\begin{table}

\begin{center}

    \begin{tabular}
    { | >{\centering\arraybackslash} p{3em} || >{\centering\arraybackslash} p{7em}  | >{\centering\arraybackslash} p{7em} | >{\centering\arraybackslash} p{7em} |}
    \hline
    &(i) no gradients  & (ii) gradients & [(i)-(ii)]/(i) \\ \hline\hline
 
    $\bar{B}^{(1)}_{\rm S}$  & $-\,0.583$  & $-\,0.585$ & $0.34\%$ \\ \hline
    
    $\bar{B}^{(1)}_{\rm spec}$  & $-\,0.320$  & $-\,0.324$ & $1.25\%$ \\ \hline
    
  $\bar{B}^{(1)}_{\rm D}$ & $1.278$  & $1.345$ & $5.24\%$ \\
    \hline
    
   $\bar{B}^{(1)}$ & $0.375$ & $0.436$ & $16.3\%$ \\ \hline
    
    $\bar{B}^{(2)}$ & $5.085\times 10^{-4}$  &  $-\,5.719\times 10^{-3}$  & \\ \hline

    \end{tabular}
    
\end{center}

\caption{Comparison of $\bar{B}^{(1)}$ and $\bar{B}^{(2)}$ (i) without gradients (i.e.~based on the
Green's functions in homogeneous backgrounds) and (ii) with gradients (i.e.~based on the Green's function in the background of the tree-level soliton). These quantities are computed for the benchmark point~\eqref{benchmarkpoint}. {We draw attention to the fact that the values of $\bar{B}^{(2)}$ differ in sign for cases (i) and (ii), leading to a relative increase in the tunneling rate when the gradients are included.} For completeness, the value of $\bar{B}$ --- which does not differ between cases (i) and (ii) --- is approximately 2.828.}
    \label{TableBB}
    
\end{table}

We now investigate the gradient effects on the action and the tunneling rate.
For this purpose, we define the contributions to the action per unit area of bubble surface
\begin{align}
\bar B\ =\ \frac{B}{2\pi^2 R^3}\;,\qquad \bar B^{(1,2)}\ =\ \frac{B^{(1,2)}}{2\pi^2 R^3}\;,
\end{align}
and accordingly for the individual contributions from the Higgs, Dirac and scalar spectator fields,
as well as for
the loop corrections $\bar B^{(1,2)\rm hom}$ derived from the Green's functions in the homogeneous background.
For the benchmark parameters~\eqref{benchmarkpoint}, the results for the various corrections to the action
are compared in Table~\ref{TableBB}.

With or without gradient effects, we see that the
corrections $B^{(1)}$ for the scalar loops are negative (leading to an enhanced decay rate)
and for fermion loops, these are positive (leading to a suppressed decay rate). Since our renormalization
conditions fix $\lambda$ at the false vacuum, fermion loops lead to a running toward smaller $|\lambda|$ and scalar
loops toward larger $|\lambda|$ around the center of the bounce, where $\varphi=0$. From the tree-level action~\eqref{B:tree}, 
we see that larger (smaller) $\lambda$ lead to a faster (slower) decay rate, which qualitatively
explains this numerical observation.
As for the relative contributions from field gradients
to $B^{(1)}$ shown in  Table~\ref{TableBB}, we note that for Dirac spectators,
these are larger by a factor of about four in comparison
with the contribution from scalar spectators.

\begin{figure}
\centering
\includegraphics[scale=0.6]{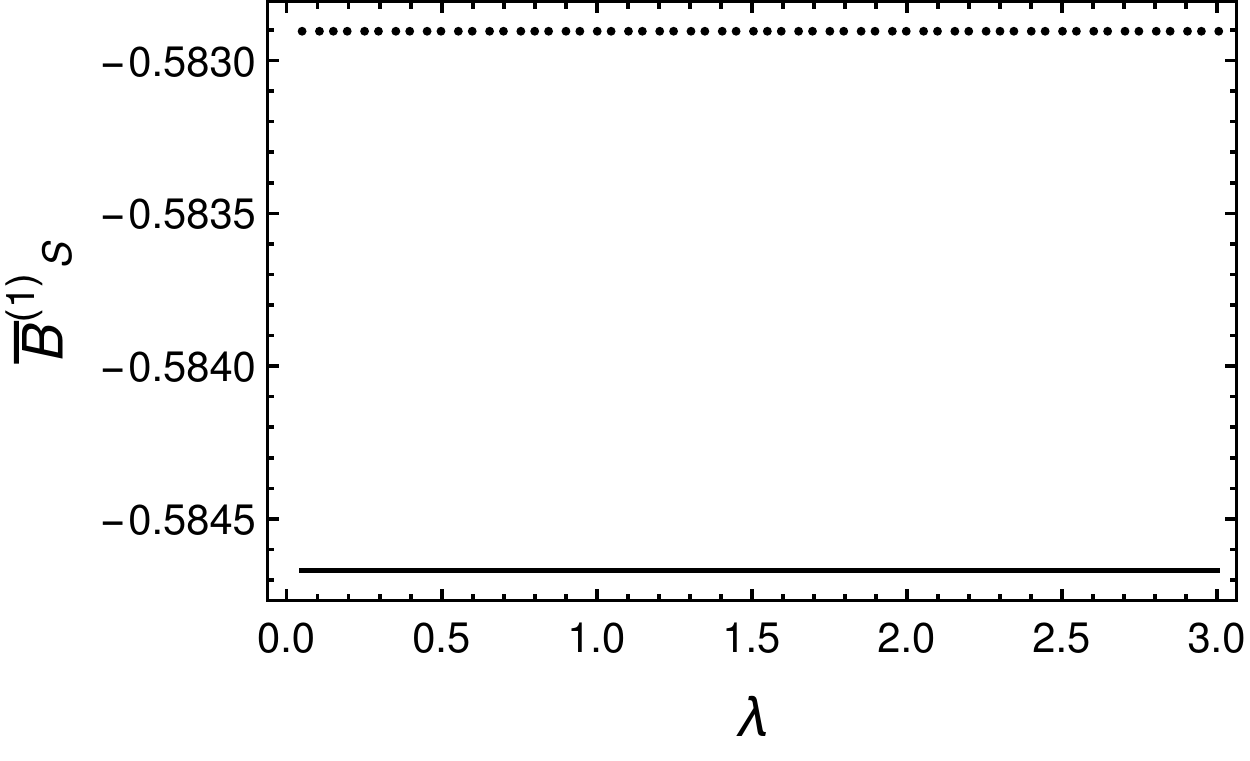}\\ \vspace{1em}

\includegraphics[scale=0.6]{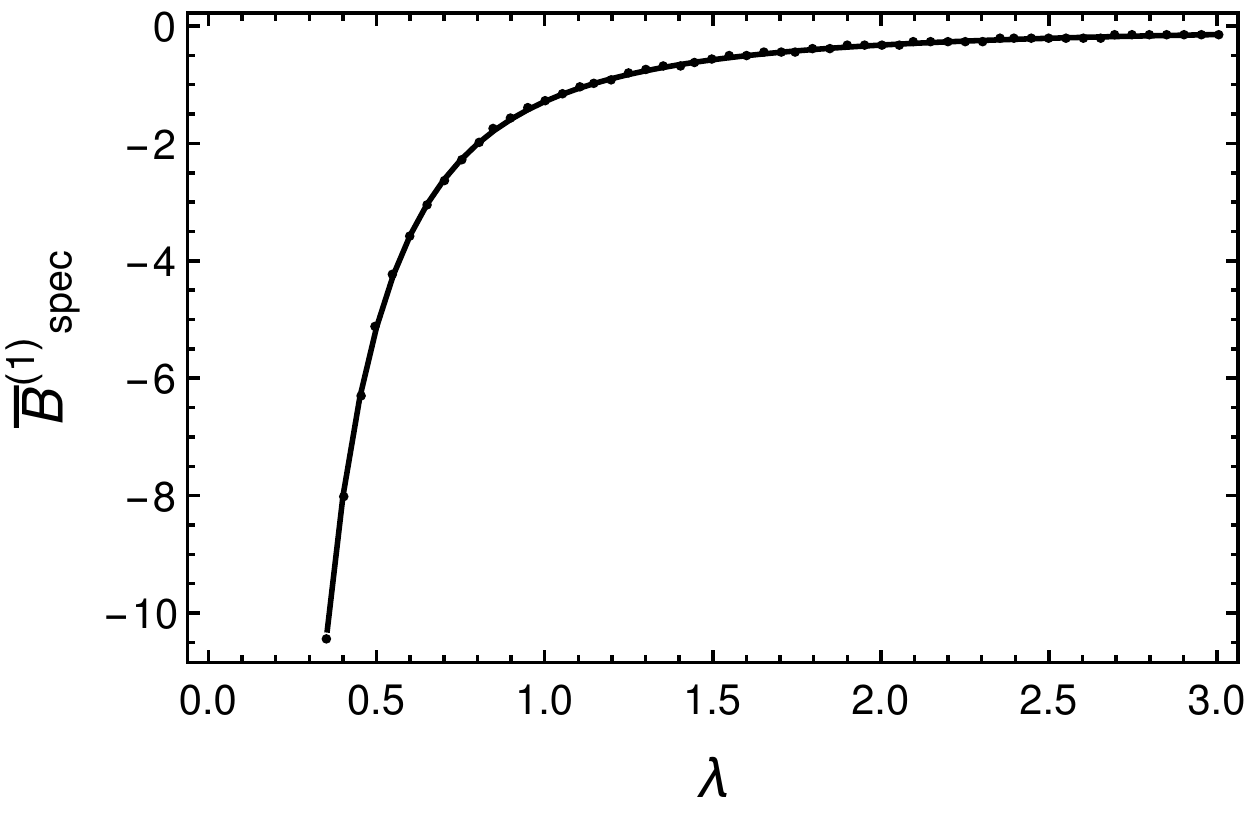}\\ \vspace{1em}

\includegraphics[scale=0.6]{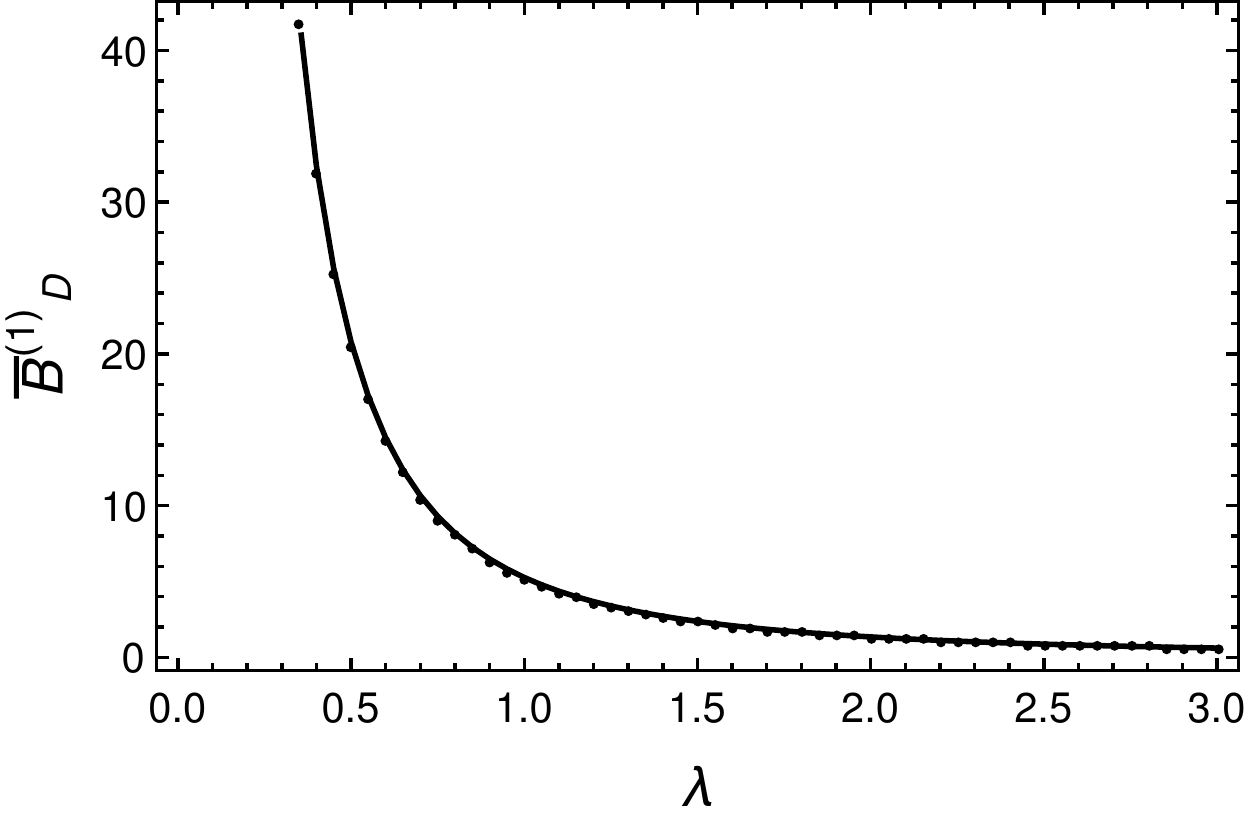}

\caption{Dependence of the one-loop corrections to the bounce action on $\lambda$.
Solid: $\bar{B}^{(1)}$ (i.e.~with gradients). Dotted:  $\bar{B}^{(1)\rm hom}$ (i.e.~without gradients). The remaining parameters besides $\lambda$ are given in Eq.~\eqref{benchmarkpoint}.} 
\label{fig:B1lambda} 
\end{figure} 

\begin{figure}
\centering

\includegraphics[scale=0.6]{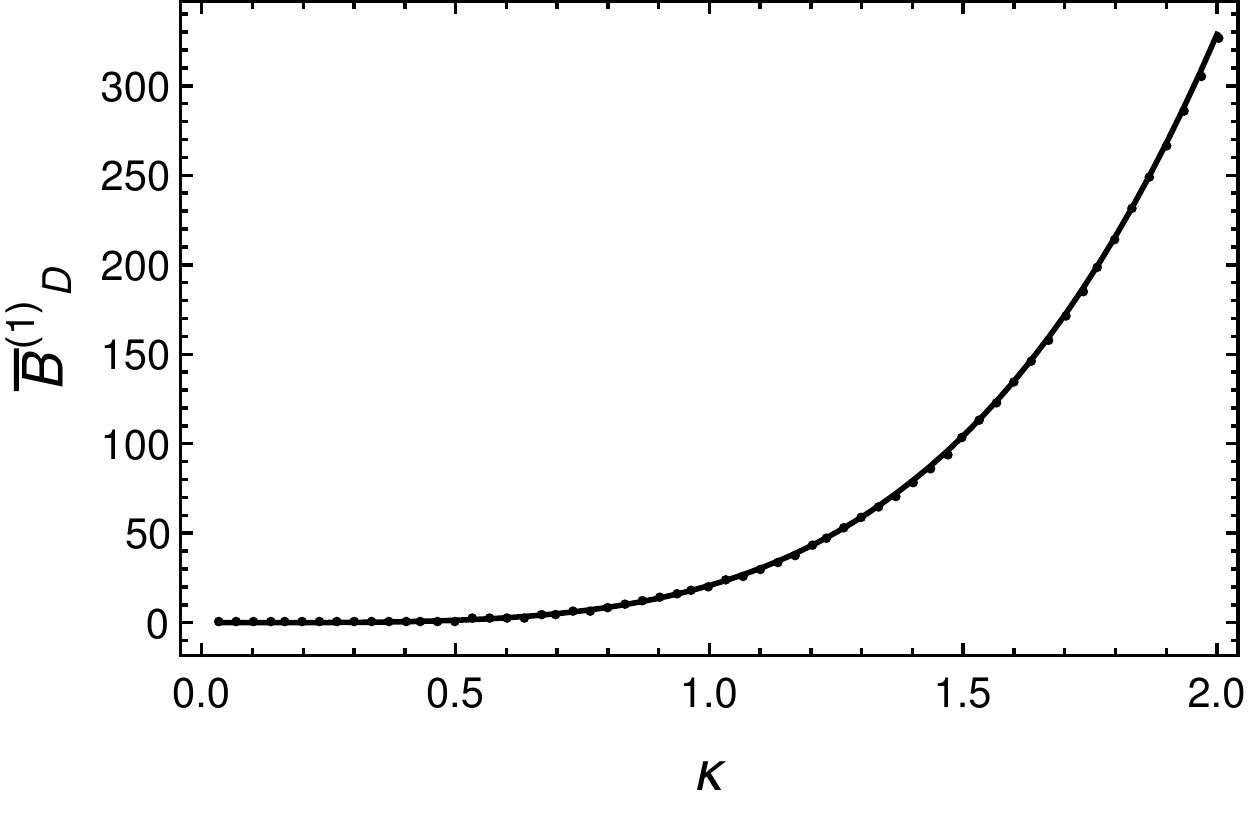}

\caption{Dependence of the one-loop corrections to the bounce action on $\kappa$.
Solid: $\bar{B}^{(1)}_D$ (i.e.~with gradients). Dotted:  $\bar{B}^{(1)\rm hom}_D$ (i.e.~without gradients). The remaining parameters besides $\kappa$ are given in Eq.~\eqref{benchmarkpoint}.} 
\label{fig:B1kappa} 
\end{figure}

\begin{figure}
\centering

\includegraphics[scale=0.6]{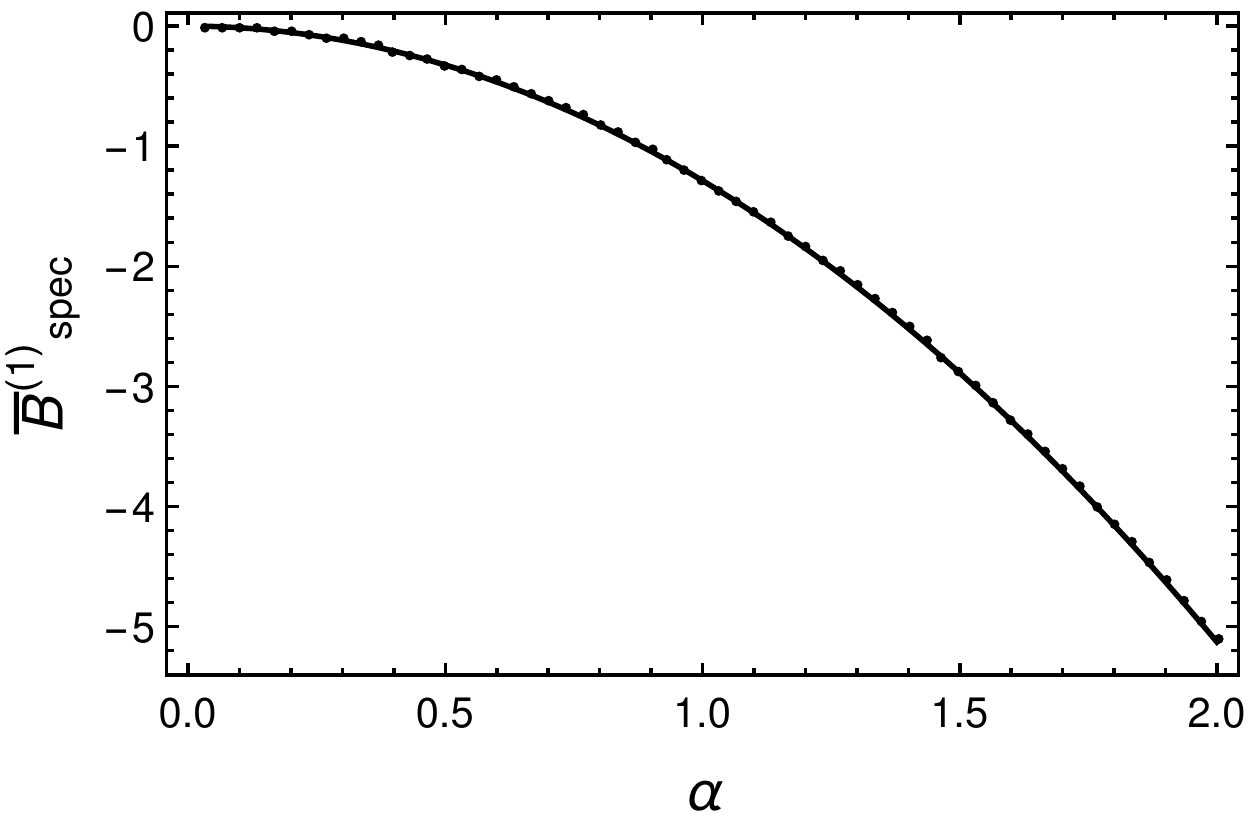}

\caption{Dependence of the one-loop corrections to the bounce action on $\alpha$.
Solid: $\bar{B}^{(1)}_{\rm spec}$ (i.e.~with gradients). Dotted:  $\bar{B}^{(1)\rm hom}_{\rm spec}$ (i.e.~without gradients). The remaining parameters besides $\alpha$ are given in Eq.~\eqref{benchmarkpoint}.} 
\label{fig:B1alpha} 
\end{figure}

\begin{figure}
\centering
\includegraphics[scale=0.6]{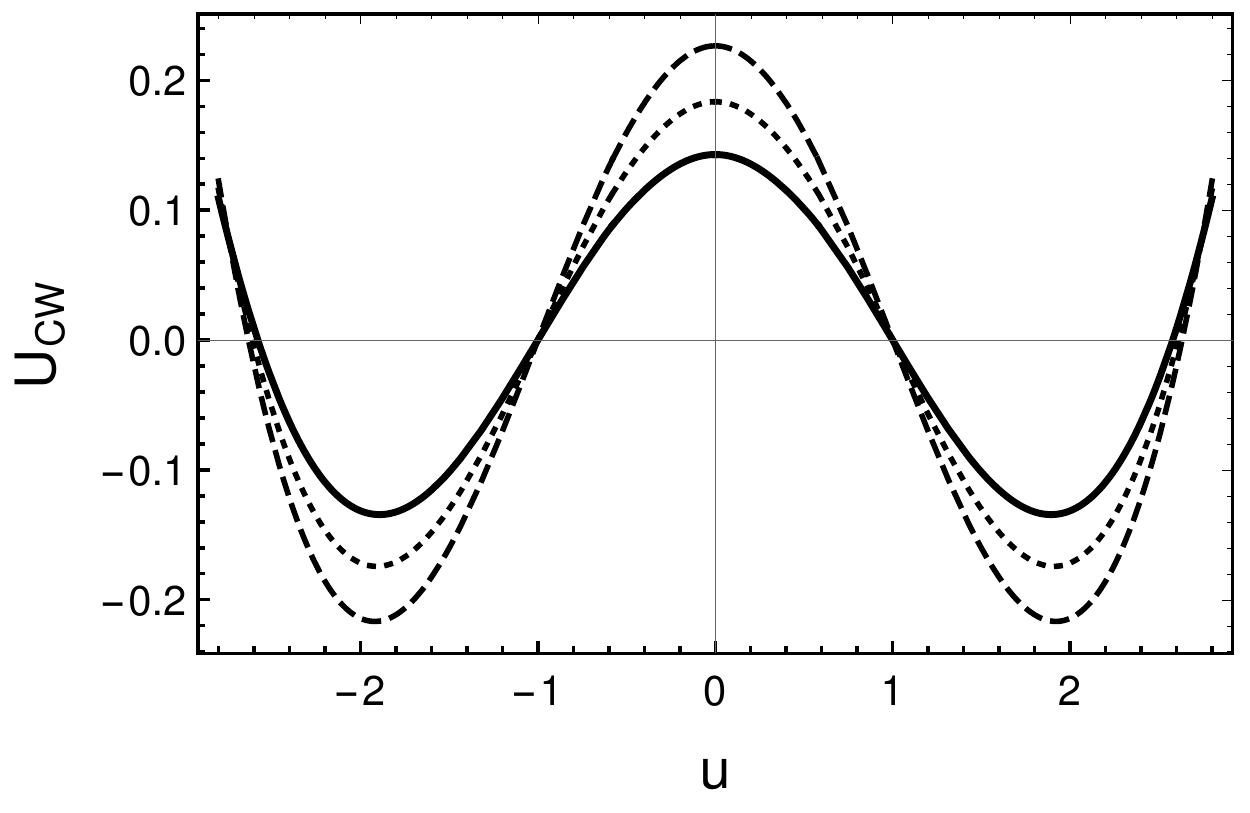}\\ \vspace{1em}

\includegraphics[scale=0.6]{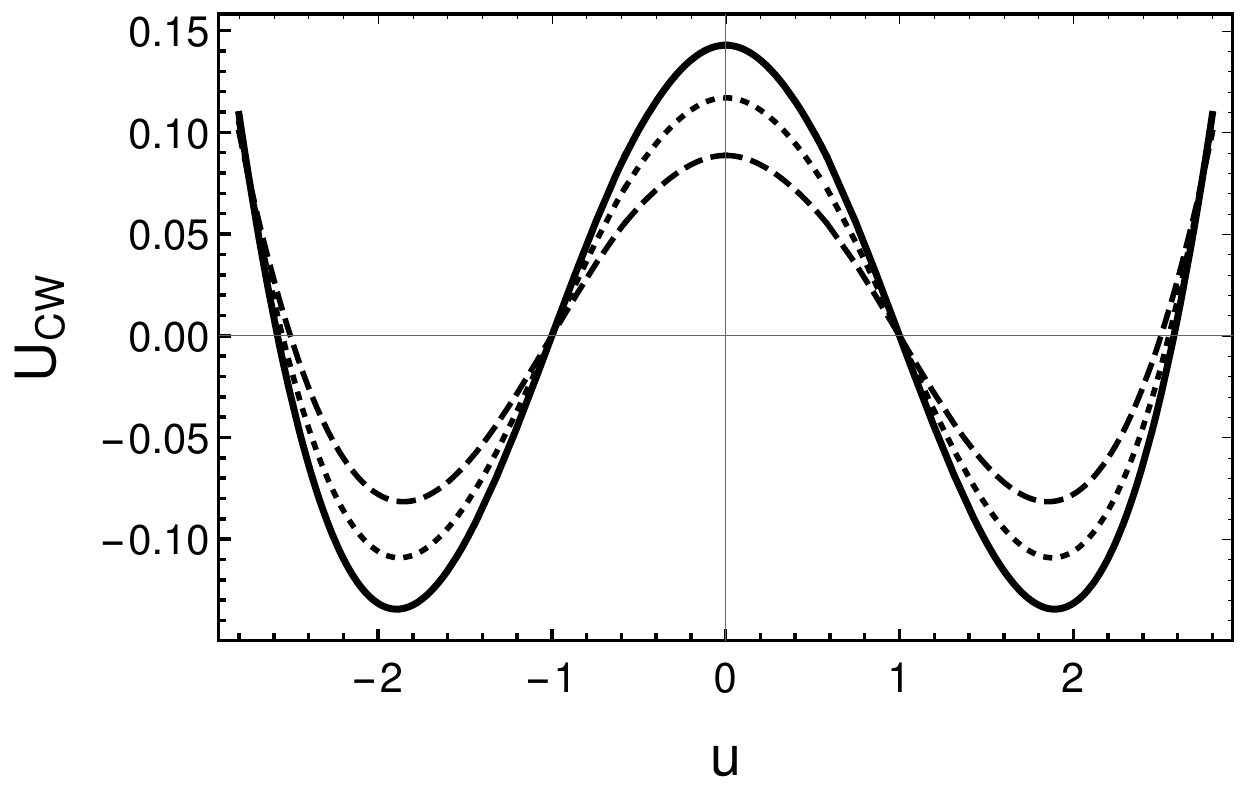}

\caption{The upper plot shows the shape of the barrier in the Coleman-Weinberg potential
$U_{\rm CW}$ for $\kappa=0.5$ (solid), $\kappa=0.51$ (dotted) and $\kappa=0.52$ (dashed), while the lower is for $\alpha=0.5$ (solid), $\alpha=0.55$  (dotted) and $\alpha=0.6$ (dashed). In both cases, the remaining parameters not being varied are chosen as in Eq.~\eqref{benchmarkpoint}.}
\label{fig:CWpotential} 
\end{figure}

\begin{figure}
\centering

\includegraphics[scale=0.6]{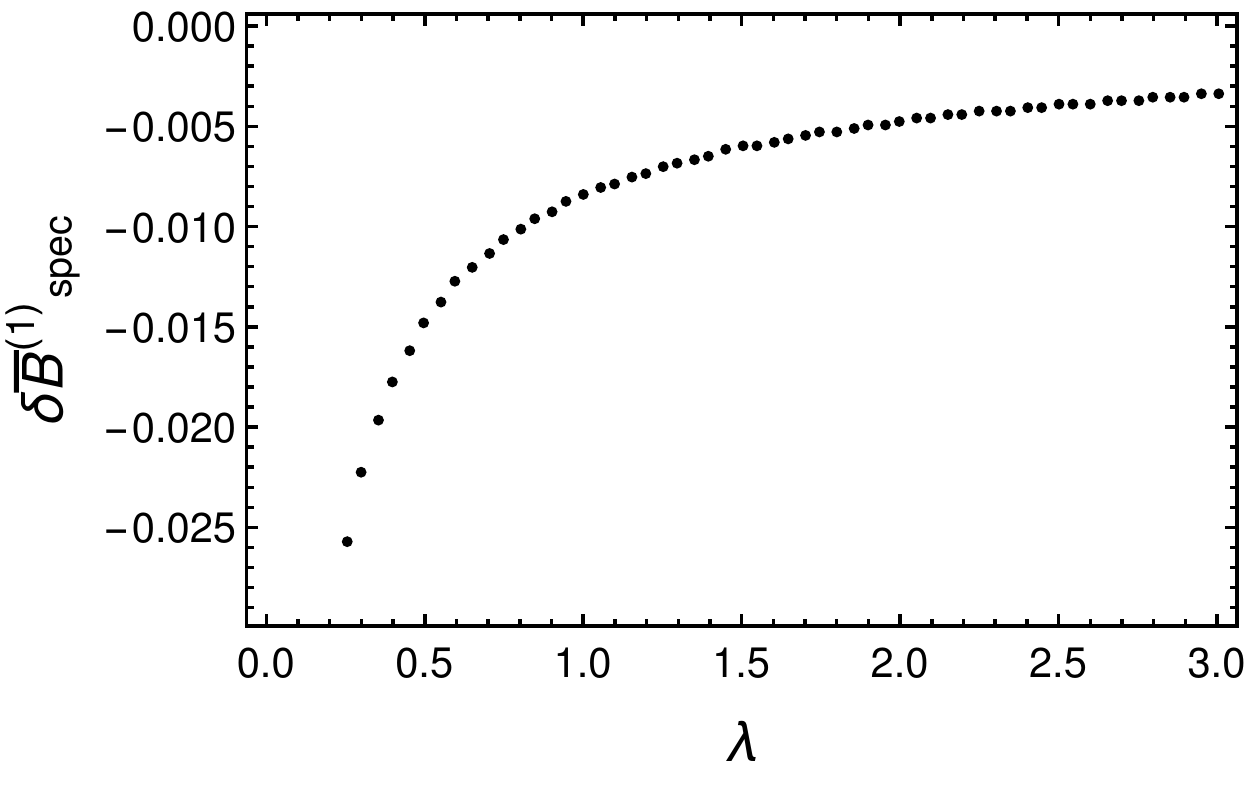}\\ \vspace{1em}

\includegraphics[scale=0.6]{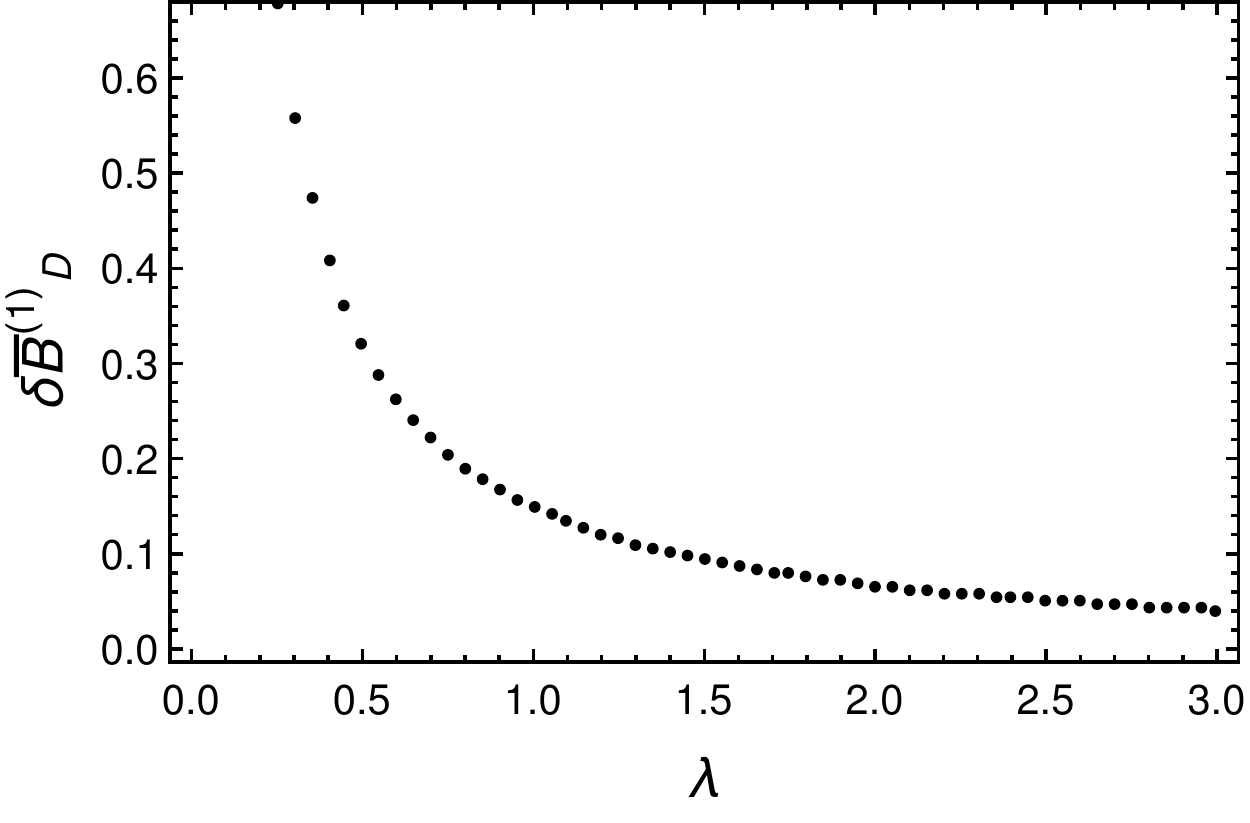}\\ \vspace{1em}

\includegraphics[scale=0.6]{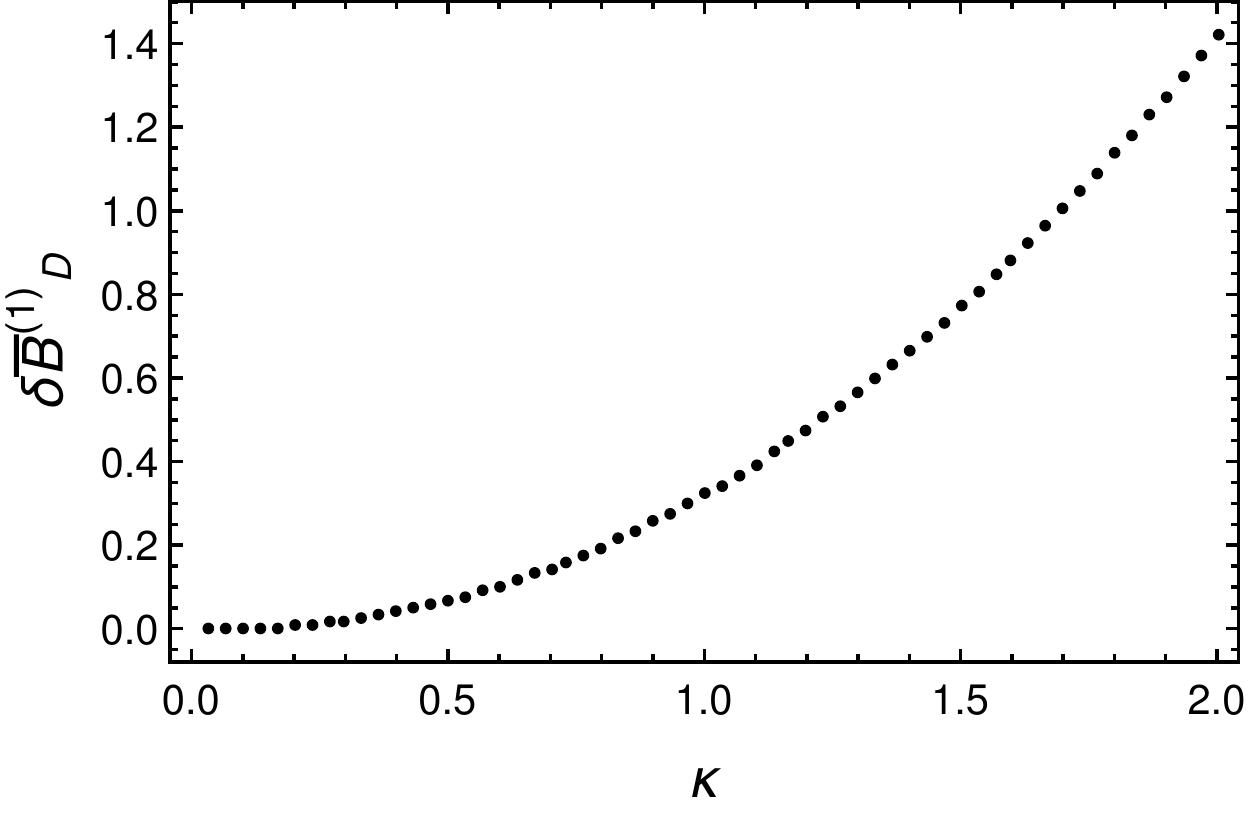}\\ \vspace{1em}

\includegraphics[scale=0.6]{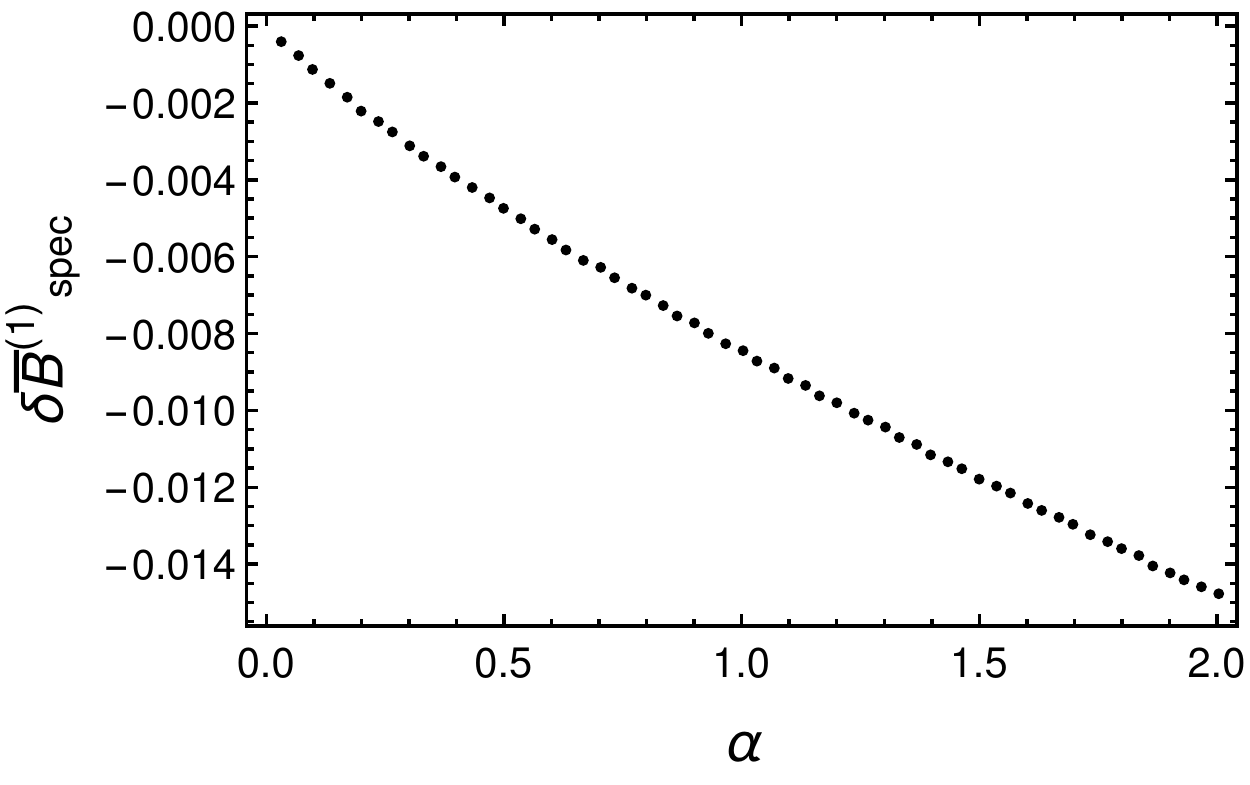}

\caption{The differences $\delta\bar{B}^{(1)}\equiv\bar{B}^{(1)}-\bar{B}^{(1){\rm hom}}$, which isolate the gradient contributions to the one-loop action. The remaining parameters besides those
varied on the horizontal axes are chosen as in Eq.~\eqref{benchmarkpoint}.} 
\label{fig:dletaB1parameter} 
\end{figure}

In Figs.~\ref{fig:B1lambda},~\ref{fig:B1kappa} and~\ref{fig:B1alpha}, we show the
results of varying the dimensionless couplings $\lambda$, $\kappa$ and $\alpha$ separately
about the benchmark point given in Eq.~\eqref{benchmarkpoint}. Only the Dirac spectator correction
\smash{$\bar B^{(1)}_D$} depends on $\kappa$ and only
the scalar spectator correction
\smash{$\bar B^{(1)}_{\rm spec}$} depends on $\alpha$, which is why these are the functions that
we present in Figs.~\ref{fig:B1kappa}
and~\ref{fig:B1alpha}. When comparing the Dirac and scalar spectator contributions in Figs.~\ref{fig:B1kappa} and~\ref{fig:B1alpha},
one should bear in mind that the squared masses depend on the couplings as
$\kappa^2 \varphi^2(z)$ and $\alpha\, \varphi^2(z)/2$, respectively. This explains the
stronger curvature in the plot for Dirac spectators.

Since a change in the self-coupling $\lambda$ of the Higgs field changes the shape of the bounce,
this is of relevance for all three types of one-loop corrections: \smash{$\bar B^{(1)}_S$}, \smash{$\bar B^{(1)}_D$}, and \smash{$\bar B^{(1)}_{\rm spec}$}. From Fig.~\ref{fig:B1lambda}, we see, however, that \smash{$\bar B^{(1)}_S$} is insensitive to the value of
$\lambda$. This is because, in our perturbation expansion, one loop corresponds to one power in $\lambda$,
such that \smash{$\bar B^{(1)}_S$} is one order higher in $\lambda$ compared to $\bar B$, which itself scales as $1/\lambda$, cf.~Eq.~\eqref{B:tree}. 
In principle, there are further logarithmic dependencies on $\lambda$ [cf., e.g., the effective potential~\eqref{UCW}]
but these turn out to cancel when performing the integral over $\D z$, as is shown
analytically in Refs.~\cite{Konoplich:1987yd,Garbrecht:2015oea,Bezuglov:2018qpq}. 

In accordance with the above remarks on the effect
of fermion and scalar loops on the decay rate, we can understand the dependence of
\smash{$\bar{B}^{(1)}_{\rm D}$} on $\kappa$ and of \smash{$\bar{B}^{(1)}_{\rm spec}$} on $\alpha$ in Figs.~\ref{fig:B1kappa} and~\ref{fig:B1alpha}. As $\kappa$ increases, \smash{$\bar{B}^{(1)}_D$} increases and therefore the decay rate decreases. This is due to a higher barrier from fermion fluctuations
in the effective potential for larger Yukawa couplings. Similarly, as $\alpha$ increases, \smash{$\bar{B}^{(1)}_{\rm spec}$} decreases and therefore the decay rate increases. This implies that there is a lower barrier in the effective potential for larger couplings $\alpha$.
To illustrate this point,
the dependence of the barrier in the Coleman-Weinberg potential~\eqref{UCW} on $\kappa$ and $\alpha$ is shown in Fig~\ref{fig:CWpotential}. 
We emphasize, however, that, while the Coleman-Weinberg effective potential can be used in order to
interpret the leading effects from radiative corrections, it does not include the subleading
gradient effects on the particles running in the loops.

We note that the bounce evolves rapidly about $z=0$, where the gradient effects on the tadpoles are largest, such
that their relative impact on the one-loop action, which is an integral quantity over $z$, is comparably small.
In Fig.~\ref{fig:dletaB1parameter}, we explicitly isolate the gradient effects by
plotting \smash{$\delta \bar B^{(1)}= \bar B^{(1)}-\bar B^{(1)\rm hom}$} for the one-loop contributions from the particular species.

\begin{figure}
\centering

\includegraphics[scale=0.6]{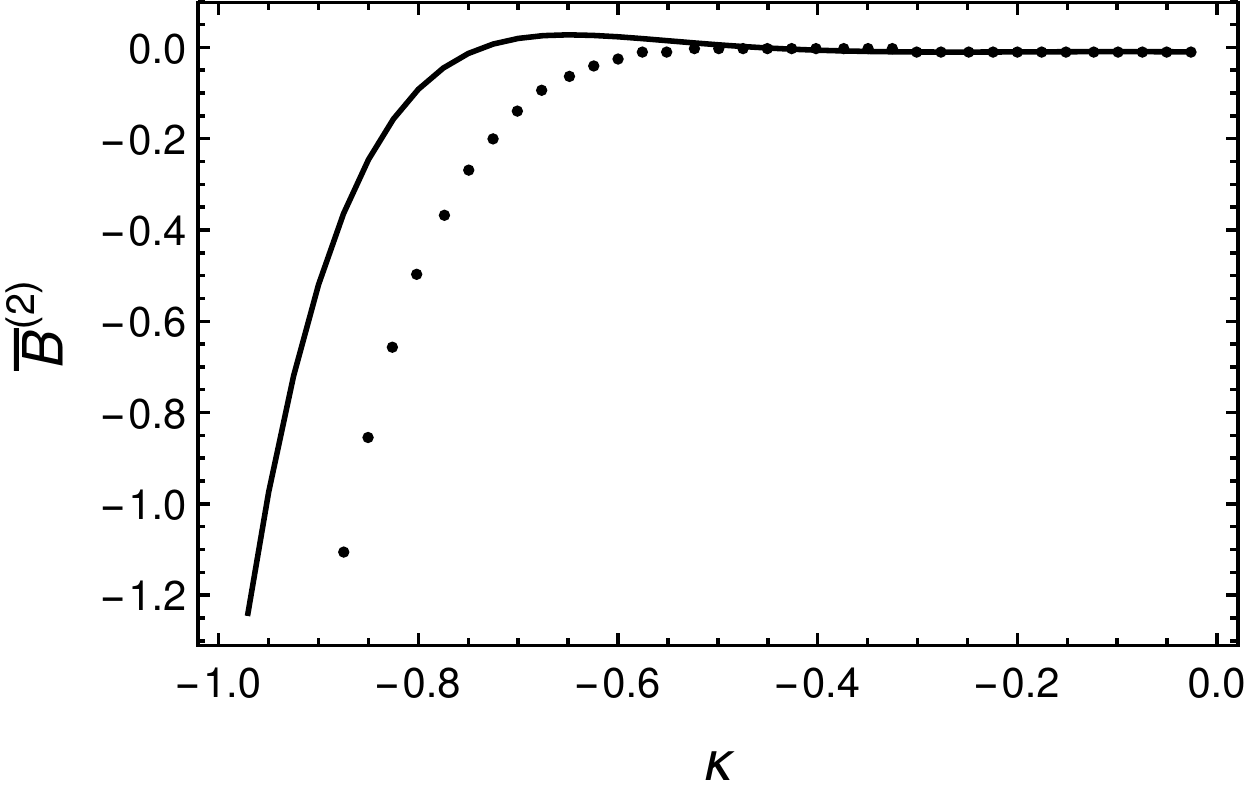}\\ \vspace{1em}

\includegraphics[scale=0.6]{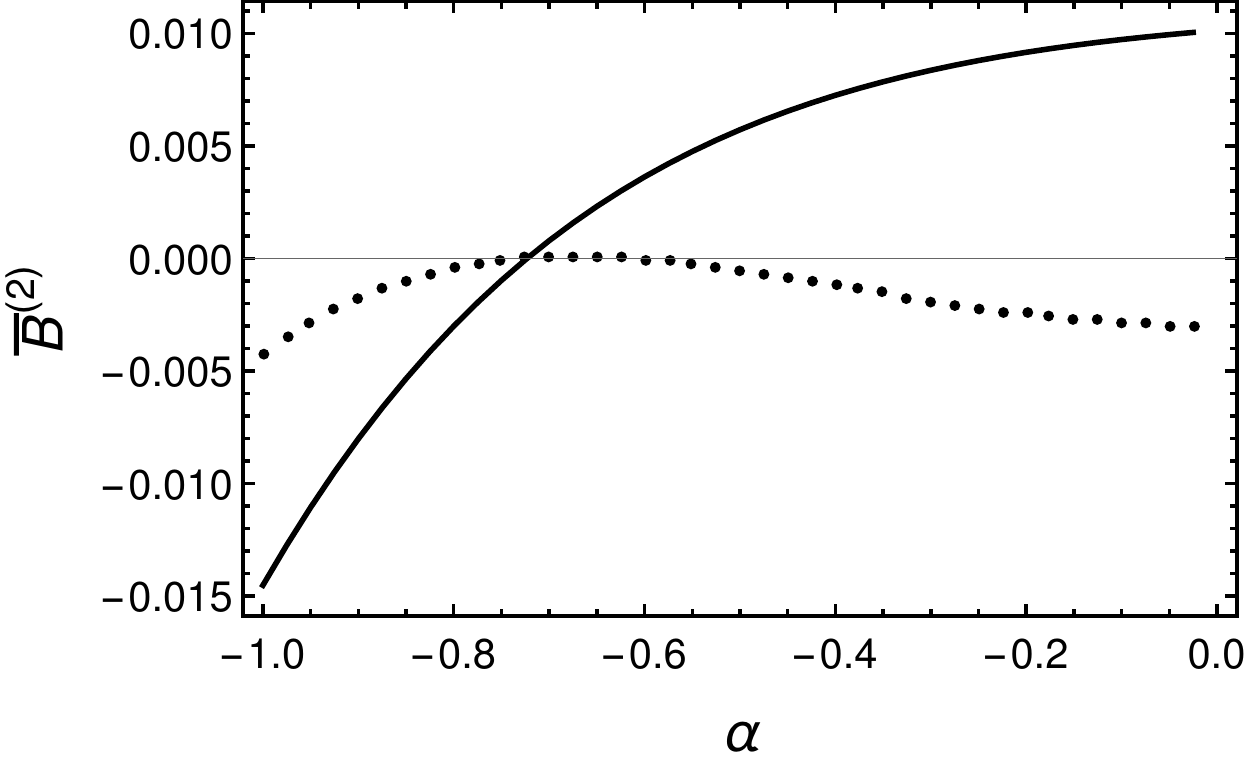}

\caption{Parametric dependencies of $\bar{B}^{(2)}$
when varied about the benchmark point~\eqref{benchmarkpoint}.
Solid: $\bar{B}^{(2)}$ with gradient effects. Dotted:
$\bar{B}^{(2)\rm hom}$ without gradient effects.} 
\label{fig:B2parameter} 
\end{figure}

In Fig.~\ref{fig:B2parameter}, we compare the dependence of $B^{(2)}$
on the various coupling constants
with and without gradient effects. The relative difference between the cases with and without gradient effects is
of order one. When recalling the dependence of $B^{(2)}$ on the tadpole functions given in Eq.~\eqref{B2},
we see that this sizable difference is because of the large relative impact of gradient effects on $\Pi$ due to
the cancellation from fermion and scalar loop effects, cf.~Fig.~\ref{fig:Pi}.
We reiterate, however, that this cancellation is coincidental due to the parameter choices in
Eq.~\eqref{benchmarkpoint}.

%%%%%%%%%%%%%%%%%%%%%%%%%%%%%%%%%%%%%%%%%%%%%%%%%%%%%%%%%%%%%%%%%%%%%%%%%%%%%

\section{Conclusions}
\label{sec:conc}

We have extended the Green's function method
proposed in Ref.~\cite{Garbrecht:2015oea} to fermion fields in order to deal with the radiative effects
on false vacuum decay in Higgs-Yukawa theory. The bounce background provides an inhomogeneous mass term $m_D=\kappa\varphi^{(0)}$ for the fermions via their Yukawa coupling.
We have therefore constructed the Green's functions for the Dirac operator in a Euclidean spacetime-dependent background with ${O}(4)$ spherical symmetry. In the thin-wall approximation, however, we have found that it is possible to reduce the functional determinant of the Dirac operator to functional determinants of two scalar operators of second order in derivatives, which simplifies explicit calculations. We
have then used this procedure to calculate the radiative corrections to the bubble profile and the decay rate numerically in the thin-wall approximation.

The results show that, while the quantum corrections from scalar loops broaden the bubble wall and lead to a faster decay rate, fermion loops, on the other hand, have the opposite effect. These conclusions could have been reached also when ignoring the
gradient effects. However, essential for this work is that we numerically calculate these quantum corrections, fully capturing the gradient effects beyond the Coleman-Weinberg effective potential. In particular, the profile of the radiatively improved bounce solutions are found to be smoother when the gradient effects are included. In the planar-wall approximation, the  corrections to the bounce profile from the gradient effects are small. This is because these corrections are proportional to $(\Pi-\Pi^{\rm hom})\varphi^{(0)}$. While $\Pi-\Pi^{\rm hom}$ peaks at the bubble wall $u=0$, the classical bounce $\varphi^{(0)}$ vanishes there, leading to a suppression of the effect on the decay rates~\cite{Garbrecht:2015oea}. However,
these conclusions reached for the archetypal example of tunneling between quasi-degenerate vacua
in scalar theory with quartic interactions may change when considering phenomenological models leading
to different bubble profiles. One interesting application of this work would therefore be a study of the
gradient effects on top-quark loops in electroweak phase transitions and the decay of the electroweak vacuum.
These tasks require a numerical calculation and renormalization of the fermionic Green's functions
in a spherical background, based on the methods developed in the present work, and further, in the case of a thermal phase transition,
the inclusion of finite-temperature effects.
The differences in the coupling of the scalars and fermion fields to the inhomogeneous background may also be of interest
in the study of solitons in supersymmetric theories.

\begin{acknowledgments}

W.Y.A. is supported by the China Scholarship Council (CSC). The work of P.M. was supported
by the Science and Technology Facilities Council (STFC) under Grant No.~ST/L000393/1 and a Leverhulme Trust 
Research Leadership Award.

\end{acknowledgments}

%%%%%%%%%%%%%%%%%%%%%%%%%%%%%%%%%%%%%%%%%%%%%%%%%%%%%%%%%%%%%%%%%
\appendix

\section{Fermionic Green's function}
\label{app:greenfunction}
In this appendix, we describe in detail the derivation of the spherically symmetric, fermionic Green's function in the presence of an inhomogeneous mass $m(r)\equiv \kappa\varphi^{(0)}(r)$.

\subsection{Angular-momentum recoupling}

We begin by reviewing the recoupling of spin and orbital angular momenta in four dimensions (see, e.g., Refs.~\cite{Biedenharn:1961drs,VanIsacker1991,Schwinger}). We closely follow the particularly lucid explanation presented in Ref.~\cite{VanIsacker1991}, and a discussion comparable to our own can be found in Ref.~\cite{Avan:1985eg} in the context of the fermion determinant.

Rotations on $\mathbb{R}^4$ are generated by the operators
\begin{equation}
M_{\mu\nu}\ =\ -\,i\bigg(x_\mu\frac{\partial}{\partial_\nu}-x_\nu\frac{\partial}{\partial_\mu}\bigg)\;.
\end{equation}
By virtue of the antisymmetry under interchange of the indices $\mu$ and $\nu$, i.e.~$M_{\mu\nu}=-M_{\nu\mu}$, six are independent:
\begin{subequations}
\begin{align}
N_1\ \equiv\ M_{23}\;,\quad N_2\ \equiv\ M_{31}\;,\quad N_3\ \equiv\ M_{12}\;,\\
N_1'\ \equiv\ M_{41}\;,\quad N_2'\ \equiv\ M_{42}\;,\quad N_3'\ \equiv\ M_{43}\;.
\end{align}
\end{subequations}
These form a basis $\{N_i,N_i'\}$ of the Lie algebra $\mathfrak{o}(4)$:
\begin{subequations}
\begin{align}
\big[N_i,N_j\big] \ &= \ i\epsilon_{ijk}N_k\;,\\
\big[N_i,N_j'\big] \ &= \ i\epsilon_{ijk}N_k'\;,\\
\big[N_i',N_j'\big] \ &= \ i\epsilon_{ijk}N_k\;,
\end{align}
\end{subequations}
where $\epsilon_{ijk}$ is the Levi-Civita tensor. 

We can introduce another basis $\{A_i,A_i'\}$:
\begin{equation}
A_i\ \equiv\ \frac{1}{2}\big(N_i+N_i'\big)\;,\qquad A_i'\ \equiv\ \frac{1}{2}\big(N_i-N_i'\big)\;,
\end{equation}
satisfying
\begin{subequations}
\begin{align}
\big[A_i,A_j\big]\ &= \ i\epsilon_{ijk}A_k\;,\\
\big[A_i,A_j'\big]\ &= \ 0\;,\\
\big[A_i',A_j'\big]\ &= \ i\epsilon_{ijk}A_k'\;.
\end{align}
\end{subequations}
The two sets of operators $\mathbf{A}\equiv\{A_i\}$ and $\mathbf{A}'\equiv \{A_i'\}$ form two commuting $\mathfrak{o}(3)$ algebras, reflecting the fact that the $\mathfrak{o}(4)$ algebra is equal to the direct sum of two $\mathfrak{o}(3)$ algebras, i.e.~$\mathfrak{o}(4)=\mathfrak{o}(3)\oplus\mathfrak{o}'(3)$, and, correspondingly, that the $O(4)$ group is locally isomorphic to the direct product of two $O(3)$ groups, i.e.~$O(4)\simeq O(3)\otimes O'(3)$. We index states in the basis $\{N_i,N_i'\}$ by the partition numbers $(p,q)$ and states in the basis $\{A_i,A_i'\}$ by the labels $\{j,j'\}$. These labels are related via
\begin{equation}
p\ =\ j\:+\:j'\;,\qquad q\ =\ j\:-\;j'\;.
\end{equation}
Eigenstates of $\mathbf{A}^2$ and $\mathbf{A}^{\prime 2}$ have eigenvalues $j(j+1)$ and $j'(j'+1)$, respectively. Eigenstates of the total orbital angular momentum operator
\begin{equation}
{\frac{1}{2}}\,M\cdot M\ \equiv\ {\frac{1}{2}}\,M_{\mu\nu}M_{\mu\nu}\ =\ \mathbf{N}^2+\mathbf{N}^{\prime 2} \ =\ 2\big(\mathbf{A}^2+\mathbf{A}^{\prime 2}\big)
\end{equation}
have eigenvalues $p(p+2)+q^2=2[j(j+1)+j'(j'+1)]$ and transform as $\{j/2,j'/2\}$ representations with partition numbers $(\frac{j+j'}{2},\frac{j-j'}{2})$. 

Under spatial reflection: $x_i\rightarrow -x_i$, $x_4\rightarrow x_4$, we have $N_i\rightarrow N_i$, $N_i'\rightarrow -N_i'$ and $A_i\leftrightarrow A_i'$.  The spatial reflection symmetry of $O(4)$ therefore forces $j=j'$ for the total angular momentum eigenstates. Thus, the total orbital angular momentum eigenstates transform as $\{j/2,j/2\}$ representations of $O(4)$ with partition numbers $(j,0)$, and $\frac{1}{2}\,M\cdot M$ has eigenvalues $4j(j+1)$.

In the case of the spin group ${\rm Spin}(4)\simeq SU_L(2) \otimes SU_R(2)$, we have the generators
\begin{equation}
\Sigma_{\mu\nu}\ =\ -\,\frac{i}{4}\,\big[\gamma_{\mu},\gamma_{\nu}\big]\;,
\end{equation}
with the six independent operators
\begin{subequations}
\begin{align}
S_1\ \equiv\ \Sigma_{23}\;,\qquad S_2\ \equiv\ \Sigma_{31}\;,\quad S_3\ \equiv\ \Sigma_{12}\;,\\
S_1'\ \equiv\ \Sigma_{41}\;,\qquad S_2'\ \equiv\ \Sigma_{42}\;,\quad S_3'\ \equiv\ \Sigma_{43}\;,
\end{align}
\end{subequations}
forming a basis of the Lie algebra $\mathfrak{so}(4)$:
\begin{subequations}
\begin{align}
\big[S_i,S_j\big] \ &= \ i\epsilon_{ijk}S_k\;,\\
\big[S_i,S_j'\big] \ &= \ i\epsilon_{ijk}S_k'\;,\\
\big[S_i',S_j'\big] \ &= \ i\epsilon_{ijk}S_k\;.
\end{align}
\end{subequations}
As before, we can introduce another basis $\{S^L_i, S^R_j\}$:
\begin{equation}
S_i^L \ \equiv\ \frac{1}{2}\big(S_i+S_i'\big)\;,\qquad S_i^R \ \equiv\ \frac{1}{2}\big(S_i-S_i'\big)\;,
\end{equation}
satisfying
\begin{subequations}
\begin{align}
\big[S_i^L,S_j^L\big]\ &= \ i\epsilon_{ijk}S_k^L\;,\\
\big[S_i^L,S_j^R\big]\ &= \ 0\;,\\
\big[S_i^R,S_j^R\big]\ &= \ i\epsilon_{ijk}S_k^R\;.
\end{align}
\end{subequations}
The sets of operators $\mathbf{S}^L\equiv\{S^L_i\}$ and $\mathbf{S}^R\equiv\{S^R_i\}$ form two commuting $\mathfrak{su}(2)$ algebras, reflecting the fact that the $\mathfrak{so}(4)$ algebra is equal to the direct sum of two $\mathfrak{su}(2)$ algebras, i.e.~$\mathfrak{so}(4)=\mathfrak{su}_L(2)\oplus\mathfrak{su}_R(2)$, and, correspondingly, that the $SO(4)$ group is locally isomorphic to the direct product of two $SU(2)$ groups, i.e.~$SO(4)\simeq SU_L(2)\otimes SU_R(2)$. The operators $\mathbf{S}^L$ act on the subspace generated by the projector $P_L=\frac{1}{2}(\mathbb{I}_4-\gamma_5)$, and the operators $\mathbf{S}^R$ act on the subspace generated by the projector $P_R=\frac{1}{2}(\mathbb{I}_4+\gamma_5)$. We call spinors that transform in these two subspaces left and right handed, respectively. Left-handed spinors transform as $\{1/2,0\}$ representations and are labeled by the partition numbers $(1/2,+1/2)$; right-handed spinors transform as $\{0,1/2\}$ representations and are labeled by the partition numbers $(1/2,-1/2)$.

In terms of the above generators, we can write the Dirac operator in the form
\begin{align}
\label{Dirac:spherical}
\gamma\cdot\partial_x\:+\:m(r)\ =\ \gamma\cdot\hat{x}\bigg[\hat{x}\cdot\partial_x\:-\:\frac{M\cdot \Sigma}{r}\:+\:\gamma\cdot \hat{x}\,m(r)\bigg]\;,
\end{align}
where $\hat{x}_{\mu}\equiv x_{\mu}/|x|\equiv x_{\mu}/r$. Hence, provided we can find the Green's function $\tilde D$ of the operator in square brackets, the Green's function of the complete Dirac operator will be given by $D=  \gamma\cdot \hat x \tilde D$. Our aim then is to find the eigenstates of $M\cdot \Sigma$, i.e.~the simultaneous eigenstates of $\frac{1}{2}M^2$, $\frac{1}{2}\Sigma^2$ and the total angular momentum $K^2\equiv\frac{1}{2}(M+\Sigma)^2$.

We begin by introducing the following basis of orbital angular momentum states, transforming under $O(4)$:
\begin{align}
\ket{(j,0),\ell,m_{\ell}}\ &=\ \sum_{m_j,m_j'}\big(j/2,m_j,j/2,m_j'\big|\ell,m_{\ell}\big)\nonumber\\&\qquad\times\:\ket{j/2,m_j}\otimes\ket{j/2,m_j'}\;,
\end{align}
where $\big(j/2,m_j,j/2,m_j'\big|\ell,m_{\ell}\big)$ are the Clebsch-Gordan coefficients for the branching $O(3)\supset O(2)$. The quantum number $\ell$ labels the irreducible representations of $O(3)$ and $m_{\ell}$ those of $O(2)$. The basis of spin states, transforming under $SO(4)$, is spanned by
\begin{align}
\ket{(s,\pm s),s,m_s}\ &=\ \sum_{m_s^L,m_s^R}\big(s^L,m_s^L,s^R,m_s^R\big|s,m_s\big)\nonumber\\&\qquad\times\:\ket{s^L,m_s^L}\otimes \ket{s^R,m_s^R}\;,
\end{align}
with $s=1/2$. Since either $s^L=m_s^L=0$ or $s^R=m_s^R=0$, the Clebsch-Gordan coefficients simplify, and we have
\begin{subequations}
\begin{align}
\ket{(s,+s),s,m_s}\ =\ \ket{s,m_s}\otimes\ket{0,0}\;,\\
\ket{(s,-s),s,m_s}\ =\ \ket{0,0}\otimes\ket{s,m_s}\;.
\end{align}
\end{subequations}

Now we consider the eigenstates of $M\cdot\Sigma$. We need to consider the coupling between the representation of $O(4)$ and that of ${\rm Spin}(4)$. We first couple the states transforming under $O(3)$ with those under $SU_L(2)$, leading to the quantum numbers $J$ and $m_J$, and then proceed similarly for $O'(3)$ and $SU_R(2)$, leading to the quantum numbers $J'$ and $m_J'$. Finally, we couple the two resulting states, replacing the quantum numbers $m_J$, $m_J'$ by $L$, $M$. There is a unitary transformation relating these two couplings~\cite{Biedenharn:1961drs}:
\begin{align}
\label{total:angular:momentum:eigenstates}
&\ket{(j,0),(s,\pm s);\{J,J'\},L,M}\nonumber\\&\quad  =\ \sum_{\ell,m_{\ell},m_s}(-i)^{\ell}\big[(2J+1)(2J'+1)(2\ell+1)(2s+1)]^{1/2}\nonumber\\&\quad\qquad \times\:\left\{\begin{matrix} j/2 & (s\pm s)/2 & J\\ j/2 & (s\mp s)/2 & J'\\ \ell & s & L\end{matrix}\right\}\big(\ell,m_{\ell},s,m_s|L,M\big)\nonumber\\&\quad\qquad \times\:\ket{(j,0);\ell,m_{\ell}}\otimes\ket{(s,\pm s);s,m_s}\;,
\end{align}
where the summations run over $\ell = j,j-1,\dots,0$, $m_{\ell} = \ell,\ell-1,\dots,-\ell$ and $m_s=\pm s$. The expression within curly braces is the Wigner $9j$-symbol. We have also introduced a phase $(-i)^{\ell}$ that depends on the explicit representation of the individual product states. The various quantum numbers take values
\begin{subequations}
\begin{align}
J\ &=\ j/2+(s\pm s)/2,|j/2-(s\pm s)/2|\;,\\
J'\ &=\ j/2+(s\mp s)/2, |j/2-(s\mp s)/2|\;,\\
L\ &=\ J+J',J+J'-1,\dots,|J-J'|\;,\\
M\ &=\ L,L-1,\dots, -L\;,
\end{align}
\end{subequations}
and, correspondingly, the partition numbers take values
\begin{subequations}
\begin{align}
P\ &\equiv J+J'=\ j\:+\:s,|j\:-\:s|\;,\\
Q\ &\equiv J-J'=\ \pm\,s,\mp\,s\;.
\end{align}
\end{subequations}
The states have the following eigenvalues of the total angular momentum:
\begin{align}
K^2\ &\rightarrow\ P(P+2)+Q^2\ =\ 2[J(J+1)+J'(J'+1)]\nonumber\\ &=\ \begin{cases} j^2\:+\:3j\:+\:3/2\;,&\quad P\ =\ j\:+\:s\;,\ j\ > \ 0\;,\\ j^2\:+\:j\:-\:1/2\;,&\quad P\ =\ j\:-\:s\;,\ j\ >\ 0\;,\\
3/2\;,&\quad P\ =\ s\;,\ j\ =\ 0\;.\end{cases}
\end{align}
It follows then that
\begin{align}
\label{evs:spinorbit}
&M\cdot \Sigma\ =\ K^2\:-\:{\frac{1}{2}}\,M^2\:-\:{\frac{1}{2}}\,\Sigma^2\nonumber\\&
\rightarrow \ \mathcal{J}\ \equiv\ P(P+2)\:+\:Q^2\:-\:j(j+2)\:-\:2s(s+1)\nonumber\\&
=\ \begin{cases}
j\;, &\quad P\ =\ j\:+\:s\;,\ j\ >\ 0\;,\\
-\,(j+2)\;,&\quad P\ =\ j\:-\:s\;,\ j\ >\ 0\;,\\
0\;,&\quad P\ =\ s\;,\ j\ =\ 0\;.
\end{cases}
\end{align}

In the coordinate representation, the orbital angular momentum states are given by the four-dimensional hyperspherical harmonics
\begin{equation}
Y_{j\ell m_{\ell}}(\mathbf{e}_r)\ =\ \braket{\mathbf{e}_r |(j,0),\ell, m_{\ell}}\;,
\end{equation}
where the four-dimensional unit vector is
\begin{equation}
\mathbf{e}_r\ =\ (\cos\chi,\sin\chi\cos\theta,\sin\chi\sin\theta\sin\varphi,\sin\chi\sin\theta\cos\varphi)\;.
\end{equation}
In terms of these coordinates, the hyperspherical harmonics in four dimensions are given by~\cite{WenAvery:1983}
\begin{align}
\label{Y4D}
Y_{j\ell m_{\ell}}(\mathbf{e}_r)\ &=\ 2^\ell\,\Gamma(\ell+1)\left(\frac{2(j+1)(j-\ell)!}{\pi(j+\ell+1)!}\right)^{\frac12}\notag\\
&\qquad \times\:\sin^\ell(\chi)C^{\ell+1}_{j-\ell}(\cos\chi)Y_{\ell m_{\ell}}(\theta,\varphi)\;,
\end{align}
where $Y_{\ell m_{\ell}}$ are the usual three-dimensional spherical harmonics. The spin states can be written
\begin{equation}
\xi_{\pm,m_s}\ =\ \frac{1}{2}(\mathbb{I}\mp\gamma_5)\begin{pmatrix}\xi_{m_s} \\ \xi_{m_s}\end{pmatrix}\ =\ \braket{\xi |(s,\pm s),s,m_s}\;,
\end{equation}
where we have defined $\xi_{+s}=(1\;0)^{\mathsf{T}}$, $\xi_{-s}=(0\;1)^{\mathsf{T}}$. We can then introduce the {\it spin hyperspherical harmonic}
\begin{align}
&Y^{j,\pm}_{JJ'LM}(\mathbf{e}_r)\ = \ \braket{\xi;\mathbf{e}_r|(j,0),(s,\pm s);\{J,J'\},L,M}\nonumber \\ &=\ \sum_{\ell,m_{\ell},m_s}(-i)^{\ell}\big[(2J+1)(2J'+1)(2\ell+1)(2s+1)]^{1/2}\nonumber\\&\quad \times\:\left\{\begin{matrix} j/2 & (s\pm s)/2 & J\\ j/2 & (s\mp s)/2 & J'\\ \ell & s & L\end{matrix}\right\}\big(\ell,m_{\ell},s,m_s|L,M\big)\nonumber\\&\quad \times\:\xi_{\pm,m_s}Y_{j\ell m_{\ell}}(\mathbf{e}_r)\;.
\end{align}

\subsection{Green's function: spherical problem}

We now turn our attention to the problem of finding the fermionic Green's function for a radially varying mass $m(r)$. The Euclidean-space Dirac equation takes the form
\begin{align}
\big[\gamma\cdot\partial_x\:+\:m(r)\big]D(x,x^\prime)\ =\ \delta^4 (x-x^\prime)\;.
\end{align}
We can proceed by making the following ansatz for the solution:
\begin{equation}
D(x,x^\prime)\ =\ \sum_{\lambda}\big[a_{\lambda}(r,r^\prime)\:+\:b_{\lambda}(r,r^\prime)\,\gamma\cdot x \big]\tilde{D}_{\lambda}(\mathbf{e}_r,\mathbf{e}_r')\;,
\end{equation}
where $\lambda=\{j,\pm,J,J',L,M\}$ is a multi-index and
\begin{equation}
\tilde{D}_{\lambda}(\mathbf{e}_r,\mathbf{e}_r')\ =\ [Y^{j,\pm}_{JJ'LM}(\mathbf{e}_r')]^*Y^{j,\pm}_{JJ'LM}(\mathbf{e}_r)\;.
\end{equation}
Given the completeness of the eigenstates in Eq.~\eqref{total:angular:momentum:eigenstates}, this leads to the equation
\begin{align}
&\gamma\cdot x \bigg[\frac{ x\cdot \partial-{\cal J}}{r^2}\,a_{\lambda}(r,r^\prime)\,+\,m(r) b_{\lambda}(r,r^\prime)\bigg]\tilde{D}_{\lambda}(\mathbf{e}_r,\mathbf{e}_r')
\notag\\
&+\,\bigg[m(r) a_{\lambda}(r,r^\prime)\,+\,\bigg(\frac{\partial}{\partial r}+\frac{{\cal J}+3}{r}\bigg) r\,b_{\lambda}(r,r^\prime)\bigg]\tilde{D}_{\lambda}(\mathbf{e}_r,\mathbf{e}_r')\notag\\
&=\ \frac{\delta(r-r^\prime)}{r^3}\,\tilde{D}_{\lambda}(\mathbf{e}_r,\mathbf{e}_r')\;.
\label{eveq:factorized}
\end{align}
Here, we have also used Eq.~(\ref{Dirac:spherical}) and the relation
\begin{align}
\gamma\cdot \hat{x}\,M\cdot \Sigma\,\gamma \cdot \hat{x}\ =\ -\:M\cdot \Sigma\:-\:3\;.
\end{align}
The radial equation only depends on the orbital angular momentum quantum number $j$ and the partition number $P$.
The two linearly independent components of Eq.~(\ref{eveq:factorized}) allow us to
eliminate $b_{\lambda}$ such that we obtain a second-order equation. Substituting for the allowed values for ${\cal J}$ from Eq.~(\ref{evs:spinorbit}) leads to
\begin{align}
\label{A18}
&\bigg[-\:\frac{\D^2}{\D r^2}\:-\:\frac{3}{r}\,\frac{\D}{\D r}\:+\:\frac{j(j+2)}{r^2}\:+\:m^2(r)\nonumber\\&\quad +\:\frac{\D \ln m(r)}{\D r}\bigg(\frac{\D}{\D r}-\frac{{\cal J}}{r}\bigg)
\bigg]a_{\lambda}(r,r^\prime)\ =\ \frac{m(r)\delta(r-r^\prime)}{r'^3}\;.
\end{align}
Note that, for a constant mass, this coincides with the spherical Klein-Gordon equation, as one would expect. 

We are ultimately interested in the coincident limit of the fermionic Green's function. It is therefore useful to 
evaluate the dyadic product of the eigenstates in Eq.~(\ref{total:angular:momentum:eigenstates}) and to trace
over all but $P$ and $j$. This can be simplified dramatically, since we may take the angle $\chi=0$ without loss
of generality. Doing so, $\ell=0$ in all non-vanishing contributions, and we find the result
\begin{align}
&\sum_{Q,L,M,\pm} |\braket{\mathbf{e}_r|(j,0),(s,\pm s);(P,Q),L,M}|^2\nonumber\\&\qquad =\ \frac{1}{4\pi^2}\,(j+1)\,\mathcal{K}\,\mathbb{I}_4\;,
\end{align}
where
\begin{equation}
\mathcal{K}\ =\ \begin{cases} j\:+\:2\;,&\quad P\ =\ j\:+\:s\;,\ j\ >\ 0\;,\\
 j\;,&\quad P\ =\ j\:-\:s\;,\ j\ >\ 0\;,\\
2\;,&\quad P\ =\ s\;,\ j\ =\ 0\;.
\end{cases}
\end{equation}
Note that summing over the allowed values for $P$, we
recover the corresponding result from Ref.~\cite{Garbrecht:2015oea}, appearing in the coincident Green's function
of the scalar field. Putting everything together, we arrive at the following expression for the coincident fermionic Green's function:
\begin{align}
\label{coincidentgreen}
 D(x,x)\ =\ \frac{1}{4\pi^2}\displaystyle\sum_{P,j} (j+1)\,\mathcal{K}\bigg[\mathbb{I}_4\:+\:\frac{\mathcal{J}-x\cdot\partial}{m r}\,\gamma\cdot \hat{x}\bigg] a_j(r)\;,
\end{align}
where $a_j(r)\equiv a_{\lambda}(r,r)$. 

\subsection{Green's function: planar problem}

In the thin-wall limit, we may apply the planar-wall approximation, neglecting the damping term and replacing $j/R$ by the continuous variable $|{\bf k}|$ (the three-momentum in the hyperplane of the bubble wall). In addition, we consistently replace $\delta(r-r')/r^{\prime 3}$ by $\delta(r-r')/R^3$ on the right-hand side of the radial equation~\eqref{eveq:factorized}. Doing so, Eq.~\eqref{A18} becomes
\begin{align}
&\left[-\partial^2_z+{\bf k}^2+m^2(z)+\:\frac{\D \ln m(z)}{\D z}\left(\partial_z-h\,|\mathbf{k}|\right)
\right]a_{h}({\bf k};z,z^\prime)\notag\\
&=\ \frac{m(z)\,\delta(z-z^\prime)}{R^3}\;,
\end{align}
where we have replaced $r$ by the variable $z$ and $h=\pm$, coming from the partition number $P$ in Eq.~\eqref{evs:spinorbit}. Defining
 $\tilde{a}_{h}({\bf k};z,z^\prime)\equiv R^3\,a_{h}({\bf k};z,z^\prime)$, we arrive at the final form
\begin{align}
\label{hyperradialfermionic}
&\left[-\partial^2_z+{\bf k}^2+m^2(z)+\:\frac{\D \ln m(z)}{\D z}\left(\partial_z-h\,|{\bf k}|\right)
\right]\tilde{a}_{h}({\bf k};z,z^\prime)\notag\\
&=\ m(z)\,\delta(z-z^\prime)\;.
\end{align}

Substituting Eq.~\eqref{coincidentgreen} into Eq.~\eqref{tadpole:fermi} with $a_{\lambda}(r,r')$ replaced by $a_h(\mathbf{k};z,z')$ and employing the representations of the gamma matrices in Eq.~\eqref{gamma}, the fermion tadpole can be written as
\begin{align}
\label{eq:fermiontadatilde}
 \,\Pi_D(\varphi^{(0)};x)\,\varphi^{(0)}(x) &=\ -\,\kappa\:\mbox{tr}_{{\rm S}}\,D(\varphi^{(0)};x,x)\notag\\
 &=\ -\:\frac{\kappa}{\pi^2}\displaystyle\sum_{h\,=\,\pm}\int \D |\mathbf{k}|\,\mathbf{k}^2
 \,\tilde{a}_{h}({\bf k};z,z^\prime)\;.
\end{align}
The numerical results from the main part of this work have been checked
by simultaneously solving Eq.~\eqref{hyperradialfermionic} and using the expression for the fermion tadpole in Eq.~\eqref{eq:fermiontadatilde} to compute the radiative effects.

\section{Fermion contributions to the tadpole from the one-loop effective action}
\label{app:fermitadpole}

From the one-loop correction~\eqref{B1:Dirac} to the effective action,
the fermion contribution to the tadpole corrections can be derived as
\begin{align}
\label{PiD:functional}
\Pi_{D}(\varphi^{(0)};x)\, \varphi^{(0)}(x)\ &=\ \frac{\delta B^{(1)}_D[\varphi]}{\delta\varphi_x}\bigg|_{\varphi^{(0)}}\nonumber\\ &=\ -\:\frac{\delta}{\delta\varphi_x}{\rm tr}\ln D^{-1}(\varphi)\bigg|_{\varphi^{(0)}}\notag\\
&=\ -\:D^{\alpha\beta}_{yz}(\varphi^{(0)})\,\frac{\delta}{\delta\varphi_x}\,{D_{zy}^{-1}(\varphi)}^{\beta\alpha}\bigg|_{\varphi^{(0)}}
\nonumber\\&=\ -\:D^{\alpha\beta}_{yz}(\varphi^{(0)})\,\kappa\,\delta_{zx}\,\delta_{zy}\,
\delta^{\beta\alpha}\notag\\
&=\ -\:\kappa\, {\rm tr}_{\mbox{s}}\, D_{xx}(\varphi^{(0)})\;.
\end{align}
Herein, $\alpha$, $\beta$ represent the spinor indices; $x$, $y$, $z$ represent the spacetime coordinates. In addition, the Kronecker symbol on the spacetime coordinates should be understood as the four-dimensional Dirac delta function. We have used Eq.~\eqref{operator:Dirac:twopoint}, and the Einstein and DeWitt conventions for repeated indices and coordinates throughout.

%%%%%%%%%%%%%%%%%%%%%%%%%%%%%%%%%%%%%%%%%%%%%%%%%%%%%%%%%%%%%%%%%%%%%%

%%%%%%%%%%%%%%%%%%%%%%%%%%%%%%%%%%%%%%%%%%%%%%%%%%%%%%%%%%%%%%%%%%%%%%

%%%%%%%%%%%%%%%%%%%%%%%%%%%%%%%%%%%%%%%%%%%%%%%%%%%%%%%%%%%%%%

\end{document}